\documentclass[showpacs,aps,pra,twocolumn,superscriptaddress]{revtex4-2}

\usepackage[english]{babel}
\usepackage{graphicx}
\usepackage{comment}
\usepackage{hyperref}
\usepackage{epsfig}
\usepackage{color}
\usepackage{xcolor}
\usepackage{physics}
\usepackage{amssymb}
\usepackage{amsmath}
\usepackage{mathtools}
\usepackage{stmaryrd}
\usepackage{array}
\usepackage{soul}
\usepackage{flushend}
\usepackage{enumitem}

\hypersetup{
	colorlinks,
	linkcolor={red!95!black},
	citecolor={green!60!black},
	urlcolor={blue!95!black}
}

\newcommand{\Hef}{H_{\mbox{\scriptsize{eff}}}}
\newcommand{\orb}{\mbox{\scriptsize{orb}}}

\newcommand{\Hhop}{H_{\mbox{\scriptsize{hop}}}}
\newcommand{\Hint}{H_{\mbox{\scriptsize{int}}}}
\newcommand{\qone}{{q_1}}
\newcommand{\qtwo}{{q_2}}
\newcommand{\Lconti}{{\ell}}
\DeclareMathOperator\erfi{erfi}

\newcommand{\dc}{{\delta^{*}}}
\newcolumntype{P}[1]{>{\hspace{0pt}}m{#1}}
\newcolumntype{M}[1]{>{\centering\arraybackslash}m{#1}}
\begin{document}
%
\title{Universality and two-body losses: lessons from the effective non-Hermitian dynamics of two particles}
%
%
\author{Alice Marché}
\affiliation{Université Paris-Saclay, CNRS, LPTMS, 91405, Orsay, France}
\author{Hironobu Yoshida}
\affiliation{Department of Physics, Graduate School of Science, The University of Tokyo, 7-3-1 Hongo, Tokyo 113-0033, Japan}
\author{Alberto Nardin}
\affiliation{Université Paris-Saclay, CNRS, LPTMS, 91405, Orsay, France}
\author{Hosho Katsura}
\affiliation{Department of Physics, Graduate School of Science, The University of Tokyo, 7-3-1 Hongo, Tokyo 113-0033, Japan}
\affiliation{Institute for Physics of Intelligence, The University of Tokyo, 7-3-1 Hongo, Tokyo 113-0033, Japan}
\affiliation{Trans-scale Quantum Science Institute, The University of Tokyo, 7-3-1, Hongo, Tokyo 113-0033, Japan}
\author{Leonardo Mazza}
\affiliation{Université Paris-Saclay, CNRS, LPTMS, 91405, Orsay, France}
\date{\today}
%
\begin{abstract}
We study the late-time dynamics of two particles 
confined in one spatial dimension and subject to two-body losses.
The dynamics is exactly described by a non-Hermitian Hamiltonian that can be analytically studied both in the continuum and on a lattice.
The asymptotic decay rate and the universal power-law form of the decay of the number of particles are exactly computed in the whole parameter space of the problem.
When in the initial state the two particles are far apart, the average number of particles in the setup decays with time $t$ as $t^{-1/2}$; 
a different power law, $t^{-3/2}$, is found when the two particles overlap in the initial state.
These results are valid both in the continuum and on a lattice, but in the latter case a logarithmic correction appears.

\end{abstract}
%
\maketitle 
%
%
\section{Introduction}
\label{Intro}
Universality is a key notion in physics related to the possibility of performing the transition from the microscopic to the macroscopic scale by systematically discarding irrelevant information~\cite{Hohenberg_1977, Goldenfeld_1992, ZinnJustin_2002}.
In recent years significant interest has been sparked by the discovery of emergent universal behaviors in open quantum many-body systems~\cite{Sieberer_2016, sieberer2023universality}.
Among the many setups where they have been identified, we list here 
polariton condensates~\cite{altman_2015, Fontaine_2022}, lossy gases~\cite{Barontini_2013}, quantum reactive systems~\cite{perfetto_2023,  perfetto_2023bis, rowlands2023quantum},
 Rydberg atoms~\cite{buchhold_2017, Wintermantel_2021} or quantum spin chains~\cite{cai2013algebraic, begg2024quantum}.
Our focus here is on two-body losses, one of the most investigated mechanisms that can induce a many-body correlated open quantum dynamics in a bosonic or fermionic degenerate gas~\cite{Durr_2009, GarciaRipoll_2009, Baur_2010, FossFeig_2012, Grisins_2016, Johnson_2017, Yamamoto1_2019, Goto_2020, Bouchoule_2020b, Rossini_2020, Nakagawa_2020, Booker_2020, Bouchoule_2021, Bouchoule_2021PRA, Rosso_2021, Nakagawa_2021, Yamamoto2_2021, iskin2021non, Rosso_2022, Yamamoto3_2022, wang2022complex, Yoshida_2023, Huang_2023, Mazza_2023, Hamanaka_2023,  nagao2023semiclassical, xiao2023local, tajima2023non, maki2024loss}, and that is also of particular experimental relevance in the framework of ultra-cold gases~\cite{Syassen_2008, Yan_2013, Zhu_2014, Rauer_2016, Tomita_2017, Schemmer_2018, Sponselee_2018, Tomita_2019, Bouchoule_Schemmer_2020, Honda_2023}. 
Additionally, some articles have highlighted potential metrological and quantum information applications~\cite{FossFeig_2012, Rosso_2022}.
So far, theoretical studies have focused on two universal features: 
(i) the scaling of the asymptotic exponential decay rate (ADR) with the linear system size, $\lambda_{\rm ADR} \sim L^{-\alpha}$, and 
(ii) the late-time decay of the gas density, $n(t) \sim t^{-\beta}$.
The calculation of the exponent $\alpha$ benefits from several simplifications that are inherent to lossy gases, and that have allowed for studies in wide ranges of parameters~\cite{Nakagawa_2021}. 
On the other hand, the exponent $\beta$ has proven more difficult to compute, and it has been characterized only under certain restrictive assumptions. 
In the bosonic context, for instance, analytical expressions have been derived only for weakly-interacting gases with adiabatic losses, or for hard-core gases~\cite{Rossini_2020, Yoshida_2023, Rosso_2023, Huang_2023, gerbino2023largescale}; remarkably, it has been possible to perform the study also in the presence of a harmonic confinement~\cite{Rosso_2021bis, Riggio_2024, gerbino2023largescale, maki2024loss}.
However, identifying the value of $\beta$ for generic values of interparticle interactions or for arbitrary loss rates remains an open and challenging task.
In this article, we address the problem of computing $\alpha$ and $\beta$ for generic model parameters
by considering a simplified initial condition, namely that of a quantum system consisting of just two particles.
In this situation, only one two-body loss process can take place and this allows for important technical simplifications~\cite{wang2022complex}. 
Indeed, for open quantum systems, it is always possible to consider the \textit{no-click dynamics} by performing a postselection, so that the Lindblad master equation maps onto a simpler non-Hermitian dynamics~\cite{Ashida_2020}. In the situation that we are considering, the number of particles can be deduced exactly from this effective non-Hermitian dynamics~\cite{Yoshida_2023}.
Since we consider local losses, in order to have a nontrivial loss dynamics we need to consider an orbital wavefunction that is symmetric under exchange, so that two particles can occupy the very same position in space.
Notice that such a state can be realized both with bosons (be they spinless or spinful, in which case the spin part of the wavefunction needs to be symmetric under spin exchange) or spin-$1/2$ fermions (provided the spin part of the wavefunction is a singlet).
We will discuss two spatial geometries in which the two atoms are confined: a translationally invariant one-dimensional (1D) tube and a uniform tight-binding chain with a single energy level per spatial site.
In both cases, we revisit the known results for the ADR exponent $\alpha$ and compute analytically the value of the density power-law exponent $\beta$ for different initial conditions.
Our approach is based on an effective ``linearization" of the eigenvalue equations of the non-Hermitian Hamiltonian around the most long-lived modes in the limit $L \to \infty$. 
This allows us to describe correctly the late-time dynamics of the gas in the whole space of parameters, also for intermediate interaction strengths or loss rates.
We present numerical exact results that are well described by our analytical approximation, that exhibits a universal scaling collapse.
The approach has the additional merit of highlighting the link existing between the exponents $\alpha$ and $\beta$, which has been overlooked so far.
In the continuum limit, we compute the value $\alpha = 3$;
on a lattice, not only we find a set of modes with $\alpha = 3$, but also additional branches characterized by $\alpha = 5$.
The value of the exponent $\beta$ depends non-trivially on the initial conditions.
We consider the case of two overlapping particles, for which $\beta=3/2$, and the case of two well-separated particles, for which $\beta=1/2$; 
we will also demonstrate that by changing initial conditions we can create situations where the dynamics of the mean number of particles crosses over between these two algebraic decays; no other exponents $\beta$ appear in the problem.
These universal results are valid for the entire parameter space and provide clear scaling functions that allow one to collapse the data onto universal curves.
The branches characterized by $\alpha=5$ are responsible for the appearance of a logarithmic correction in the lattice.
It is important to observe that it is difficult to identify the logarithmic correction by pure numerical means, as in the situations that we studied it could be interpreted as a different and non-universal scaling exponent $\beta'$.
The possibility of studying the problem in a limit where analytical calculations could be performed has the merit of clarifying the difficulties associated with numerical studies of $\beta$ in lattice setups.
These analytical insights result from the simplified initial condition with two particles.
We believe, however, that the linearization technique that we introduce here, based on the approximation of the dynamics around the most long-lived modes, can be employed in an appropriate form also for the more demanding situations in which a many-particle gas at finite density is initially prepared.
One such technique would allow to investigate whether the results presented in this article capture the dynamics also in finite-density situations when the time is sufficiently late for the system to be close to the stationary state.
The article is organized as follows.
In Sec.~\ref{Sec:Models:New}
we introduce the models for a continuum setup and on a lattice, and
we give an overview of the mapping to a non-Hermitian Hamiltonian that we employ in the study of the dynamics.
In Sec.~\ref{SecSummary}, we present a brief summary of our results and describe the general behavior of the lossy dynamics. 
In Sec.~\ref{SecSpectrum}, we discuss the spectrum of the non-Hermitian Hamiltonian and the linearization procedure. In Sec.~\ref{SecScaling}, we derive universal expressions for the mean number of particles at late times; we comment also on the appearance of the quantum Zeno effect in this system.
The conclusions and perspectives of our results are presented in Sec.~\ref{Conclusion}.
The article is concluded by nine appendices.

\section{Models and effective non-Hermitian dynamics}\label{Sec:Models:New}

We begin by presenting the two models (in the continuum and on a lattice) that will be studied in the article;
we will only consider the case of local losses, meaning that a loss event can occur only if the two particles have a non-zero probability of being at the same position. 
Our focus is on initial states with two identical spinless bosons and in the final part of this section, we describe the exact mapping of its Lindblad master-equation time-evolution onto a non-Hermitian dynamics.

\subsection{Two spinless bosons in the continuum}\label{sec:twoBosonsCont}

We consider a one-dimensional continuous system of length $\Lconti$ (we assume periodic boundary conditions) populated by bosons with contact interactions described by the Lieb-Liniger model. 
We denote by $M$ the mass of the bosons and by $\sigma$ the typical spatial extension of their individual initial wavefunctions;
the typical initial energy of each boson is $E_0 = \hbar^2/(2 M \sigma^2)$. 
In the following, lengths are measured in units of~$\sigma$, energies in units of~$E_0$, and times in units of~$\hbar/E_0$. 

Let $\Psi^\dag(x)$ and $\Psi(x)$ be the creation and annihilation operators of a boson at position~$x$. In this case, the Lindblad master equation~\cite{Breuer_2007} for the density matrix of the system $\rho$ reads~\cite{Durr_2009,Bouchoule_2020b}
\begin{equation}
    \frac{d \rho}{dt} = - i [ H, \rho] + \int_0^\Lconti \Bigl[ L(x)  \rho L^\dag(x) - \frac{1}{2} \{ L^\dag(x) L(x) , \rho \} \Bigr] dx, \label{MEconti}
\end{equation}
where $L(x) = \sqrt{\frac{\gamma}{2}} \Psi(x)^2$ are the jump operators modeling the local two-body losses and $\gamma > 0$ is the loss rate. 
The Lieb-Liniger Hamiltonian $H$ is defined by~\cite{LiebLiniger1963,Franchini_2017}
\begin{equation}
    H = \int_0^\Lconti \Psi^\dag(x) \left( - \frac{\partial^2}{\partial x ^2} + \frac{g}{2} \Psi^\dag(x) \Psi(x) \right) \Psi(x) dx,
    \label{Hconti}
\end{equation}
where $g >0$ is the repulsive interaction strength; 
this Hamiltonian can be diagonalized by the Bethe ansatz \cite{Gaudin, korepin1997quantum, Zvonarev_2010, vsamaj2013introduction}.
The number operator in this context reads $N = \int_0^\Lconti \Psi^\dag(x) \Psi(x) dx$. 

\subsection{Two spinless bosons on a lattice} \label{sec:twoBosonsLatt}

We will also consider a one-dimensional lattice of $L$ sites with periodic boundary conditions populated by bosons that can hop between neighboring lattice sites and feature contact interaction; this system is described by the Bose-Hubbard model. 
We set the lattice spacing $a$ to $1$. We denote $b_{j}^{\dag}$ and $b_{j}$ the creation and annihilation operators of a boson at site $j \in \{ 1 , \ldots ,L \}$. In the following, energies are measured in units of $J$, the hopping amplitude between neighboring sites, and times are measured in units of $\hbar/J$. In this case, the master equation reads~\cite{GarciaRipoll_2009,Rossini_2020}
\begin{equation}
    \frac{d \rho}{ dt} = - i \left[ H , \rho\right] + \sum_{j = 1}^L \left( L_j \rho L_j^\dag - \frac{1}{2} \{ L_j^\dag L_j , \rho \} \right). \label{MElattice}
\end{equation}
Here  $H$ is  the single-band Bose-Hubbard Hamiltonian, $
    H = \Hhop + \Hint $,
which comprises a hopping term and a local repulsive interaction with strength $U >0$
\begin{equation}
    \Hhop = - \sum_{j =1}^L \left( b_{j+1}^{\dag} b_{j} +  h.c. \right) ,
    \quad 
    \Hint = \frac{U}{2} \sum_{j = 1}^L b_{j}^{\dag 2} b_{j}^2. \label{Hintlattice}
\end{equation}
On the other hand, $L_j$ is a jump operator modeling a two-body loss event at site~$j$,
$
L_j = \sqrt{\frac{\gamma}{2}} b_{j}^2 \label{Llattice}
$.
The operator counting the total number of particles is $N = \sum_{j=1}^L  n_{j}$ where $n_{j} = b_{j}^{\dag} b_{j}$ counts the particles at site $j$. 
We note in passing that, unlike the Lieb-Liniger model, the Bose-Hubbard model is no longer solvable via the Bethe ansatz for $N>2$ \cite{choy1982failure}.

\subsection{Remarks on the fermionic problem}

Dealing with two bosonic particles forces the wavefunction of the system written in first quantization to be symmetric under exchange of particles.
Two-fermion states are characterized by a symmetric orbital wavefunction under the condition that the spin wavefunction is antisymmetric under exchange.
In the case of spin-$1/2$ fermions, this is what happens when we consider the two particles in a singlet state.
As we further elaborate in Appendix~\ref{Ap0}, the results that we are going to derive in the rest of the paper also apply to fermionic problems with two particles in a singlet state and local two-body losses. We note in passing that some of our results also apply to the two-point correlation function in the tight-binding chain of spinless fermions with dephasing \cite{medvedyeva2016exact, alba2023free}.

\subsection{Effective non-Hermitian dynamics} \label{Sec:EffDyn}

In general terms,
the Lindblad master equations that we introduced in Eqs.~\eqref{MEconti} and~\eqref{MElattice} can be written using $\mathcal L$, the Liouvillian superoperator in Lindblad form acting on the
density matrix of the system $\rho$:
\begin{equation}
    \frac{d \rho}{ dt} = \mathcal{L}[\rho] ,\qquad \mathcal{L} = \mathcal{K} + \mathcal{J}. \label{GeneME}
\end{equation}
We divide $\mathcal{L}$ into two superoperators $\mathcal{K}$ and $\mathcal{J}$: the former conserves the number of particles whereas the latter decreases it by enforcing the two-body loss. 
The action of $\mathcal{K}$ can be written by introducing an effective non-hermitian Hamiltonian~$\Hef$ 
\begin{equation}
    \mathcal{K}[\rho] = - i \left( \Hef \rho - \rho \Hef^\dag \right), \label{GeneK}
\end{equation}
which commutes with $N$, the total number operator, $[\Hef , N] = 0$.

In the continuum case,
by comparing Eqs.~(\ref{MEconti}) and (\ref{Hconti}) with Eqs.~(\ref{GeneME}) and (\ref{GeneK}), we identify
\begin{equation}
    \Hef = \int_0^\Lconti \Psi^\dag(x) \left( - \frac{\partial^2}{\partial x ^2} + c \Psi^\dag(x) \Psi(x) \right) \Psi(x) dx
    \label{Hefconti}
\end{equation}
and $\mathcal{J[\rho]} = \int_0^\Lconti L(x) \rho L^\dag(x) dx$.
The effective Hamiltonian $\Hef$ is a Lieb-Liniger Hamiltonian with a complex two-body interaction strength $c = g/2 - i \gamma/4$.
Analogously, in the lattice case we identify
\begin{equation}
    \Hef = H - \frac{i}{2} \sum_{j=1}^L L_j^\dag L_j = \Hhop + 2u \sum_{j = 1}^L b_{j}^{\dag 2} b_{j}^2
\end{equation}
and $ \mathcal{J}[\rho] = \sum_{j=1}^L L_j \rho L_j^\dag$.
The effective Hamiltonian is a single-band Bose-Hubbard Hamiltonian with an effective complex on-site interaction parameter $2 u = U/2-i\gamma/4$.
In both the lattice and continuum cases, we observe that $\mathcal J$ satisfies:
\begin{equation}
      N \mathcal J [\rho] = \mathcal J[ N \rho ] - 2 \mathcal J[\rho].
\end{equation}
When $\rho$ has a well-defined number of particles, this equality can be interpreted as a mathematical statement of the fact that $\mathcal J$ decreases $N$ by two.

Let us now focus on the dynamics of the mean number of particles in the system 
\begin{equation}
\mathcal{N}(t) = \Tr[\rho(t) N].
\end{equation}
We assume that initially the system is prepared in a pure two-particle state $\ket{\Psi_0}$: 
only one loss event can take place and once it occurs the system is described by the vacuum, a state that contributes trivially to the observable of our interest. Therefore, $\mathcal{N}(t)$ can be computed via the non-Hermitian and number-conserving dynamics of $\Hef$ (see Ref.~\cite{Yoshida_2023} and Appendix~\ref{sec:derivation_particle_number} for a proof): 
\begin{equation}
 \mathcal{N}(t) = 2 \braket{\Psi(t)},  \qquad \ket{\Psi(t)} = e^{- i \Hef t} \ket{\Psi_0} .\label{eqN2}
\end{equation}
The effective non-Hermitian Hamiltonian $\Hef$ is responsible for reducing the squared norm of $\ket{\Psi(t)}$ in Eq.~\eqref{eqN2}, and thus gives a nontrivial time dependence to the average particle number, even though it commutes with the total particle number, i.e., $[H_{\rm eff},N]=0$. 
On the other hand, as we anticipated, the operator $\mathcal{J}$ enforces a two-body loss event since the jump operators satisfy $[N, L(x)] = -2L(x)$ in the continuum and $[N,L_j] = -2 L_j$ in the lattice setup.

The dynamics of $\mathcal N(t)$ depends on the spectral properties of $\Hef$.
Since we are focusing on models with local contact interactions or local losses, we can label the eigenstates with two quasimomenta $k_1$ and $k_2$; however, since $H_{\rm eff}$ is non-Hermitian, these quasimomenta are complex.
We denote $\ket{\Psi^{R}_{k_1 k_2}}$ and $\ket{\Psi^{L}_{k_1 k_2}}$ the right and the left eigenvectors of $\Hef$, respectively (see Ref.~\cite{Ashida_2020} for a broad overview of non-Hermitian Hamiltonians):
\begin{subequations}
\begin{align}
\Hef \ket{\Psi_{k_1 k_2}^R} =& E_{k_1 k_2} \ket{\Psi_{k_1 k_2}^R} , \label{eigeneqR} \\
 \Hef^\dag \ket{\Psi_{k_1 k_2}^L} =& E_{k_1 k_2}^* \ket{\Psi_{k_1 k_2}^L}. \label{eigeneqL}
\end{align}
\end{subequations}
The eigenenergies are complex and will be parametrized as:
\begin{equation}
E_{k_1 k_2} = \epsilon_{k_1 k_2} -\frac{i}{2} \Gamma_{k_1 k_2}
\label{Eq:Notation:Energy}
\end{equation}
with $\epsilon_{k_1 k_2}$ and $\Gamma_{k_1 k_2}$ real-valued. 
In our model, which includes only lossy events and no gain, $\Gamma_{k_1,k_2}\geq 0$.

We remark here that in the rest of the article we will also make use of the alternative parametrization in terms of the total momentum $K = k_1 + k_2$ and of the relative quasimomentum $\frac \delta 2 = \frac{k_1 - k_2}{2}$ \cite{essler_2005, zhi1991pseudospin, zhang2021eta, Yoshida_2023}. Here $K$ will be real because we are considering Hamiltonians $\Hef$ that are translationally invariant; it is expected to take the form $2 \pi \mathbb{Z} / \ell$ in the continuum and $2 \pi p /L$ on the lattice, with $p \in \{ 0 , \cdots , L-1 \}$.

With a few algebraic passages that make use of the completeness relation $ \sum_{(k_1,k_2)} \ket{\Psi^{L}_{k_1 k_2}} 
      \hspace{-0.08cm}
      \bra{\Psi^{R}_{k_1 k_2}} = \mathbb{I} 
$, we obtain    
 \begin{equation}
               \mathcal{N}(t)  = \sum_{(k_1,k_2)} \sum_{(\qone,\qtwo) }   f\left( k_1,k_2,\qone,\qtwo \right) 
               e^{-i \left( E_{\qone \qtwo} - E_{k_1 k_2}^* \right) t } 
 \label{Evol}
\end{equation}
with
\begin{equation}
    f \left( k_1,k_2,\qone,\qtwo \right) = 2  \bra{\Psi_0} \ket{\Psi^{L}_{k_1 k_2}} \bra{\Psi^{R}_{k_1 k_2}} \ket{\Psi^{R}_{\qone \qtwo}}  \bra{\Psi^{L}_{\qone \qtwo}} \ket{\Psi_0}  .
  \label{eqf}
\end{equation}
This equation cannot be further simplified because, unlike in Hermitian systems, the right eigenvectors are not orthogonal.

In the long-time limit, the expression~\eqref{Evol} can be further simplified: complex exponentials dephase in time and one can simply retain a single sum: 
\begin{equation}
               \mathcal{N}(t)  \simeq 
               \sum_{(k_1,k_2)} \tilde{f}\left( k_1,k_2 \right) e^{- \Gamma_{k_1 k_2}t } ,
 \label{Evol:2}
\end{equation}
where we have introduced the notations $\tilde{f}\left( k_1,k_2 \right) = f \left( k_1,k_2,k_1,k_2 \right)$.
The decay of the particles is the sum of many exponential decays, 
and the most long-lived states correspond to the energy eigenvalues of $\Hef$ with the smallest absolute value of the imaginary part, i.e.,~closest to zero; this is the definition of the ADR: 
\begin{equation}
    \lambda_{\rm ADR} =  \min_{\substack{k_1, k_2 \\ \text{such that }
    \Gamma_{k_1 k_2} \neq 0}} \frac{\Gamma _{k_1 k_2}}{2}. 
    \label{Eq:lambdaADR:Def}
\end{equation}
The ADR is often called the Liouvillian gap in the context of master equations. Our goal is to show that there is an intermediate time regime where the decaying exponentials conjure an effective universal power-law decay susceptible to a scaling collapse for all parameters.

\section{Summary of results}
\label{SecSummary}

Before entering into the details of our calculations, we present here an overview of our results. 

First of all, in this article we revisit the study of the spectral properties of the non-Hermitian Hamiltonian $\Hef$, which has complex eigenvalues with negative imaginary part. 
Eigenvalues can be characterized by two quantum numbers, and for this discussion the best choice is to consider the total momentum and the relative quasimomentum of the two particles. 

\begin{table}[t]
\begingroup
\renewcommand{\arraystretch}{2}
\begin{tabular}{l||c|c|c}
   & continuum, $\forall K$ & lattice, $K \sim 0 \;$ &  lattice, $K \sim \pi \;$\\
   \hline \hline
  $\lambda_{\rm ADR}$ & $\ell^{-3}$&   $L^{-3}$ & $L^{-5} $\\
  $\alpha$ & 3 & 3 & 5
\end{tabular}
\endgroup
\caption{
Universal scaling of the asymptotic decay rate $\lambda_{\rm ADR} $ with the system size in the continuum $\sim \ell^{- \alpha}$ or on a lattice $L^{- \alpha}$ in the symmetry sector with total momentum $K$; on a lattice, the scaling depends on whether $K$ is close to $0$ or $\pi$.}
\label{Table:2}
\end{table}

In the continuum limit, for any value of $K$ we find states with $\delta \sim 0$ which have an imaginary part of the energy, $-\Gamma_{K, \delta}/2$, scaling as $\ell^{-3}$.
In the lattice, depending on the value of $K$, two scalings appear:
for $K \sim 0$ 
the scaling is $L^{-3}$,
for $K \sim \pi$ 
the scaling is $L^{-5}$; this is summarized in Table~\ref{Table:2}.
To the best of our knowledge, the existence of a $\lambda_{\rm ADR}$ in the lattice scaling as $L^{-5}$ has never been observed before. 
In Appendix~\ref{sec:many_body} we show that this scaling persists also for larger numbers of particles, $N>2$, in the context of the non-Hermitian fermionic Hubbard model, that can be studied using the techniques of integrability~\cite{Nakagawa_2021}.

\begin{figure}[t]
\includegraphics[width=1.0\linewidth]{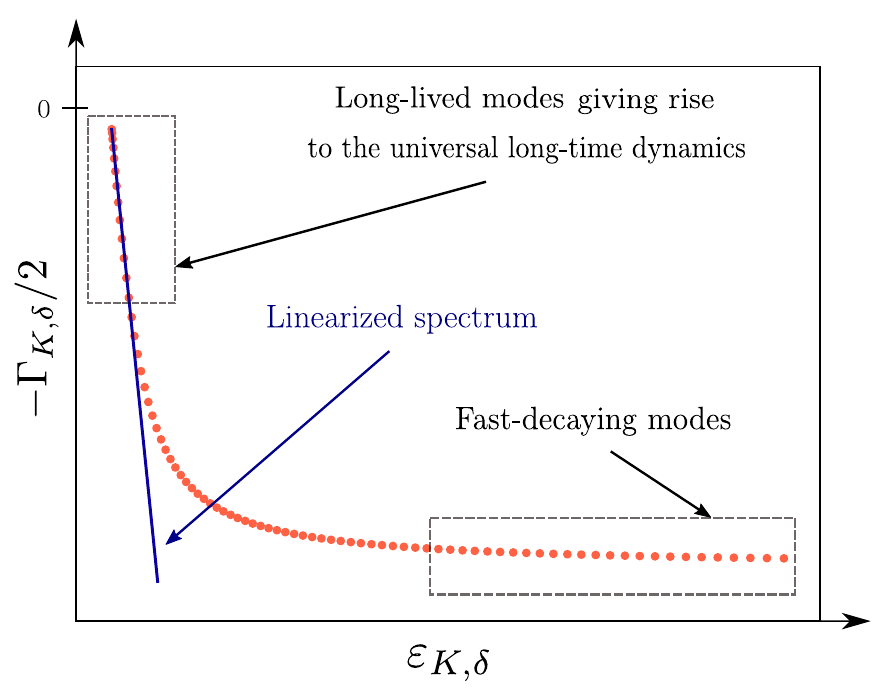}
\caption{Sketch of the spectrum of $\Hef$ plotted in the complex plane, $E_{K,\delta} = \varepsilon_{K,\delta}- i \Gamma_{K, \delta}/2$. 
We consider the continuum Hamiltonian~\eqref{Hefconti} and fix $K=0$.
The long-lived modes correspond to the smallest $\Gamma_{K=0, \delta}$ while the fast-decaying modes correspond to the largest $\Gamma_{K=0, \delta}$. The linearized spectrum drawn with a dark blue line captures well the long-lived modes.}
\label{fig_spec}
\end{figure}

The general form of the spectrum is sketched in Fig.~\ref{fig_spec}, where we plot a branch of complex eigenvalues $E_{K,\delta}$ for the continuum non-Hermitian Hamiltonian in Eq.~\eqref{Hefconti}.
The sketch highlights the different role played by the eigenvalues in the dynamics, as it can be intuitively deduced from Eq.~\eqref{Evol:2}.
Those characterized by a large $\Gamma_{K=0, \delta}$ are labeled as fast-decaying modes and they influence the initial transient dynamics.
The universal long-time dynamics is instead determined by the long-lived modes with small $\Gamma_{K=0, \delta}$.
The exact description of the universal dynamics therefore does not necessitate the use of the correct expression for all eigenvalues.
The technique that we propose in this paper is to approximate the spectrum with a linearized branch that coincides with the exact one only in the region with $\Gamma_{K,\delta} \sim 0$.
The poor approximation of the fast decaying modes is solely responsible 
for the fact that we produce a poor description of the initial transient dynamics, which is non-universal and 
is not of interest here.

\begin{figure}[t]
\centering
\includegraphics[width=0.75\linewidth]{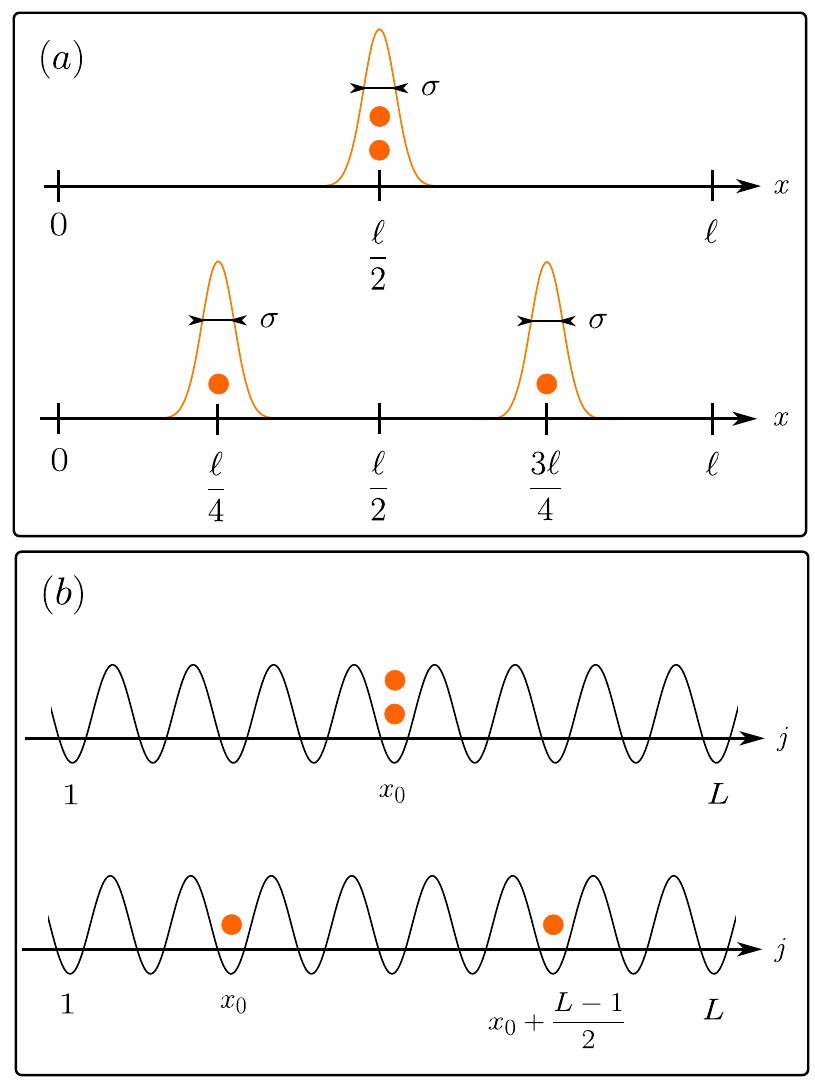}
\caption{Sketch of the four initial states studied in this article. 
(a) Continuum case. In the top sketch both particles are localized around $\frac{\Lconti}{2}$; in the bottom one, one particle is localized around $\frac{\Lconti}{4}$ and the other one around $\frac{3 \Lconti}{4}$. 
We set $\sigma = 1$ as unit of length. 
(b) Lattice case. In the top sketch both particles are at the site $x_0$; in the bottom one, one particle is at the site $x_0$ and the other one is at the site $x_0 + \frac{L-1}{2}$ (in this article $L$ is odd). 
}
\label{fig_IC}
\end{figure}

Another set of results concerns the dynamics of the system.
Figure~\ref{fig_IC} shows a sketch of the initial conditions that we consider, both in the continuum (panel (a)) and on the lattice (panel (b)). In each case, we consider two scenarios: two particles spatially localized at the same position (top sketch) and two particles separated by half the system size (bottom sketch).

\begin{figure}[t]
\includegraphics[width=1.0\linewidth]{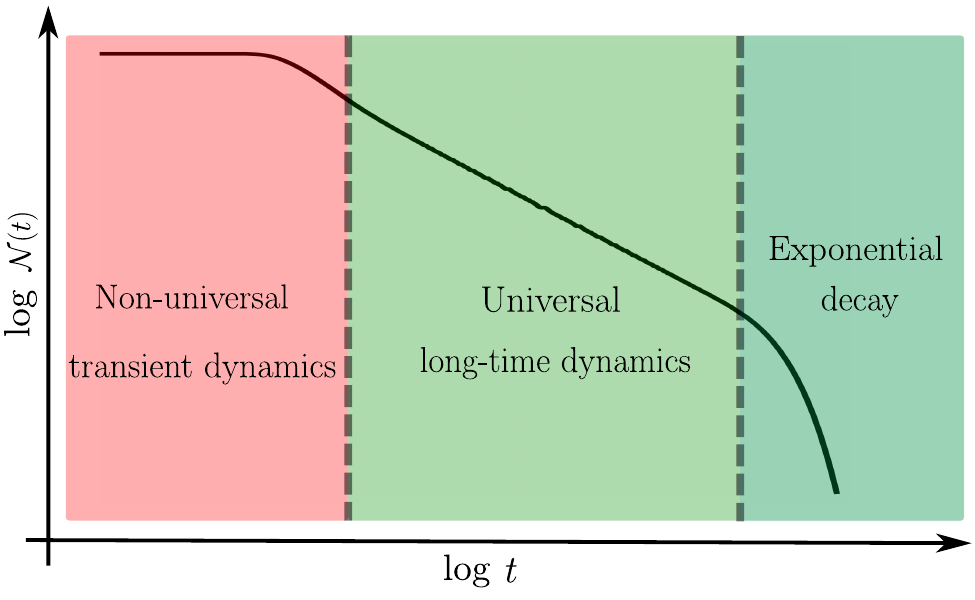}
\caption{Sketch of the dynamics of the mean number of particles as a function of time $\mathcal{N}(t)$ in log-log scale. The dynamics in the green area is predicted analytically for any value of interaction strength, loss rate and system size.}
\label{fig_dyn}
\end{figure}

For these initial conditions, we find that the behavior of the average number of particles $\mathcal{N}(t)$ in Eq.~(\ref{eqN2}) evolves through three different regimes, as we sketch in Fig.~\ref{fig_dyn} and explain below.
First, a non-universal transient dynamics takes place; this transient depends both on the initial condition and on the specific parameters of the master equation.
Since the goal of this article is to look for universal behaviors of the lossy gas, in this study we will not consider it.

After this initial transient, the system enters a regime in which the universal features of $\mathcal{N}(t)$ do emerge. The main result of this paper is the characterization of such an intermediate universal regime as summarized in Table~\ref{Table:1}. In the continuum, we find a power-law decay $\mathcal{N}(t) \sim t^{-\beta}$ where $\beta = 3/2$ if the two particles are initially at the same position and $\beta = 1/2$ if the two particles are initially separated by half of the system size. 
On the lattice, a logarithmic correction appears: we find $\mathcal{N}(t) \sim t^{-\frac{3}{2}} \left( \log(\kappa t^{-\frac{1}{2}}) + {C} \right)$ if the two particles are initially at the same position and $\mathcal{N}(t) \sim t^{-\frac{1}{2}}  \left( \log(\kappa t^{-\frac{1}{2}}) + {C^\prime} \right)$ if the two particles are initially separated by half of the system size. 
Here $\kappa$ is a known function of $U$, $\gamma$ and $L$ while {$C$} and ${C^\prime}$ are constants; their explicit expressions are presented in Sec.~\ref{dynLattice}. 

Finally, after this universal long-time dynamics,  an exponential decay $\mathcal{N}(t) \sim e^{-2 \lambda_{\rm ADR} t}$ appears that is related to the ADR. This takes place at a time scaling as $\lambda_{\rm ADR}^{-1}$ that in our problem increases polynomially with the system size.

The estimated typical time scales of transition between non-universal and universal regimes, as well as from power law to exponential decays, are summarized in Table~\ref{Table:3}.

\begin{table}[t]
\begingroup
\renewcommand{\arraystretch}{2}
\begin{tabular}{l||c|c}
   & \shortstack{Initially at the\\same position} & \shortstack{Initially\\far apart} \\
   \hline \hline
    Continuum & $t^{-\frac{3}{2}}$ & $t^{-\frac{1}{2}}$ \\
 \hline
  Lattice &   $t^{-\frac{3}{2}} \left( \log(\kappa t^{-\frac{1}{2}}) + C \right)$ & $t^{-\frac{1}{2}}  \left( \log(\kappa t^{-\frac{1}{2}}) + {C^\prime} \right) $\\[0.1cm]
\end{tabular}
\endgroup
\caption{Summary of the universal long-time dynamics of $\mathcal{N}(t)$. Here $\kappa$ is a known function of $U$, $\gamma$ and $L$ while $C$ and $C^\prime$ are known constants. In the continuum, the number of particles decays as a power-law. In the lattice, a log correction appears.}
\label{Table:1}
\end{table}

\begin{table}[h!]
\begingroup
\renewcommand{\arraystretch}{2}
\begin{tabular}{l||c|c}
   & \shortstack{Non-universal\\ to universal regimes} & \shortstack{Power law to\\ exponential decay regimes} \\
   \hline \hline
    Continuum & $ \sim   \frac{\ell}{\gamma}$ & $ \sim \left( \frac{g^2}{4} + \frac{\gamma^2}{16}\right) \frac{ \ell^3}{4 \pi^2 \gamma} $ \\[0.1cm]
 \hline
  Lattice &   $\sim \frac{4 U^2 +\gamma^2}{1024 \pi^2 \gamma} L^3$ & $ \sim \frac{4 U^2 +\gamma^2}{1024 \pi^2 \gamma} L^5$\\[0.1cm]
\end{tabular}
\endgroup
\caption{Typical time scales of transition between non-universal and universal regimes and from power law to exponential decays.}
\label{Table:3}
\end{table}

\section{Structure of the spectrum, linearization procedure, and asymptotic decay rate}
\label{SecSpectrum}

In this Section, we diagonalize the effective non-Hermitian Hamiltonian in the $N=2$ sector. We characterize its eigenvectors, 
analyze the structure of its eigenvalues, and linearize the Bethe equations around the most long-lived modes to compute the ADR.

\subsection{Right and left eigenvectors of $\Hef$ and associated eigenvalues}

We begin by discussing the eigenvectors of $\Hef$.
In the continuum, we write~\cite{Gaudin, korepin1997quantum, Zvonarev_2010, vsamaj2013introduction}
\begin{subequations}
\begin{equation}
    \ket{\Psi^{R}_{k_1 k_2}} =  \int_0^\Lconti  \int_0^\Lconti dx_1 dx_2 \Psi^{R}_{k_1 k_2}(x_1,x_2) \Psi^\dag(x_1) \Psi^\dag(x_2) \ket{v} \label{Rcont}
\end{equation}
where $\ket{v}$ denotes the vacuum state.
On the lattice, we analogously have~\cite{Oelkers_2007,Li_2022}
\begin{equation}
        \ket{\Psi_{k_1 k_2}^R} =  \sum_{x_1 = 1}^L \sum_{x_2 = 1}^L  \Psi^{R}_{k_1 k_2}(x_1,x_2) b_{x_1}^\dag b_{x_2}^\dag \ket{v}. \label{Rlatt}
\end{equation}
\end{subequations}
In both cases, we define
\begin{align}
    \Psi^{R}_{k_1 k_2}&(x_1,x_2) \propto \nonumber \\ 
    & 
\begin{cases}
     e^{i \left( k_1 x_1 + k_2 x_2 \right)} + s_{k_1 k_2} e^{i \left( k_2 x_1 + k_1 x_2 \right)}, &  x_1 \leq x_2 \\
     e^{i \left( k_1 x_2 + k_2 x_1 \right)} +  s_{k_1 k_2} e^{i \left( k_1 x_1 + k_2 x_2 \right)}           ,   &  x_1 > x_2
\end{cases}
\label{Eq:Wavefunction:Def}
\end{align}
where the two-particle S-matrix $s_{k_1 k_2}$ in the continuum case takes the value:
\begin{subequations}
\begin{equation}
    s_{k_1 k_2}  =   
    \dfrac{k_1 -  k_2  -  i c}{ k_1 - k_2 + i c}, 
\end{equation}
while on a lattice it reads:
\begin{equation}
      s_{k_1 k_2}  =    
      \dfrac{\sin k_1 - \sin k_2  - 2 i u}{\sin k_1 - \sin k_2 + 2 i u}. \label{squatPhaseLatt}
\end{equation}
\end{subequations}

Note that $\Psi^{R}_{k_1 k_2}(x_1,x_2)$ in Eq.~\eqref{Eq:Wavefunction:Def} can be expressed as a separable function of the center-of-mass position $\frac{x_1 + x_2}{2}$ and of the relative distance $x_1 - x_2$. 
One then obtains a natural writing in terms of the center-of-mass momentum $K=k_1+k_2$ and the relative quasimomentum $\frac{\delta}{2}=\frac{k_1-k_2}{2}$, as we anticipated in Sec.~\ref{Sec:EffDyn}:
\begin{equation}
\begin{split}
    \Psi^{R}_{k_1 k_2}&(x_1,x_2) \propto 
    e^{ i \left( k_1 + k_2 \right) \frac{ x_1 + x_2 }{2}} \times \\ & \Bigl( e^{ - i \frac{ k_1 - k_2 }{2} | x_1 - x_2 |}  + s_{k_1 k_2} e^{ i \frac{ k_1 - k_2 }{2} | x_1 - x_2 |}\Bigr).
\end{split}
\label{PsiR}
\end{equation}

Because of the periodic boundary conditions, any pair of quasimomenta $(k_1,k_2)$ should satisfy the Bethe equations, which in the continuum read:
\begin{subequations}
 \begin{equation}
     e^{i k_1 \Lconti} = s_{k_2 k_1} = - \frac{c - i \left( k_1 - k_2 \right)}{c + i \left( k_1 - k_2 \right)}, \qquad  k_1 + k_2 = \frac{2 \pi}{\Lconti} \mathbb{Z}. \label{BEconti}
 \end{equation}
On the lattice, they read:
  \begin{equation}
     e^{i k_1 L} = s_{k_2 k_1} = - \frac{2 u - i \left( \sin{k_1} - \sin{k_2} \right)}{2 u + i \left( \sin{k_1} - \sin{k_2} \right)}, \quad  k_1 + k_2 = \frac{2 \pi}{L} p ;
     \label{BeLattice}
 \end{equation}
\end{subequations}
 with $p \in \{ 0 , \cdots , L-1 \}$. 
 Once the quasimomenta $(k_1,k_2)$ are determined with these equations, one eventually obtains the eigenenergies of the Hamiltonian, whose right eigenvalues in the continuum are:
\begin{subequations}
\begin{equation}
 E_{k_1 k_2} =  k_1^2 + k_2^2, 
 \label{eqEconti} 
 \end{equation}
 and on a lattice read:
 \begin{equation}
    E_{k_1 k_2} = -2 \left( \cos k_1  + \cos k_2 \right).
    \label{ELattice}
\end{equation}
\end{subequations}

For both setups, we can show that the left eigenvectors of $H_{\rm eff}$ can be related to the functional form of the right ones (see Appendix~\ref{ApLeftEV} and Ref.~\cite{Tanaka_2013} for more details):
\begin{equation}
\ket{\Psi^{L}_{k_1 k_2}} = \frac{1}{\bra{\Psi^{R}_{k_1 k_2}} \ket{\Psi^{R}_{k_1^* k_2^*}}} \ket{\Psi^{R}_{k_1^* k_2^*}}.
\label{PsiL}
\end{equation}

In the discrete case, when $L$ is even, an eigenstate of total momentum $K=\pi$ appears that does not take the Bethe-ansatz form in Eq.~\eqref{Eq:Wavefunction:Def}. 
It is of the form $\frac{1}{\sqrt{2L}} \sum_{j = 1}^L e^{i \pi j}  (b_{j}^\dag)^2 \ket{v}$. 
The state can be interpreted as a bosonic analogue of the $\eta$-pairing states introduced in Ref.~\cite{Yang_1989} in the context of the fermionic Hubbard model.
This state has energy $U - i \gamma/2$ and corresponds to a state that loses particles at a rate that does not depend on $L$ and contributes to the initial non-universal transient dynamics.

When $L$ is even and $K=\pi$, 
the Bethe equation~\eqref{BeLattice} has solutions $k_1=(2m-1)\pi/L$ and $k_2= \pi-k_1$ $(m \in \mathbb{Z})$. 
The corresponding energy and wavefunction are $E=0$ and 
$\Psi^{R}_{k_1 k_2}(x_1,x_2) \propto e^{ i \pi \frac{ x_1 + x_2 }{2}}  \sin \left(\frac{ k_1 - k_2 }{2} | x_1 - x_2 |\right)$, respectively. Note that the wavefunction is nonvanishing only when ${4m-2\neq (2n-1)L}$ $(n \in \mathbb{Z})$.
Such a state is an eigenstate of the Bose-Hubbard model with zero interaction energy; as a consequence, it will not experience two-body losses and it will play the role of a nontrivial steady state of the dynamics (the trivial steady state is the vacuum).
Since we are interested in the universal features of the decaying density, we will mostly
work with odd $L$, so that the system does not have any nontrivial steady state and we are certified that, asymptotically, the system will be empty because the vacuum is the only stationary state.

\subsection{Linearization of the spectrum in the continuum and asymptotic decay rate}

If we use the center-of-mass momentum and the relative quasimomentum in the continuum, Eq.~\eqref{BEconti} becomes
 \begin{equation}
  e^{\frac{i}{2} \delta \Lconti} = (-1)^{p+1} \frac{c-i\delta}{c+i\delta}, \qquad K = \frac{2 \pi}{\Lconti} p, \qquad p \in \mathbb{Z} , \delta \in \mathbb C; \label{Beconti2}
\end{equation} 
and accordingly the energy in Eq.~\eqref{eqEconti} reads
 \begin{equation}
    E_{p  \delta} =  \frac{1}{2}  \delta^2 + \frac{1}{2}  \left( \frac{2 \pi}{\Lconti} p \right)^2 . \label{Econti}
\end{equation}

\begin{figure}[t]
\includegraphics[width=1.0\linewidth, trim = {0 10.38cm 0 0}, clip]{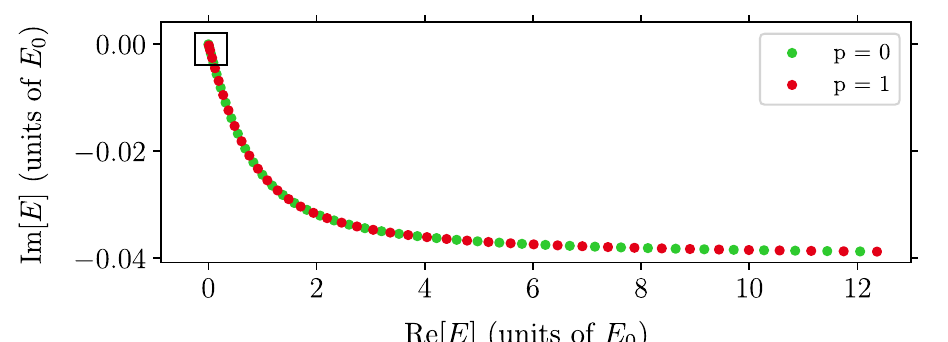}
\caption{Eigenenergies $E_{p\delta}$ of $\Hef$ for two bosons in the continuum for $p = 0$, $1$ and $c = 1 - i$ and $\Lconti = 100$. 
The other values of $p$ can be obtained with a rigid translation along the $\Re[E]$-axis. 
The spectrum continues towards infinite $\Re[E]$, with the $\Im[E]$ tending asymptotically towards the value $4 \Im [c] / \ell$. }
\label{spectrumconti}
\end{figure}

The sets of relative quasimomenta $\delta$ which are solutions of Eq.~(\ref{Beconti2}) depend on $p$ only via its parity. 
Therefore, the full spectrum can be deduced from the two branches corresponding to $p=0$ and $p=1$ only; the other branches are obtained by translation along the $\Re[E]$-axis. 
We solve Eq.~(\ref{Beconti2}) with an iterative numerical method described in Ref.~\cite{giam} and we use Eq.~(\ref{Econti}) to obtain the spectrum of $\Hef$. 
In Fig~\ref{spectrumconti}, we show the energy branches for $p=0,1$.

\begin{figure}[t]
\includegraphics[width=1.0\linewidth, trim = {0 0 0 8.85cm}, clip]{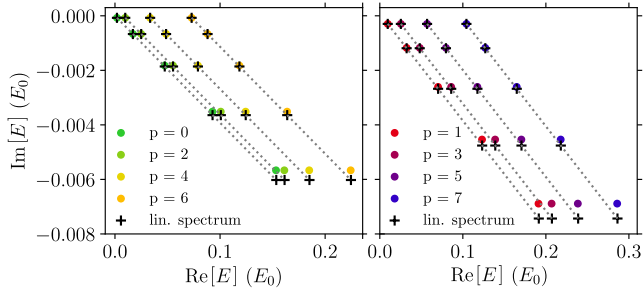}
\caption{Eigenenergies $E_{p\delta}$ of $\Hef$ for two  bosons in the continuum for $p \in \{ 0 , \ldots ,7 \}$ and $c = 1 - i$ and $\Lconti = 100$. 
The colored points are obtained by a numerical solution of Eq.~(\ref{Beconti2}) and the black crosses are the linearized spectrum $\mathcal E_{p,q}$.}
\label{spectrumconti_2}
\end{figure}

By inspecting the previous equations in certain regimes, we are also able to extract some analytic asymptotic behavior.
From Eq.~(\ref{Econti}), and recalling the definition in Eq.~\eqref{Eq:Notation:Energy}, one can easily show that $\Gamma_{p\delta} = -2 \Re[\delta] \Im[\delta]  $
and hence in order to characterize the properties of the spectrum for long-lived states we perform
a Taylor expansion of Eq.~(\ref{Beconti2}) for small $\delta$, namely $|\delta/c| \ll 1$. 
We obtain the following linearized quasimomenta (see Appendix~\ref{Ap2}) 
\begin{equation}
    \delta = \begin{cases} \dfrac{4 \pi q + 4 \pi}{\Lconti + 4/c} \quad& \mbox{if } p \mbox{ is odd}, \qquad \\ \\ \dfrac{4 \pi q + 2 \pi }{\Lconti + 4/c} \quad& \mbox{if } p \mbox{ is even} ; \label{dconti} \end{cases}
    \quad q \in \mathbb N = \{ 0,1,2,\ldots \}.
\end{equation}
We substitute this analytical approximate expression into Eq.~(\ref{Econti}) to obtain an approximate expression of the spectrum that we call \emph{linearized spectrum} $\mathcal E_{p,q}$. 
We observe that in the complex energy plane, the linearized spectrum occupies a line because (see Appendix~\ref{Ap2})
\begin{equation}
\frac{\Im[\mathcal E_{p=0,q}]}{\Re[\mathcal E_{p=0,q}]} = 
\frac{ 8 \Im[c] \left( |c|^2 \ell + 4 \Re[c] \right) }
{ \left( |c|^2 \ell + 4 \Re[c]\right)^2 + 16 \Im[c]^2 } \label{eq:lin_conti}
\end{equation}
is a constant that does not depend on $q$.
The linearized spectrum is compared to the exact spectrum in Fig.~\ref{spectrumconti_2}, where it is plotted with black crosses. 

Only the eigenvalues corresponding to the smallest $\Gamma_{p\delta}$, and thus to the largest decay times, are well captured by the linearized spectrum. The linearization is accurate in the limit $\Lconti |c| \gg 1 $ because the most long-lived modes become accumulation points for the eigenvalues of $\Hef$. 
By substituting Eq.~(\ref{dconti}) for $p$ even and $q=0$ into Eq.~(\ref{Econti}) we find an analytical expression for the asymptotic decay rate $\lambda_{\rm ADR}$ that we defined in Eq.~\eqref{Eq:lambdaADR:Def}:
    \begin{equation}
    \lambda_{\rm ADR} = \frac{4 \pi^2 \gamma}{\left( \frac{g^2}{4} + \frac{\gamma^2}{16}\right)  \Lconti^3} + o ( \Lconti^{-3}).
    \label{ADR:continuum:LM}
\end{equation}
A second regime that we can characterize analytically is the opposite one of $|\delta/c| \gg 1$, corresponding to the short-lived states which will ultimately be responsible for the initial transient dynamics.
By doing an analysis of Eq.~(\ref{Beconti2}) for $|\delta/c| \gg 1$, we show that asymptotically the imaginary part of the energy tends to the value $ 4 |\Im[c]|/\Lconti$ 
(see Appendix~\ref{NewAppendix}).

\subsection{Linearization of the spectrum on the lattice  and asymptotic decay rate} \label{SubSec:Linearization:Lattice}

\begin{figure}[t]
\includegraphics[width=1.0\linewidth, trim = {0 7cm 0 0}, clip]{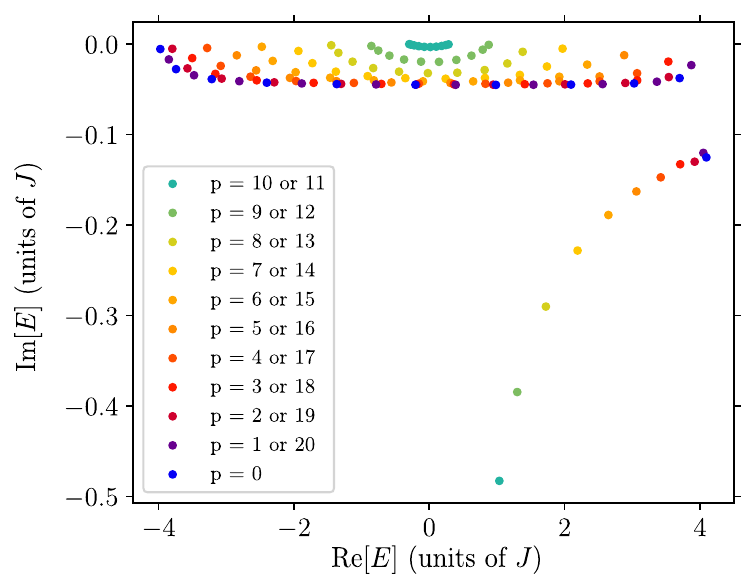}
\caption{Eigenenergies $E_{p\delta}$ of $\Hef$ for two bosons on a lattice of $L = 21$ sites for $U = 1$ and $\gamma=1$. 
The branches characterized by a total momentum $K = 2 \pi p/L$ are represented with the same color. 
The plot highlights the symmetry $p \leftrightarrow L-p$, for which the eigenvalues coincide exactly.}
\label{spectrumlatt}
\end{figure}

\begin{figure}[t]
\includegraphics[width=1.0\linewidth, trim = {0 0 0 10cm}, clip]{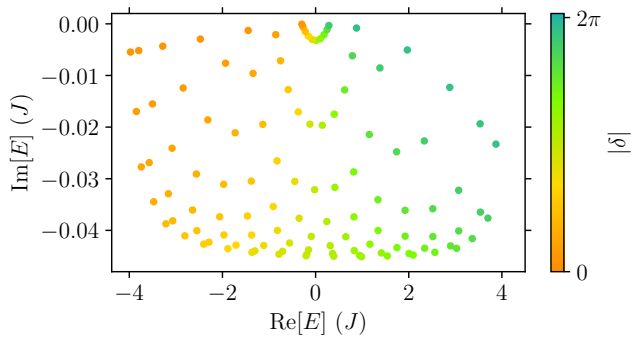}
\caption{Eigenenergies $E_{p \delta}$ of $\Hef$ for two bosons on a lattice of $L = 21$ sites for $U = 1$ and $\gamma=1$; 
the modes corresponding to the same value of $|\delta|$ are represented with the same color.
We do not display the branch of short-lived states and focus only on the branches with $0 \leq p \leq 10$ (for the other values of $p$ the value of $|\delta|$ is reverted).
}
\label{spectrumlatt:bis}
\end{figure}

Using the total momentum and the relative quasimomentum for the lattice problem, Eq.~\eqref{BeLattice} becomes
\begin{align}
    e^{\frac{i}{2} \delta L} =& (-1)^{p+1} \frac{2u - 2 i \sin(\frac{\delta}{2}) \cos(\frac{\pi}{L} p)}{2u + 2 i \sin(\frac{\delta}{2}) \cos(\frac{\pi}{L} p)}  , \qquad \delta \in \mathbb C; \nonumber \\
    \quad K =& \frac{2 \pi}{L} p, \qquad  p \in \{ 0 , \ldots ,L-1 \}.
    \label{BeLattice2}
\end{align}
The energy reads
\begin{equation}
    E_{p \delta} = - 4 \cos(\frac{\delta}{2}) \cos(\frac{\pi}{L}p).
    \label{ELattice2}
\end{equation}
We first solve numerically the Eq.~(\ref{BeLattice2}) for each $p$ and we substitute these solutions into Eq.~(\ref{ELattice2}); the resulting spectrum is plotted in Fig.~\ref{spectrumlatt}.
We observe the appearance of a short-lived branch that does not have an analogue in the continuum setup.  This branch is composed of repulsive bound-states with different total momenta; we derive an expression for their energies in Appendix~\ref{Appendix_BoundStates}. These short-lived states are well-separated from the other eigenstates, which have a smaller imaginary part, and that can be grouped into branches with a fixed $p$. 
Figure~\ref{spectrumlatt:bis} illustrates that
the eigenenergies having the smallest $\Gamma$ are associated to a $\delta$ close to $0$ on one side of the branch and $\delta$ close to $2 \pi$ on the other side.

\begin{figure}[t]
\includegraphics[width=1.0\linewidth]{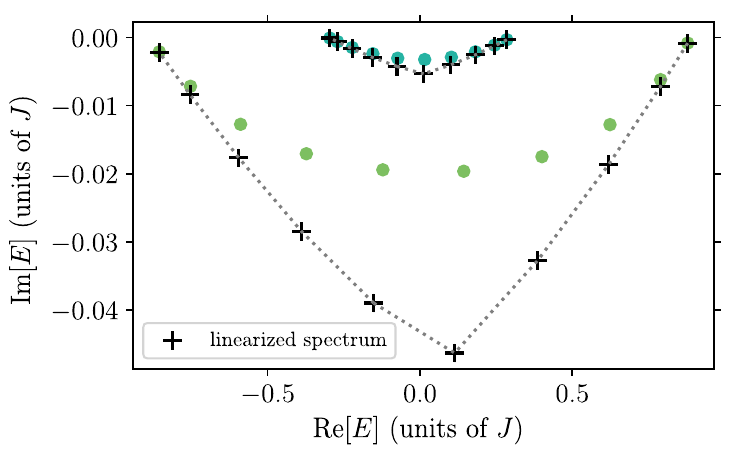}
\caption{Eigenenergies $E_{p\delta}$ of $\Hef$ for two bosons on a lattice of $L = 21$ sites for $U = 1$ and $\gamma=1$.     
The colored points are obtained with a numerical solution of Eq.~(\ref{BeLattice2}) whereas the black crosses correspond the linearized spectrum represented only for~$p = 9 ,10 $.}
\label{spectrumlatt2}
\end{figure}

Since our goal is to produce a simple theory that captures well the long-lived modes, we proceed in a way that is analogous to the previous subsection
and perform a Taylor expansion of Eq.~(\ref{BeLattice2}) for $|\delta| \ll 1$ or $|\delta - 2 \pi| \ll 1$.
From a simple observation of Fig.~\ref{spectrumlatt:bis}, 
we should expect a significant dependence on $p$ of the linearized quasimomenta $\delta$, and indeed we find the following expressions for $\delta \sim 0$:
\begin{subequations}
\label{Eq:Lattice:Linearized:delta}
\begin{equation}
    \delta = (q + D_p)\frac{4 \pi}{L + \frac{2}{u} \cos(\frac{\pi}{L} p)} , \qquad q \in \mathbb N;
    \label{d0} 
\end{equation}
and for $\delta \sim 2 \pi$
\begin{equation}
    \delta = \left( q + D_p - \frac{1}{u} \cos(\frac{\pi}{L} p) \right)\frac{4 \pi}{L - \frac{2}{u} \cos(\frac{\pi}{L} p)}, \quad q \in \mathbb N .
    \label{d2pi}
\end{equation}
\end{subequations}
Here, $D_p = \frac{1}{2}$ if $p$ is even and $D_p = 1$ if $p$ is odd. 
In the case of $\delta \sim 0$, the most long-lived modes are described by $q =0$.
In the case of $\delta \sim 2 \pi$, the most long-lived modes correspond to  $q \sim L/2$, and more precisely to $ q = \left( L - ( L \bmod{2}) \right)/2 -1 - \left( p \bmod{2}\right) \left( 1 - (L \bmod{2}) \right)$; this latter expression is valid both for even and odd $L$.
In Fig.~\ref{spectrumlatt2}
we plot the linearized spectrum $\mathcal E_{p,\delta}$ obtained by substituting the expressions in Eqs.~\eqref{Eq:Lattice:Linearized:delta} into Eq.~\eqref{ELattice2} for two values of $p$. 

Finally, we notice that the asymptotic decay rate $\lambda_{\rm ADR}$ takes the following simple form: 
\begin{equation}
    \lambda_{\rm ADR}  = \frac{4 \pi^4 \Delta_L^2 \gamma}{ \left( \frac{U^2}{4} + \frac{\gamma^2}{16} \right) L^5} + o(L^{-5}) \label{ADRconti}
\end{equation}
with 
\begin{equation}
    \Delta_L = \begin{cases}
    \frac 12 & \text{ for } L \bmod{2} = 1; \\
    2 & \text { for } L \bmod{4} =0; \\
    1 & \text { for } L \bmod 4 = 2.
    \end{cases}
\end{equation}
The ADR corresponds to an eigenvalue with $p =\pm(\frac{L}{2} - \Delta_L)$, hence to $K \sim \pi$;
Eq.~\eqref{ADRconti} is obtained by
substituting this value of $p$ and Eq.~(\ref{d0}), with $D_p=\frac{1}{2}$ and $q=0$, into Eq.~(\ref{ELattice2}).
In contrast to the continuum case, here the scaling is $\lambda_{\rm ADR} \sim L^{-5}$. 

However, if we consider the branches characterized by $p$ close to $0$, and hence by $K \sim 0$, we obtain:
\begin{equation}
    \lambda_{\rm ADR}^{(p)} = \min\limits_{\delta} | \Im [E_{p \delta}] |  
    = \frac{4 \pi^2 \gamma}{\left( \frac{U^2}{4} + \frac{\gamma^2}{16}\right) L^3} + o(L^{-3}). \label{ADRbisconti}
\end{equation}
This latter equation is the exact analogue of the expression for $\lambda_{\rm ADR}$ found in the continuum, reported in Eq.~\eqref{ADR:continuum:LM}. In both cases, we have a scaling with the inverse system size to the third power. The ADR scaling as $L^{-5}$ is associated to a total momentum at the edge of the Brillouin zone, and it is thus a lattice effect. 
The formulas that we just discussed on the ADR are summarized in Table~\ref{Table:2}.
For completeness, we mention that in Ref.~\cite{Nakagawa_2021} the authors compute the ADR for a fermionic lattice system at finite density and show that $\lambda_{\rm ADR} \sim L^{-2}$.

\section{Universal long-time scalings} \label{SecScaling}

In this section, we derive the universal expressions for the mean number of particles at late times for the two initial conditions illustrated in Fig.~\ref{fig_IC}, both for the continuum system and the lattice. 

\subsection{Universal long-time scaling in the continuum}

By using the total momentum and the relative quasimomentum, 
Eq.~\eqref{Evol:2} can be rewritten as
\begin{equation}
    \mathcal{N}(t)  = \sum_{p=-\infty}^\infty \sum_{\delta \in \mathcal{S}_p}   \tilde f \left( p, \delta \right)  e^{-\frac{i}{2} ({\delta}^2 - {\delta^{*}}^2)t} ,
    \label{eq1}
\end{equation}
where $\mathcal{S}_p$ is the set of solutions of the Eqs.~\eqref{Beconti2}. If $\delta$ is a solution, then $-\delta$ is also so; thus, when performing actual calculations, in order to avoid double counting, we always take $\Re[\delta]>0$. If we consider a time much larger than the typical fast-decay time, $\tau_{\rm fd} = \left( 4 \frac{| \Im [c]|}{\ell} \right)^{-1}$, then the dynamics is well captured by the linearized spectrum. Indeed, $\tau_{\rm fd} $ is the typical decay-time of the majority of eigenmodes, which are characterized by $|\delta/c| \gg 1$ and that are solely responsible for the short-time transient dynamics. 
We can thus simply sum over the $\delta$ in the set $\mathcal S^{(0)}_p$, corresponding to the values given in Eq.~\eqref{dconti}; 
for consistency, $\tilde f(p, \delta)$ has to be expanded to the leading order in $\delta$.

We detail this procedure for the case where the initial state is $\Psi_0(x_1,x_2) \propto e^{-\left( x_1 - \frac{\Lconti}{2} \right)^2}e^{-\left( x_2 - \frac{\Lconti}{2} \right)^2}$, illustrated in the top sketch of Fig.~\ref{fig_IC}(a), with both particles localized around the position $\Lconti/2$ and corresponding to two particles initially at the same position. 
The explicit expression for $\tilde f(p, \delta)$ is obtained using the definitions given in Sec.~\ref{SecSpectrum} and is reported in Appendix~\ref{Expr_f};
an expansion for $\delta$ close to zero, corresponding to the most long-lived modes, gives: 
\begin{equation}
    \tilde f_0(p,\delta) \approx g_0(\Lconti,c) 
    \frac{|\delta|^4 }{{\delta}^2 - {\delta^{*}}^2} 
    e^{-\frac{ 1}{4}(\frac{2\pi p }{\Lconti})^2} .
\end{equation}
The explicit expression of $g_0(\ell, c)$ is in Eq.~\eqref{Eq:g0:Appendix} (see Appendix~\ref{ApSumtoInt} for details).
We obtain:
\begin{equation}
    \mathcal{N}(t) \approx g_0(\Lconti,c)  \sum_{p=-\infty}^\infty \sum_{\delta \in \mathcal{S}_p^{(0)}}  \frac{|\delta|^4 \; 
    e^{-\frac{1}{4}(\frac{2\pi p }{\Lconti})^2}}{{\delta}^2 - {\delta^{*}}^2} e^{-\frac{i}{2} ({\delta}^2 - {\delta^{*}}^2)t} .\label{Nsamepos}
 \end{equation}
 By transforming the sums into integrals, it can be shown that long after the time $\tau_{\rm fd}$ has passed, 
 corresponding to the typical decay time of the modes characterized by $|\delta / c| \gg 1$,
  but before the asymptotic exponential decay is reached (the Liouvillian gap being small but finite) the average number of particles decays as a power law according to
\begin{equation}
     \mathcal{N}(t)  \approx  h_0(\Lconti,c) \left( \Gamma_c t \right)^{-3/2}, 
      \qquad \Gamma_c = \frac{|\Im[c]|}{|c|^2}.
     \label{ana1}
\end{equation}
The explicit expression of the universal rescaling function $h_0(\ell, c)$ is given in Eq.~(\ref{eq:h0lc}) (see Appendix~\ref{ApSumtoInt} for details). 
This is one of the main results of this article and it shows the existence of a universal behavior in the time-dependence of the density of the system. Remarkably, the power law $t^{-3/2}$ and the exponent $\beta = 3/2$ do not depend on the complex interaction constant $c$.
This result was not easily anticipated as in previous works looking for universal behaviors in gases at finite density, the gas was shown to have two different universal decays for weak and strong interactions (for instance, for the fermionic case, see Refs.~\cite{Rosso_2021, Rosso_2023}).
 
\begin{figure}[t]
\includegraphics[width=1.0\linewidth]{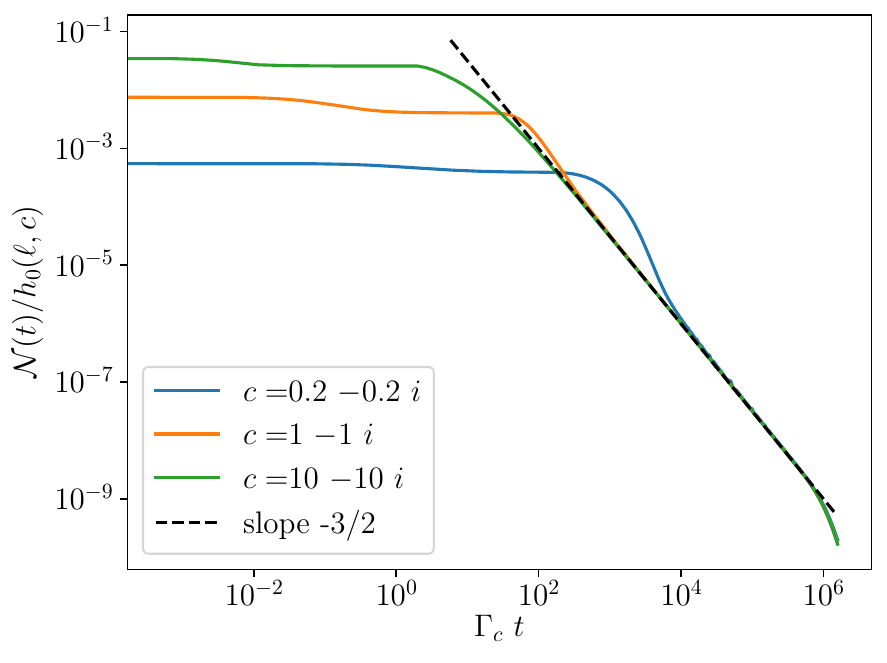}
\caption{Dynamics of the mean number of particles $\mathcal N(t)$ rescaled by the scaling function $h_0(\Lconti, c)$ for the initial state $\Psi_0(x_1,x_2) \propto e^{-\left( x_1 - \frac{\Lconti}{2} \right)^2}e^{-\left( x_2 - \frac{\Lconti}{2} \right)^2}$ and for $\Lconti=500$. 
The solid lines correspond to the exact computations from Eq.~(\ref{Evol}) for three different values of $c$. 
The dashed black line corresponds to the universal expression given by Eq.~(\ref{ana1});
the collapse of the data is excellent.}
\label{figsamesiteconti}
\end{figure}

In Fig.~\ref{figsamesiteconti},
we compare the approximate result we just derived with 
the exact expression of $\mathcal{N}(t)$ computed using the exact formula in Eq.~\eqref{Evol}.
The comparison is excellent and shows that the linearization of the spectrum is a powerful technique for extracting the universal decay of the density as $t^{-3/2}$. Moreover, thanks to the rescaling function $h_0(\ell, c)$, we observe a perfect collapse of the exact numerical data.

We now consider a different initial state $\Psi_1(x_1,x_2) \propto e^{-(x_1-\frac{\Lconti}{4})^2} e^{-(x_2-\frac{3 \Lconti}{4})^2} + e^{-(x_2-\frac{\Lconti}{4})^2} e^{-(x_1- \frac{3 \Lconti}{4})^2}$, illustrated in the bottom sketch of Fig.~\ref{fig_IC}(a). One particle is localized around $\frac{\Lconti}{4}$ and the other one is localized around $\frac{3 \Lconti}{4}$, so that the particles are separated by half of the system size. 
For this initial condition, we obtain the following expansion of $\tilde f(p,\delta)$: 
\begin{equation}
   \tilde  f_1 \left( p, \delta \right) = g_1(\Lconti,c) \frac{|\delta|^2}{{\delta}^2 - {\delta^{*}}^2} e^{-\frac{1}{4} (\frac{2\pi p }{\Lconti})^2}. 
\end{equation}
 The different behavior of $\tilde f_1$ for small $\delta$ with respect to $\tilde f_0$ is responsible for a different universal long-time decay.
Following the same steps outlined previously, we find, for the same range of~$t$ corresponding to the intermediate algebraic regime,
\begin{equation}
     \mathcal{N}(t)  \approx  h_1(\Lconti) \left( \Gamma_c t \right)^{-1/2} 
     , \qquad \Gamma_c = \frac{|\Im[c]|}{|c|^2}.
     \label{ana2}
 \end{equation} 
The explicit expressions of $g_1(\ell, c)$  and $h_1(\ell)$ are respectively presented in Eqs.~(\ref{eq:g1lc}) and (\ref{eq:h1l}) in Appendix~\ref{ApSumtoInt}. 
In Fig.~\ref{figdiffsiteconti} we plot the function $ \mathcal{N}(t)$ given by the exact expression in Eq.~(\ref{Evol}) and we compare it with the approximate analytical expression given by Eq.~(\ref{ana2}) to highlight the universal power-law scaling $t^{-1/2}$. 
The collapse of the data onto the universal curve is apparent.

\begin{figure}[t]
\includegraphics[width=1.0\linewidth]{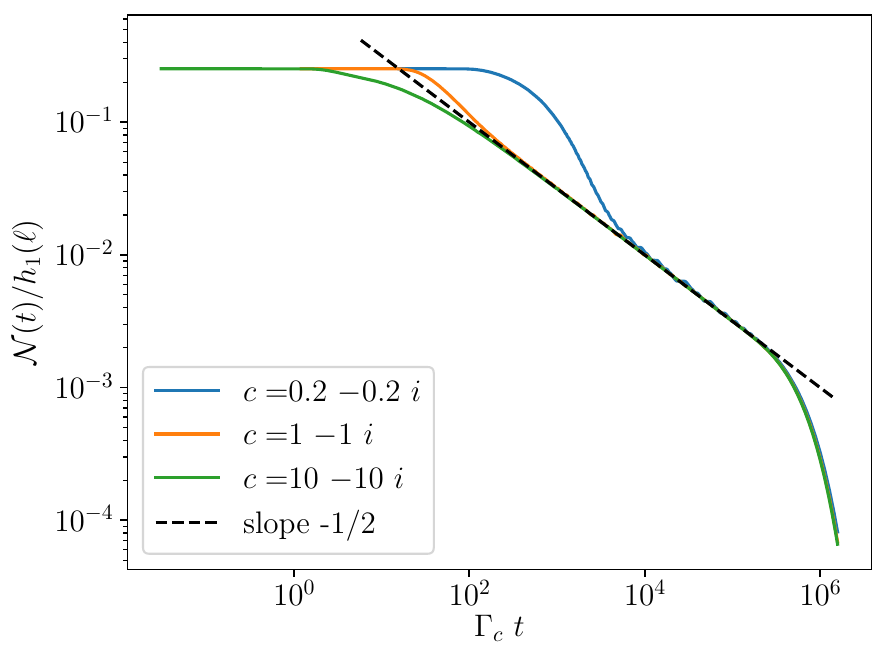}
\caption{Dynamics of the mean number of particles $\mathcal N(t)$ rescaled by the scaling function $h_1(\Lconti, c)$ for the initial state $\Psi_1(x_1,x_2) \propto e^{-(x_1-\frac{\Lconti}{4})^2} e^{-(x_2-\frac{3 \Lconti}{4})^2}  + x_1 \leftrightarrow x_2$ and for $\Lconti=500$. The solid lines correspond to the exact computations from Eq.~(\ref{Evol}) for three different values of $c$. 
The dashed black line corresponds to the universal expression given by Eq.~(\ref{ana2}); the collapse of the data is excellent.}
\label{figdiffsiteconti}
\end{figure}

We can briefly estimate for which range of times the universal scalings that we just presented are valid.
In general, the decay of particles is characterized by a first non-universal dynamics that is determined by the fast-decaying modes, which decay on a typical time scale $\tau_{\rm fd}$, and that constitute a lower bound for the appearance of the universal behavior.
At very late times, the universal power-law scaling is then followed by an exponential decay $\mathcal{N}(t) \sim e^{- 2 \lambda_{\rm ADR} t}$. 
Concluding, we can thus estimate that the expressions in Eqs.~\eqref{ana1} and~\eqref{ana2} are valid for $  \tau_{\rm fd} \ll t \ll \frac{1}{\lambda_{\rm ADR}}$.

\subsubsection{Interpolating the two initial conditions}

\begin{figure}[t]
\includegraphics[width=1.0\linewidth]{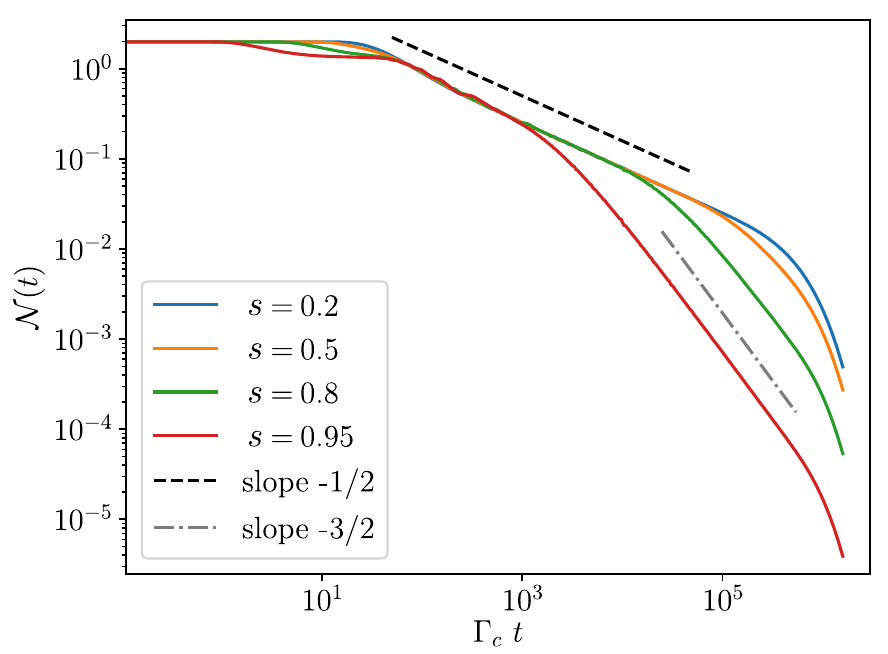}
\caption{Dynamics of the mean number of particles $\mathcal N(t)$ for the initial state $\Psi_{1-s}(x_1,x_2) \propto e^{-(x_1-\frac{\Lconti}{4} -s \frac{\Lconti}{4})^2} e^{-(x_2- \frac{3 \Lconti}{4}+ s \frac{\Lconti}{4})^2}  + x_1 \leftrightarrow x_2$ for $\Lconti=500$. 
We consider one interaction strength $c=1 - i$
and various $s \in \{ 0.2,0.5,0.8,0.95 \} $. 
The solid curves correspond to the exact computation from Eq.~(\ref{Evol}); the dashed and dashed-dotted lines correspond to the algebraic decays discussed in the main text. For $s \sim 1$ the density of bosons $\mathcal N(t)$ crosses over between two algebraic decays.}
\label{figalphaconti}
\end{figure}

As it can be understood intuitively, the scaling $\mathcal{N}(t) \sim t^{-1/2}$ is not peculiar to the situation where the two particles are initially separated by exactly half of the system size. The same scaling emerges for all initial states consisting of two non-overlapping Gaussians, i.e.~separated by a distance greater than their typical width. 
In Fig.~\ref{figalphaconti}, we show the mean number of particles computed with the exact expression given by Eq.~(\ref{Evol}) for the initial states 
\begin{equation}
\Psi_{1-s}(x_1,x_2) \propto  e^{-(x_1-\frac{\Lconti}{4} -s \frac{\Lconti}{4})^2} e^{-(x_2- \frac{3 \Lconti}{4}+ s \frac{\Lconti}{4})^2} + x_1 \leftrightarrow x_2
\end{equation} 
for various values of the parameter $s$ in the range $[0,1]$. 
This wavefunction places the two particles at an initial distance of $\frac{\ell}{2}\left( 1- s \right)$ and thus
interpolates between the two particles at the same position ($s=1$) and the two particles at their largest distance ($s=0$).

At small $s$, the particles are initially far apart, their overlap is negligible and the initial condition to any purpose completely indistinguishable from the $s=0$ one. 
We observe numerically that
$\mathcal{N}(t)$ decays as $t^{-1/2}$. 
We can thus conclude that this power law characterizes the algebraic decay of $\mathcal{N}(t)$ as long as the two particles are initially far apart.

The situation is different
when $s$ is close to $1$; initially the two particles are not far apart (as compared to their typical widths); due to the unitary dynamics, the two wave packets spread and can develop a large overlap, leading to a situation which will not be distinguishable from the one consisting of two perfectly overlapping Gaussian wave packets at the initial time. 
We numerically observe this to be the case in Fig.~\ref{figalphaconti}: after an initial $\mathcal{N}(t)\propto t^{-1/2}$ decay (which gets shorter as $s$ increases), a second algebraic region $\mathcal{N}(t)\propto t^{-3/2}$ emerges.

Importantly, even though as we discussed there may be a competition between the two asymptotic algebraic decays, no new power law emerges in this case, showcasing the robustness of the power laws we characterized to a perturbation of the system's initial conditions.

\subsubsection{Experimental relevance}

\begin{figure}[t]
\includegraphics[width=1.0\linewidth]{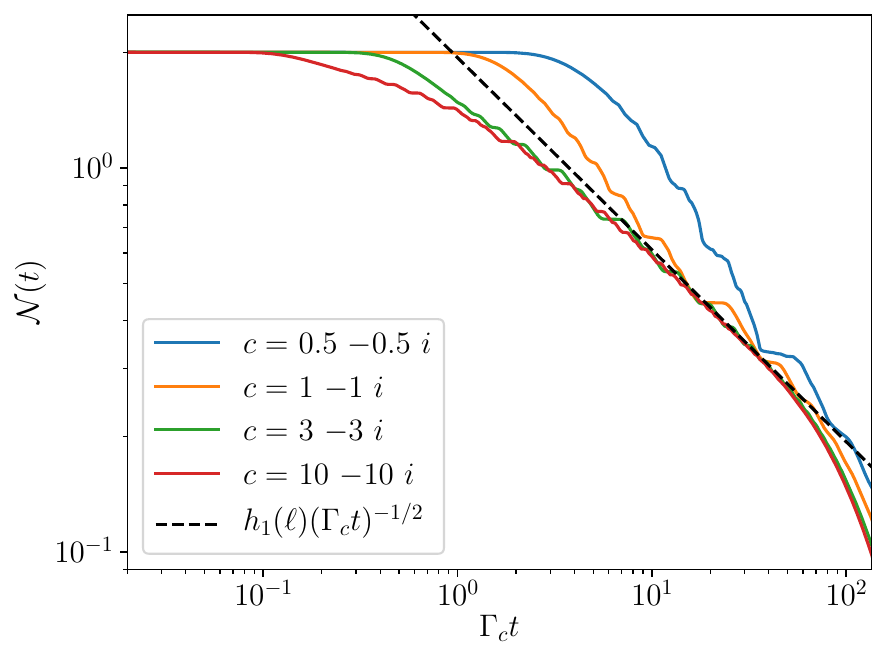}
\caption{Dynamics of the mean number of particles $\mathcal N(t)$ for $\Psi_0(x_1,x_2) \propto e^{-(x_1-\frac{\Lconti}{4})^2} e^{-(x_2-\frac{3 \Lconti}{4})^2}  + x_1 \leftrightarrow x_2$ for a small setup of length $\Lconti=30$. 
The solid lines correspond to the exact computation from Eq.~(\ref{Evol}). The dashed black line corresponds to the universal expression given by Eq.~(\ref{ana2}). 
Even if the size of the system is rather small, the universal behavior can be clearly distinguished, especially for large values of $|c|$.}
\label{figdiffsiteconti_Smallsize}
\end{figure}

So far we studied very large system sizes in order to let the universal decay emerge in full clarity;
however, since we are considering only two particles, a smaller~$\Lconti$ could be more experimentally relevant. 
For the linearization to be valid, the inequality $\Lconti |c|\gg1$ should be satisfied; 
we can therefore reduce the system size by introducing larger dissipation rates.
In Fig.~\ref{figdiffsiteconti_Smallsize}, we show that the scaling $\mathcal{N}(t) \sim t^{-1/2}$ still emerges for a smaller system of length $\Lconti = 30$; the universal behavior is more visible for large values of $|c|$, as expected.
Since in this situation the universal decay appears for significantly shorter times and for higher values of the mean number of particles with respect to the examples studied so far, we believe that the universal decay that we are proposing is not completely outside the possibilities of an experimental observation.

\subsection{Universal long-time scaling on the lattice}\label{dynLattice}

We now move to the lattice geometry and compute $\mathcal{N}(t)$ for two initial states paralleling those discussed in the continuum: in the first initial condition, the two particles are at the same site; in the second initial condition, the two particles are at distant sites.
As we already said, we take $L$ odd in order to avoid the presence of nontrivial steady states. 

In order to see the universal long-time scaling, we assume that $L |u| \gg 1$ and we consider times such that $L^3 \ll \Gamma_u t \ll L^5$ with
\begin{equation}
    \Gamma_u = \frac{ 256 \pi^2 |\Im(2 u)|}{|2 u|^2}= \frac{1024 \pi^2 \gamma}{ 4 U^2 +\gamma^2}.
\end{equation}
Under this assumption, the dynamics is well captured by the long-lived modes and the relative quasimomenta can be linearized as in Eqs.~(\ref{d0}) and~(\ref{d2pi}). 

We first consider the state in which two particles are initially on the same site. In this case, we compute 
\begin{equation}
    \mathcal{N}(t) \approx  \sqrt{\frac{2 L^3}{\pi \left( \Gamma_u t \right)^3 }} \frac{2 \Gamma_u}{\gamma} \times \left[ \log{ \left( 2 \sqrt{\frac{L^5}{\Gamma_u t}} \right)} +  C \right].
    \label{ana_same}
\end{equation}
We  find that $\mathcal{N}(t)$ scales as $t^{-3/2}$, exactly as in the continuum setup that we presented in Eq.~\eqref{ana1}. 
However, unlike the continuum case, we find a logarithmic correction to the scaling.
$C$ is a universal constant that evaluates approximately to $1.75$,
whose exact expression is in Eq.~\eqref{eq:constant_C} in Appendix~\ref{ApDerLatt}.

On the other hand, if we consider an initial state where one particle is at site $x_0$ and the other is at $x_0+(L-1)/2$, we obtain
\begin{equation}
    \mathcal{N}(t)\approx 16 \sqrt{\frac{L}{ 2\pi \Gamma_u t}} \times \left[ \log{ \left( 2 \sqrt{\frac{L^5}{\Gamma_u t}} \right)} + C^\prime \right].
    \label{ana_far}
\end{equation}
In this case $\mathcal{N}(t)$ scales as $t^{-1/2}$, analogously to the continuum problem presented in Eq.~\eqref{ana2}.
Here $C'$ is a universal constant that evaluates approximately to $1.52$, 
we refer the interested reader to Eq.~\eqref{eq:constant_Cprime}.  in Appendix~\ref{ApDerLatt} for the detailed expression. 
We sketch the derivation of Eq.~\eqref{ana_far} in Appendix~\ref{ApDerLatt}; Eq.~\eqref{ana_same} can be derived in a completely analogous way, but the calculations are slightly more involved.

The presence of the logarithmic correction is robust and is a direct consequence of the fact that the spectrum features two different ADR for the symmetry sectors with total momentum $K \sim 0$ and for $K \sim \pi$, as we discussed in Sec.~\ref{SubSec:Linearization:Lattice} and in Table~\ref{Table:2}.
It can be misleading if one tries to fit the numerical data without prior knowledge of the universal form of $\mathcal{N}(t)$. 
Indeed, a fit using a power-law decay of the form $\mathcal{N}(t) \sim t^{- \beta}$ might seem to properly describe the data; however, by doing so, one recovers a $\beta$ that depends on $U$, $\gamma$ and $L$. 
This blind approach misses the universal behavior of the decay of the density of particles and could be relevant also in other contexts, such as those discussed in Refs.~\cite{Rosso_2023, begg2024quantum}. 

\begin{figure}[t]
\includegraphics[width=1.0\linewidth]{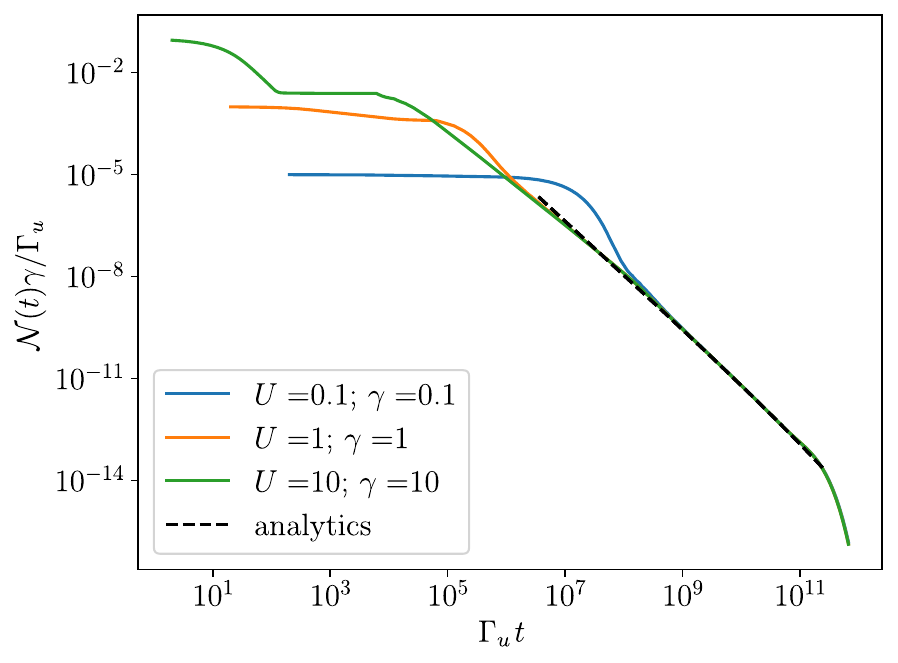}
\caption{Dynamics of the mean number of particles $\mathcal N(t)$ for the initial state 
$\ket{\Psi_2} = b_{x_0}^\dag b_{x_0}^\dag  \ket{v}/\sqrt{2}$ and for $L=121$ sites. The solid curves correspond to the exact computation. The dashed black line corresponds to the universal expression given by Eq.~(\ref{ana_same}); the collapse of the data is excellent.
}
\label{figsamepos}
\end{figure}
\begin{figure}[t]
\includegraphics[width=1.0\linewidth]{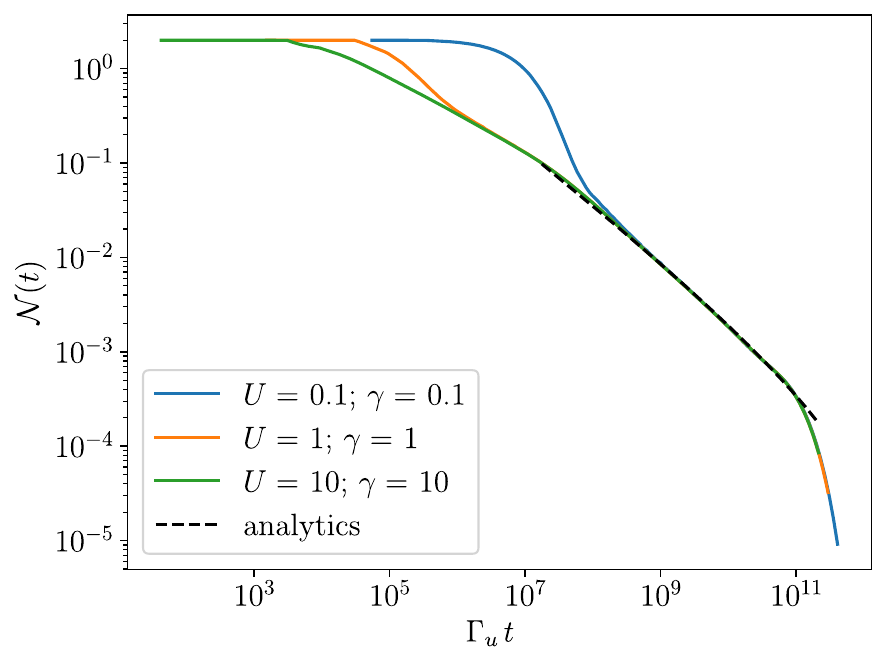}
\caption{Dynamics of the mean number of particles $\mathcal N(t)$ for the initial state $\ket{\Psi_3} = b_{x_0}^\dag b_{x_0+(L-1)/2}^\dag \ket{v}$ and for $L=121$ sites. 
The solid curves correspond to the exact computation. 
The dashed black line corresponds to the universal expression given by Eq.~(\ref{ana_far}); the collapse of the data is excellent.}
\label{figdifpos}
\end{figure}

In Figs.~\ref{figsamepos} and~\ref{figdifpos}, we compare the analytical universal scaling functions in Eqs.~(\ref{ana_same}) and~(\ref{ana_far}) with numerical results. 
For both initial conditions, we have a non-universal dynamics at first. The universal power-law scaling with logarithmic correction can be seen up to times scaling as $ \sim L^5 / \Gamma_u$, which is the scaling of $\lambda_{\rm ADR}^{-1}$ for total momentum $K \sim \pi$. For longer times, the density decays exponentially with a time scale determined by this ADR.
The collapse of the exact numerical data onto the universal scaling functions that we just derived is excellent and further proves the usefulness of the linearization of the spectrum for the long-lived modes. The results that we just discussed are summarized in Table~\ref{Table:1}.

\subsubsection{Crossover from the transient dynamics to the universal long-time behavior in the weakly-interacting and weakly-lossy regime}\label{MFLattice}

\begin{figure}[t]
\includegraphics[width=1.0\linewidth]{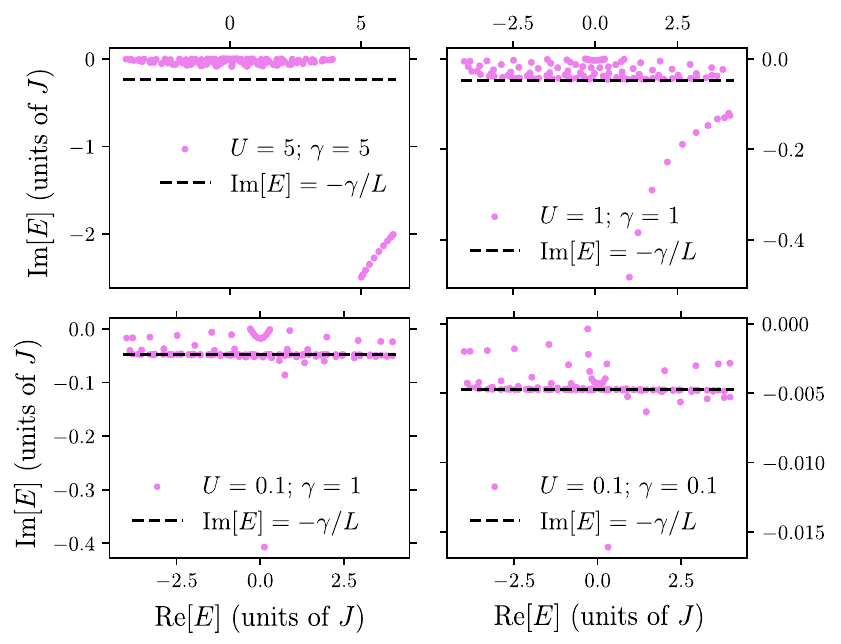}
\caption{Eigenergies of $\Hef$ for $L=21$ and various values of $U$ and~$\gamma$. The dashed black line on which the modes accumulate for $U,\gamma \sim 0$ corresponds to $\Im[E] = - \gamma/L$.
The collapse of the eigenvalues onto the black dashed line is evident, even if outlying eigenvalues describing modes decaying more slowly are always present.
}
\label{Fig_MF0}
\end{figure}

The results that we found on the universal late-time properties of $\mathcal N(t)$ are particularly surprising when we consider the weakly-interacting and weakly-lossy region, for which a previous article has found an exponential decay in time for a similar setup characterized by two-body losses~\cite{Yoshida_2023}:
\begin{equation}
    \mathcal{N}(t) = 2 e^{-2 \gamma t /L}. \label{MF_ana}
\end{equation}
The applicability of this result to our problem is corroborated by the study of the spectrum of $\Hef$.
As $\gamma$ and $U$ decrease, the eigenmodes of the effective non-Hermitian Hamiltonian accumulate on the horizontal line $\Im[E] = - \gamma/L$ as shown in Fig.~\ref{Fig_MF0} for the lattice case. 
The four panels refer to different values of $U$ and $\gamma$ and the tendency that we described appears rather clearly. The presence of other eigenvalues with an imaginary part of the energy that is closer to zero suggests that at late times the dynamics of $\mathcal N(t)$ will not be described by Eq.~\eqref{MF_ana}.

\begin{figure}[t]
\includegraphics[width=1.0\linewidth]{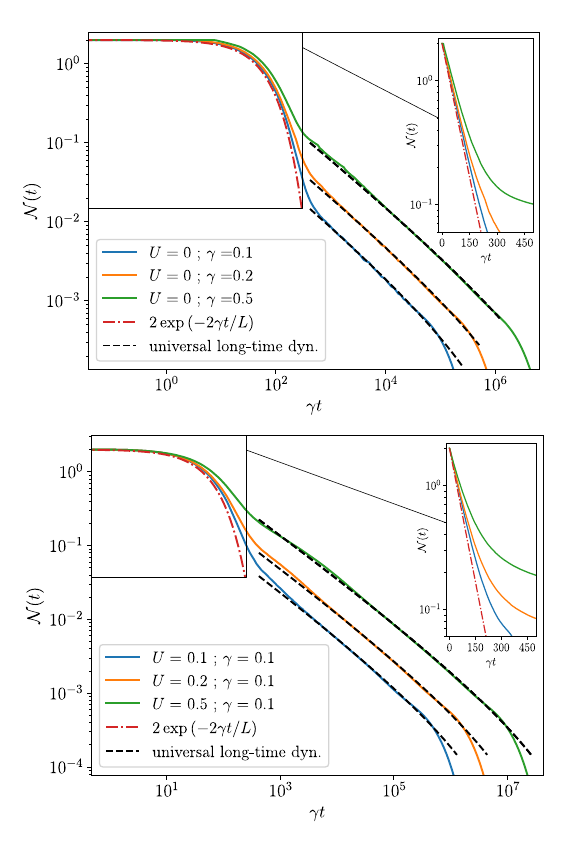}
\caption{Crossover in time from the early-time transient dynamics in Eq.~\eqref{MF_ana} to the universal late-time behaviour for both $U$ small and $\gamma$ small. 
Simulations are carried out on the lattice for $L = 121$ sites and the initial state is $\ket{\Psi_3} = b_{x_0}^\dag b_{x_0+(L-1)/2}^\dag \ket{v}$; the solid curves correspond to the exact computation. 
The black dashed lines correspond to the universal expression given by Eq.~(\ref{ana_far}); 
the red dashed-dotted line corresponds to the analytical form given by Eq.~\eqref{MF_ana}, that only describes the early-time behaviour of $\mathcal N(t)$.}
\label{Fig_MF}
\end{figure}

In Fig.~\ref{Fig_MF}, we compare the exact dynamics to the analytical form given by Eq.~\eqref{MF_ana} for the initial state consisting of two particles separated by half of the system size. 
As~$\gamma$ and $U$ decrease, not only the analytical curve of Eq.~\eqref{MF_ana} approaches the exact computation, but also the time interval over which this agreement is good increases and the crossover to the universal behavior appears later.
These plots explain the compatibility of our universal result with other approaches that could be developed to discuss the transient dynamics of the gas. 
The techniques presented in this article focus on the asymptotically late-time dynamics that precede the exponential decay in time dictated by the ADR: this does not exclude the possibility of finding other analytical behaviors as a function of time for the initial stages of the dynamics.

\subsection{Quantum Zeno effect}

 For the lattice case, the emergence of the time scale $\Gamma_u^{-1}$ in Eqs.~\eqref{ana_same} and \eqref{ana_far} can be seen as a manifestation of the continuous quantum Zeno (QZ) effect~\cite{Rossini_2020,Zhu_2014,Rosso_2023}, 
 a phenomenon that is here invoked because strong losses suppress coherent processes (in our case, the hoppings that change the number of double occupancies). 
 The suppression of motion within the lattice often slows down particle losses, 
 and in our discussion we will speak of QZ behavior whenever this takes place.
 For instance, if $\gamma \ll U$ then $\Gamma_u \sim \gamma$, but if $\gamma \gg U$, then $\Gamma_u \sim 1/\gamma$. 
 
\begin{figure}[t]
\includegraphics[width=1.0\linewidth]{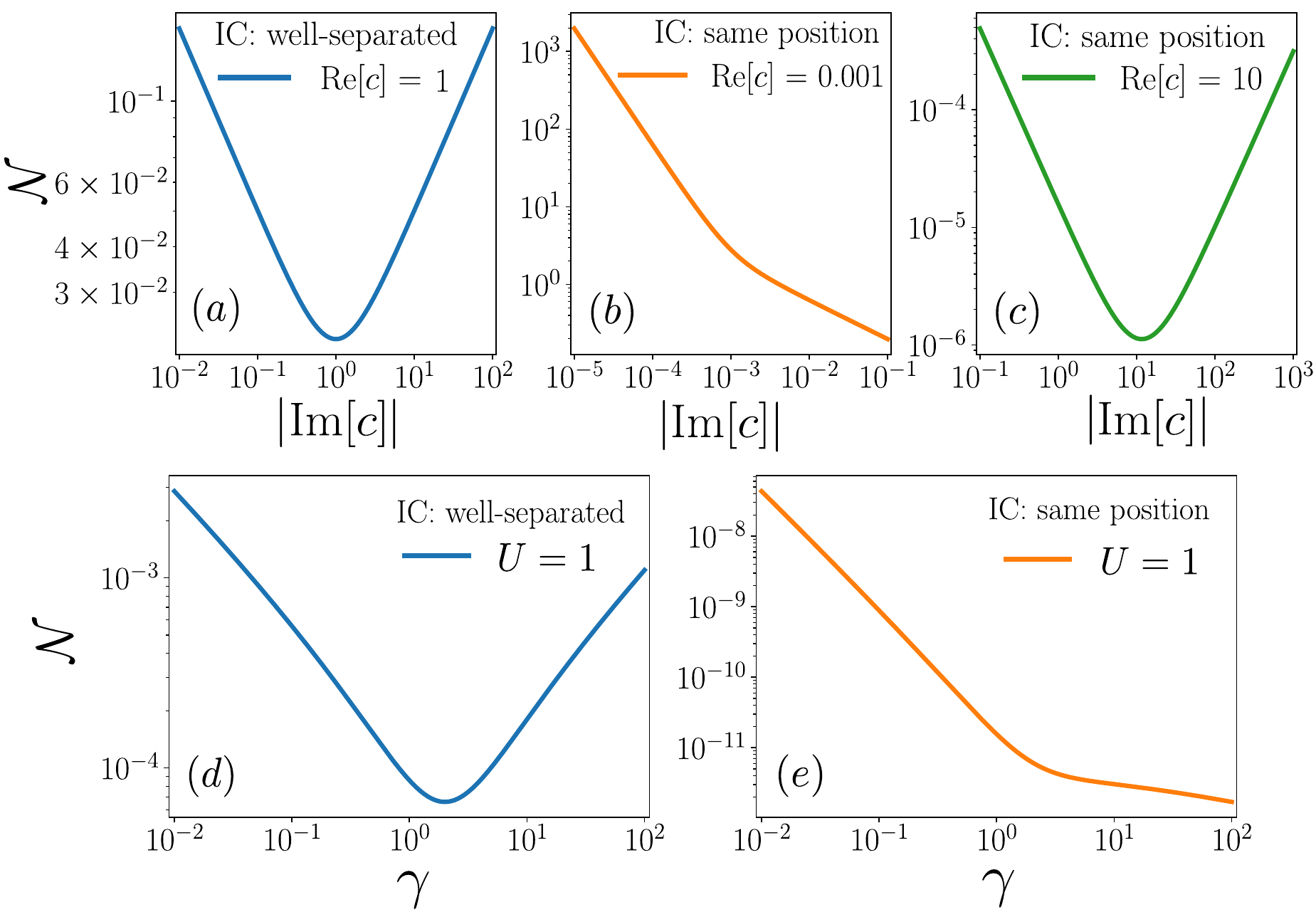}
\caption{Top panels: mean number of particles $\mathcal N(t)$ for the continuum setup as a function of $|\Im[c]|$ for $\Lconti=500$ and at time $t=\Lconti^2$; IC stands for ``initial condition''. 
(a) The particles are initially well-separated and the blue curve is obtained from Eq.~\eqref{ana2}; $\mathcal{N}(t)$ is a non-monotonic function of $|\Im[c]|$, which is a manifestation of the quantum Zeno (QZ) effect. 
(b, c) The particles are initially at the same position; the orange and green curves are obtained from Eq.~\eqref{ana1}. On the panel (b) we have $|c| \ll 1$ and no QZ effect emerges; on the panel (c) we have $|c| \gg 1$ and the QZ effect can be seen for $|\Im[c]| \gg \Re[c]$. 
Bottom panels: mean number of particles $\mathcal N(t)$ for the lattice setup as a function of $\gamma$ for $L=121$ and at time $t=L^4$. 
(d) The particles are initially well-separated and the blue curve is obtained from Eq.~\eqref{ana_far}. As in panel (a), $\mathcal{N}(t)$ is a non-monotonic function of $\gamma$, which is a manifestation of the quantum Zeno effect. 
(e) The particles are initially at the same site and the orange curve is obtained from Eq.~\eqref{ana_same}; 
$\mathcal{N}(t)$ is a decreasing function of $\gamma$ which means that no quantum Zeno effect emerges.}
\label{Fig:QZ}
\end{figure}

 In Fig.~\ref{Fig:QZ}, we illustrate this effect by studying $\mathcal N(t)$ at a fixed time $t$ and varying the two-body loss rate; we speak of QZ behaviour when the curve is increasing for increasing loss rate.
 We begin our analysis by considering the lattice situation with the two particles that are initially well-separated; as illustrated in Fig.~\ref{Fig:QZ}~(d), the number of particles has a non-monotonic behavior as a function of $\gamma$. This behavior is completely inherited from the properties of $\Gamma_u$, that is the only time scale appearing in Eq.~\eqref{ana_far}.
 
 However, if the two particles are initially on the same site, an initial exponential decay with a rate $\gamma$ is expected and the number of particles decreases rapidly at first.
 The rapid loss of particles cannot be prevented by considering large values of $\gamma$.
 Indeed, the initial state with a double occupancy is sensitive to loss and even if in the regime of large $\gamma$ the hopping is suppressed, for this initial condition no QZ behaviour is expected.
 Mathematically, the absence of QZ behaviour is enforced by the prefactor $\Gamma_u/\gamma$ in Eq.~\eqref{ana_same}, that decreases rapidly for increasing $U^2 + \gamma^2/4$. 
 With a few algebraic passages, one finds that this enforces the strictly decreasing behavior of $\mathcal{N}(t)$ as a function of $\gamma$ at a given time, as shown in Fig.~\ref{Fig:QZ}~(e). 
 In this situation, thus, increasing $\gamma$ makes the gas more dissipative and no QZ behaviour appears.

 For the continuum case, if the particles are initially well-separated, $\Gamma_c^{-1}$ is clearly the unique time-scale of the dynamics since the scaling function $h_1(\Lconti)$ is independent of $c$, as we can see in Eq.~\eqref{ana2}. 
 Thus, the situation is analogous to the lattice case; the QZ  behavior occurs when $|\Im[c]| \gg \Re[c]$, as we show in Fig.~\ref{Fig:QZ}~(a). 
 
 However, for the initial state consisting of two perfectly overlapping Gaussian wave packets, the scaling function $h_0(\Lconti,c)$, appearing in Eq.~\eqref{ana1} and explicitly written in Eq.~\eqref{eq:h0lc}, depends on~$c$; 
 this makes the discussion slightly more complicated. 
 If $|c| \ll 2 \pi$, the number of particles always decreases faster when $|\Im[c]|$ increases, similarly to the lattice case. 
 On the contrary, if $|c| \gg 2 \pi$ then the QZ behavior emerges in the usual limit $|\Im[c]| \gg \Re[c]$. We illustrate these various cases for two particles in the continuum at the same position initially in~Fig.~\ref{Fig:QZ}~(b) and~(c).

\section{Conclusions and perspectives}
\label{Conclusion}

In this article, we characterized the universal properties of the dynamics of a quantum system composed of two bosons that undergo a two-body loss.
We have focused on two quantities, the scaling of the asymptotic decay rate with the linear system size, $\lambda_{\rm ADR} \sim L^{-\alpha}$, and the late-time dynamics of the mean number of particles, $\mathcal N(t) \sim t^{- \beta}$.
Our results have been carried out with analytical techniques and have shown a nontrivial dependence both on the geometry of the system (a continuum one-dimensional setup or a lattice) and, for what concerns $\beta$, on the initial state (with overlapping or separated particles).

Our main results are summarized in Tables~\ref{Table:2} and~\ref{Table:1} and we briefly recapitulate them here.
Concerning the $\lambda_{\rm ADR}$, in the continuum $\alpha = 3$; this result is obtained also for lattice setups if we restrict the study to the symmetry sector with total momentum equal to zero. 
On the lattice, however, eigenmodes with total momentum close to the edge of the Brillouin zone display an asymptotic decay rate that decreases faster with the linear system size, characterized by $\alpha =5$.

The universal properties of the decay of the number of particles in the system depend on the initial conditions.
When particles initially occupy the same position, we compute $\beta = 3/2$, whereas when the particles are well separated we obtain $\beta = 1/2$.
No other exponent has been found by changing the relative position of the two particles, but rather a crossover between the two: 
when the particles are separated but not too distant, the decay time follows $\beta =1/2$ at short times and then turns into $\beta =3/2$.

The study of $\beta$ on the lattice has revealed the presence of a logarithmic correction which is the consequence of the aforementioned two-fold spectral structure, characterized by an asymptotic decay rate whose scaling depends on the total momentum symmetry sector.
The logarithmic correction cannot be easily fitted from the numerical data, which are compatible with a power-law decay characterized by a non-universal exponent.
Similar non-universal behaviors have been numerically found in other setups~\cite{Rosso_2023} and this work suggests investigating whether these results could be interpreted as logarithmic corrections.

The main outlook of this work is the possibility of performing these calculations for an initial number of particles that has finite density.
In this case, the dynamics cannot be described using the tools of a non-Hermitian dynamics but one has to employ the full Lindblad dynamics.
Yet, our work suggests that the study of the spectral properties of the Lindblad superoperator could be linked to the universal exponent $\beta$.
The investigation of such a connection using analytical or numerical tools is an exciting perspective that we leave for future work.

Finally, we conclude by commenting on the fact that this work is an example of the rich properties that can be associated to a many-particle non-Hermitian dynamics.
It would be interesting to extend this study to the many-particle case using the techniques of Bethe ansatz and to find the universal decay of population in the system due to two-body losses using a non-Hermitian dynamics.
Even if it is known that the study cannot describe the real dynamics of many particles under two-body losses, 
it remains of the utmost interest to assess which properties can be correctly captured with this approach, that is technically less demanding.

%
\begin{acknowledgments}
We acknowledge insightful discussions with R.~Fazio, B.~Laburthe-Tolra, B.~Pasquiou and M.~Robert-de-Saint-Vincent.
L.M.~thanks A.~Biella, J.~De Nardis and L.~Rosso for enlightening discussions on previous works on related topics.
This work was carried out in the framework of the joint Ph.~D.~program between the CNRS and the University of Tokyo.
This work is supported by the ANR project LOQUST ANR-23-CE47-0006-02. 
H.Y.~was supported by JSPS KAKENHI Grant-in-Aid for JSPS fellows Grant No. JP22J20888, the Forefront Physics and Mathematics Program to Drive Transformation, and JSR Fellowship, the University of Tokyo. 
H.K. was supported by JSPS KAKENHI Grants No. JP23H01086, No. JP23H01093, and MEXT KAKENHI Grant-in-Aid for Transformative Research Areas A “Extreme Universe” (KAKENHI Grant No. JP21H05191).
A.N. acknowledges financial support from \textit{Fondazione Angelo dalla Riccia}, LoCoMacro 805252 from the European Research Council, and LPTMS for warm hospitality. 

\end{acknowledgments}
%
%
%
%
\appendix
%
%
%

\section{Details about two fermions in a spin singlet on a lattice} 
\label{Ap0}

In this Appendix, we examine in more detail the case of two spin-$1/2$ fermions in a singlet trapped on a lattice.
We denote $c_{j,\sigma}^{\dag}$ and $c_{j,\sigma}$ the creation and annihilation operators of a fermion with spin $\sigma \in \{ \uparrow, \downarrow \}$ at site $j \in \{ 1 , \ldots ,L \}$; 
$n_{j,\sigma} = c_{j,\sigma}^{\dag} c_{j,\sigma}$ counts the number of particles at site $j$ with spin $\sigma$
and the operator counting the total number of particles is
$N = \sum_{j=1}^L \sum_{\sigma = \uparrow, \downarrow} n_{j,\sigma}$. 

The master equation of this fermionic problem is given by Eq.~(\ref{MElattice}) with $H$ 
being the ${\rm SU}(2)$ Fermi-Hubbard Hamiltonian~\cite{essler_2005} $H = \Hhop + \Hint$ with
\begin{align}
    \Hhop &= - J \sum_{j =1}^L \sum_{ \sigma = \uparrow, \downarrow} \left( c_{j+1,\sigma}^{\dag} c_{j,\sigma} +  h.c. \right), \\
    \Hint &= U \sum_{j = 1}^L n_{j, \uparrow} n_{j, \downarrow}.
\end{align}
The jump operators are
\begin{equation}
    L_j = \sqrt{\gamma} c_{j, \downarrow}  c_{j, \uparrow}. 
\end{equation}
Thus, the effective non-Hermitian Hamiltonian is  
\begin{equation}
    \Hef = H - \frac{i}{2} \sum_{j=1}^L L_j^\dag L_j = \Hhop + \Hint^\prime
    \label{eff_fermi}
\end{equation}
with
\begin{equation}
    \Hint^\prime = \left( U - i \frac{\gamma}{2} \right) \sum_{j = 1}^L n_{j, \uparrow} n_{j, \downarrow}.
\end{equation}
We also note that $\Hef$ is SU(2) symmetric, and in particular, commutes with the total spin operator $\hat S^2$.

For this fermionic system, the dynamics of the mean number of particle $\mathcal{N}(t)$ is the same as for the bosonic lattice system described in Sec.~\ref{sec:twoBosonsLatt}.
Indeed, we can always decompose a fermionic two-body initial state into a spin singlet part and a spin triplet part $\ket{\Psi_0}$ = $\ket{\Psi_0^{T}}$ + $\ket{\Psi_0^{S}}$ with $\bra{\Psi_0^{T}} \ket{\Psi_0^{S}} =0$ because the two states are eigenvectors of the Hermitian operator $S^2$ with different eigenvalues. 
The spin triplet part is associated with an antisymmetric orbital part, which cannot contain any double occupancy and thus does not yield any two-fermion decay, as exemplified by $\Hint^\prime \ket{\Psi_0^{T}} = 0$.  
As a consequence, using the spin symmetry of the Hamiltonian, 
we can single out two contributions to the time evolution of the average number of particles in Eq.~\eqref{eqN2}:
\begin{equation}
    \braket{\Psi(t)} =  \braket{\Psi_0^{T}} + \braket{\Psi^{S}(t)} \label{eqST}
\end{equation}
where we defined
\begin{equation}
\ket{\Psi^{S}(t)} = e^{-i \Hef t} \ket{\Psi_0^{S}}. \label{eqS}
\end{equation}
This shows that the spin triplet part of $\ket{\Psi_0}$ has a trivial, time-independent, contribution to $\braket{\Psi(t)}$. 
For this reason, it is more interesting to consider $\ket{\Psi_0}$ to be a spin singlet state, meaning that $\ket{\Psi_0^{T}} =0$ and $\ket{\Psi_0}  = \ket{\Psi_0^{\orb}} \otimes ( \ket{\uparrow \downarrow}  - \ket{\downarrow \uparrow} ) / \sqrt{2} $, with the orbital part $\ket{\Psi_0^{\orb}}$ being symmetric under particle exchange. 
This makes the fermionic problem completely equivalent to the study of two spinless bosons.

\section{Derivation of Eq.~\eqref{eqN2}} 
\label{sec:derivation_particle_number}
In this Appendix, we derive Eq.~\eqref{eqN2}. Using the decomposition \eqref{GeneME}, $\mathcal{N}(t)$ can be calculated as
\begin{equation}
    \mathcal{N}(t) = \Tr[N e^{\mathcal{L}t}[\rho(0)] ]=\sum_{n=0}^\infty \frac{t^n}{n!} \Tr\left[N(\mathcal{K}+\mathcal{J})^n[\rho(0)]\right].
    \label{eq:decomp}
\end{equation}
Then, $\mathcal{K}$ conserves the number of particles and $\mathcal{J}$ decreases the number of particles by two. Since $N\rho(0)=2\rho(0)$ for the initial state, only the terms in the form 
$N\mathcal{K}^n[\rho(0)]=2\mathcal{K}^n[\rho(0)]$ are nonvanishing in the right-hand side of Eq. \eqref{eq:decomp}. Therefore,
\begin{align}
   \mathcal{N}(t)
   &=2\sum_{n=0}^\infty \frac{t^n}{n!} \Tr\left[\mathcal{K}^n [\rho(0)]\right]
   =2\Tr\left[e^{\mathcal{K}t}[\rho(0)]\right] \nonumber \\
   &=2\Tr\left[e^{-iH_\mathrm{eff} t}\ket{\Psi_0}\bra{\Psi_0} e^{iH ^\dagger_\mathrm{eff} t}\right] \nonumber \\
   &=2\braket{\Psi(t)},
\end{align}
where $\ket{\Psi(t)}=e^{-iH_\mathrm{eff}t}\ket{\Psi_0}$.

\begin{widetext}
\section{The $L^{-5}$ scaling in the case of $N>2$ fermions on a lattice}
\label{sec:many_body}
In this Appendix, we show the existence of eigenvalues whose imaginary part scales with $L^{-5}$ in the case of $N>2$ spin-$1/2$ fermions with $M$ down spins on $L$ lattice sites. We consider the dilute case: we fix $N$, $M$ and take $L$ large. We also assume that $|u|=|U/4-i\gamma/8|$ is large. When $|u|$ is large enough, the spin-charge separation occurs~\cite{ogata_1990}, and the eigenvalues of $H_\mathrm{eff}$ are given by~\cite{essler_2005}
\begin{align}
 E&=-2 \sum_{j=1}^N \cos \left(k_j\right)+\frac{2 e_s}{u}\left[\frac{N}{L} \sum_{j=1}^N \sin ^2\left(k_j\right)-\frac{1}{L}\left(\sum_{j=1}^N \sin \left(k_j\right)\right)^2\right]+O(1/u^2),
\end{align}
where
\begin{equation}
    e_s=-\frac{2}{N}\sum_{l=1}^M \frac{1}{\Lambda_l^2+1}
\end{equation}
and $k_j$ and $\Lambda_l$ are charge quasimomenta and spin rapidities, respectively, which satisfy
\begin{align}
    k_j=\frac{2 \pi n_j}{L}+\frac{P_s}{L},\qquad 
    \prod_{l=1}^M \frac{\Lambda_l-i}{\Lambda_l+i} = e^{i P_s}
\end{align}
with $n_j \in \{0,\ldots, L-1\}$ and $P_s =\frac{2\pi m}{N}$ $(m \in \{0,\ldots, L-1\})$. Then, the imaginary part of $E$ reads
\begin{align}
    \Im E 
    &=\Im \left(\frac{2 e_s}{u L}\right)\sum_{1\leq j < l \leq N}(\sin k_j-\sin k_l)^2+O(1/u^2)\\
    &=\Im \left(\frac{8 e_s}{u L}\right)\sum_{1\leq j < l \leq N}\cos^2 \left(\frac{k_j+k_l}{2}\right)\sin^2 \left(\frac{k_j-k_l}{2}\right)+O(1/u^2).
    \label{eq:manybodyenergy}
\end{align}
If we choose $k_j$ as $k_j= \frac{\pi}{2}-p_j\frac{\pi}{L}$, where all $p_j$ are distinct constants, we have
\begin{align}
    \Im E 
    &=\Im \left(\frac{8 e_s}{u L}\right)\sum_{1\leq j < l \leq N}\sin^2 \left[\frac{\pi(p_j+p_l)}{2L}\right]\sin^2 \left[\frac{\pi(p_j-p_l)}{2L}\right]+O(1/u^2)\\
    &= \frac{1}{L^5}\Im \left(\frac{\pi^4 e_s}{2u}\right)\sum_{1\leq j < l \leq N} \left(p_j^2-p_l^2\right)^2+O(1/u^2)+O(1/L^7).
\end{align}
Since $p_j$, $N$ and $e_s$ are constants with respect to $L$, $\Im E$ scales with $L^{-5}$. Note that this scaling does not hold when the fermionic density $N/L$ is fixed, in which case $\Im E \propto L^{-2}$~\cite{Nakagawa_2021}.

\end{widetext}

\section{Derivation of Eq.~(\ref{PsiL})} 
\label{ApLeftEV}
In this Appendix, we derive Eq.~(\ref{PsiL}); we make also reference to Ref.~\cite{Tanaka_2013} where this link between left and right eigenvectors was already present. 
First, we note that $(k_1,k_2)$ is a solution of the Bethe equations associated to the diagonalization of $\Hef$, if and only if $(k_1^*,k_2^*)$ is a solution of the Bethe equations associated to the diagonalization of $\Hef^\dag$. 
In the continuum case, for instance, 
starting from a solution of
the Bethe equation
\begin{equation}
    e^{i k_1 \ell}  = - \frac{c - i \left( k_1 -  k_2 \right)}{c + i \left( k_1 - k_2 \right)}, \qquad k_1 + k_2 = \frac{2 \pi}{\ell} \mathbb{Z},
\end{equation}
we obtain a solution for the Bethe equation for $\Hef^\dagger$
\begin{equation*}
\begin{split}
  e^{i k_1^* \ell} & = - \frac{c^* - i \left( k_1^* - k_2^* \right)}{c^* + i \left( k_1^* - k_2^* \right)}, \qquad k_1^* + k_2^* = \frac{2 \pi}{\ell} \mathbb{Z}.
\end{split}
\end{equation*}
The solution $(k_1^*,k_2^*)$ corresponds to an energy $E_{k_1 k_2}^*$ and a total momentum $k_1^* + k_2^* = k_1 + k_2$. Therefore, we have the proportionality relation $\ket{\Psi_{k_1 k_2}^{L}} = A \ket{\Psi_{k_1^* k_2^*}^{R}}$.
The normalization constant $A$ is set by the constraint $\bra{\Psi_{k_1 k_2}^R} \ket{\Psi_{k_1 k_2}^L} = 1$.
\section{Linearized quasimomenta in the continuum} 
\label{Ap2} 

In this Appendix, we first derive Eq.~(\ref{dconti}). If $|\delta /c| \ll 1$ and $p$ is even, Eq.~(\ref{Beconti2}) becomes $e^{\frac{i}{2} \delta \Lconti}e^{- i \pi} = \frac{1 - i \frac{\delta}{c}}{1 + i \frac{\delta}{c}} \approx 1 - 2 i \frac{\delta}{c}$. Thus, we have $ i \frac{ \delta \Lconti }{2} - i \pi \approx \ln(1-2 i \frac{\delta}{c}) + 2 \pi i q \approx - 2 i \frac{\delta}{c} + 2 \pi i q$  with $q \in  \mathbb{Z}$. Finally, we obtain $\delta \approx \frac{4 \pi q + 2 \pi}{\Lconti + \frac{4}{c}}$. In Eq.~\eqref{dconti}, we only consider $q\geq 0$ to avoid double counting of the eigenstates and eigenenergies. The case $p$ odd is treated in a similar way. 

Let us now discuss Eq.~\eqref{eq:lin_conti}.
If we consider $p=0$ as an example, we have $\frac{\Im[\mathcal E_{p=0,q}]}{\Re[\mathcal E_{p=0,q}]} = \frac{\Im[\delta^2]}{\Re[\delta^2]} = \frac{\Im[\tilde \delta^2]}{\Re[\tilde \delta^2]}$ where $\tilde \delta = \frac{\delta}{4 \pi q + 2 \pi} = (\Lconti + 4/c)^{-1}$. Thus, $\frac{\Im[\mathcal E_{p=0,q}]}{\Re[\mathcal E_{p=0,q}]}$ is independent of $q$ and we obtain Eq.~\eqref{eq:lin_conti}.
\section{Asymptotic value of $|\Im[E]|$ for large $\delta$ in the continuum}
\label{NewAppendix}
In this Appendix, we derive the expression of the asymptotic value of $|\Im[E]|$ in the limit $|c/\delta| \ll 1$ in the continuum case. 
If $|c/\delta| \ll 1$, Eq.~(\ref{Beconti2}) becomes $e^{ \frac{i}{2} \delta \ell} = (-1)^p \frac{1 + i \frac{c}{\delta}}{1 - i \frac{c}{\delta}} \approx (-1)^p \left( 1 + 2 i \frac{c}{\delta}  \right)$. 
Thus, we have $e^{ i \delta \ell} \approx \left( 1 + 4 i \frac{c}{\delta}  \right)$, and $ i \delta \ell  \approx \ln ( 1 + 4 i \frac{c}{\delta}  ) + 2 \pi  i q \approx  4 i \frac{c}{\delta} + 2 \pi i q$ where $q \in \mathbb{Z}$. 
We obtain $\delta^2 \ell - 2 \pi q \delta - 4 c = 0$ which is solved by $\delta_\pm = \frac{\pi q}{\ell} \left( 1 \pm (1 + \frac{2 c \ell }{(\pi q)^2})\right)$ 
in the limit of large integers $q \gg |c| \ell$; 
in order to be consistent with the numerical solution we should select the solution $\delta_+$. 
Finally, we find $|\Im[E]| \approx |\Re[\delta_+] \Im[\delta_+]| \approx \frac{4 |\Im[c]|}{\ell}$.

\section{Energies of the repulsive bound states on the lattice}
\label{Appendix_BoundStates}
\begin{figure}[t]
\centering
\includegraphics[width=\linewidth]{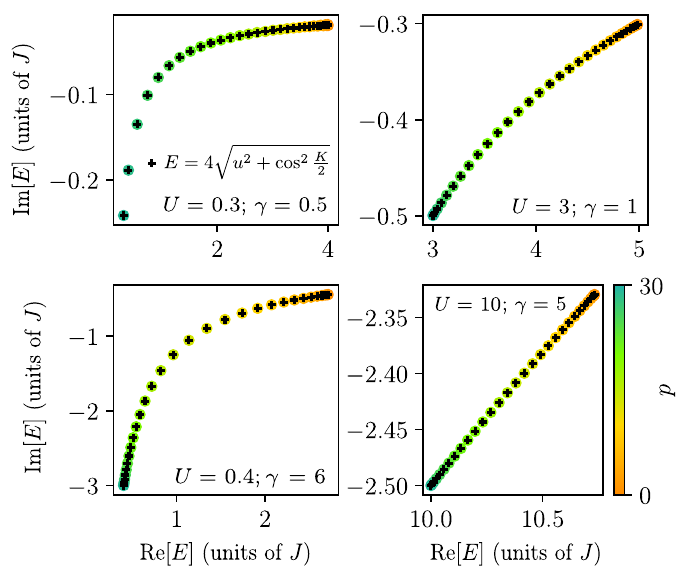}
\caption{Bound states energies of $\Hef$ on a lattice of $L=61$ sites for various values of $K =\frac{2 \pi}{L}p$, $U$ and~$\gamma$. The colored dots are obtained by a numerical solution of the Bethe equations. The black crosses correspond to the analytical expression $E_b = 4  \sqrt{ \cos^2 \frac{K}{2} + u^2}$.}
\label{BoundSates}
\end{figure}
 In this Appendix, we show that the short-lived branch on the lattice is composed of repulsive bound states and we derive the expression of their energies. In the large system size limit, bound states are associated to a vanishing or diverging scattering phase $s_{k_1 k_2}$  given by Eq.~\eqref{squatPhaseLatt} on the lattice. Thus, bound states correspond to quasimomenta $(k_1,k_2)$ satisfying $2 i u = \pm \left( \sin k_1 - \sin k_2 \right)$. By doing the change of parameters $K = k_1 + k_2$ and $\delta = k_1 - k_2$, we find $u = \mp i \cos \frac{K}{2} \sin \frac{\delta}{2}$. We substitute the latter constraint in the expression of the eigenenergies:
\begin{equation}
    \begin{split}
E &= - 4 \cos \frac{K}{2} \cos \frac{\delta}{2} = \pm 4 \cos \frac{K}{2} \sqrt{1 - \sin^2 \frac{\delta}{2}} \\ &= \pm 4  \sqrt{ \cos^2 \frac{K}{2} -  \cos^2 \frac{K}{2} \sin^2 \frac{\delta}{2}} = \pm 4  \sqrt{ \cos^2 \frac{K}{2} + u^2}.
    \end{split}
\end{equation}
 In order to be consistent with the numerical solution of the Bethe equations, we should select $E_b =  4  \sqrt{ \cos^2 \frac{K}{2} + u^2}$, with the common convention that the branch cut of square root in the complex plane is along the negative real axis $(-\infty, 0)$. In Fig.~\ref{BoundSates}, we show a comparison between the energies of the bound states found from this numerical solution and the latter analytical expression for various values of $K$, $U$ and $\gamma$. This highlights the perfect agreement between the numerical and analytical solutions.
 
The numerical solutions corresponding to $s_{k_1 k_2} = 0$ always satisfy $\Im[\delta] < 0$; meaning that the wave function is well-normalized since $\Psi^R (x_1,x_2) \propto e^{ i K \frac{ x_1 + x_2 }{2}} e^{ - i \frac{\delta}{2} | x_1 - x_2 |}$ (see Eq.~\eqref{PsiR}). Indeed, if $\Im[\delta] <0$ then the wave function vanishes exponentially fast when the distance between the two particles increases. In the same way, the numerical solutions corresponding to $|s_{k_1 k_2}| \to \infty$ always satisfy $\Im[\delta] > 0$.
\begin{widetext}
\section{Exact expressions of $\tilde f(p, \delta)$}
\label{Expr_f}
In this Appendix, we give the exact expressions of~$\tilde f(p,\delta)$ for the four initial states studied in the main text.
In the continuum, assuming $\Lconti \gg 1$, 
\begin{enumerate}[label=(\roman*)]
  \item if the initial wave-function is $\Psi_0(x_1,x_2) = \frac{1}{\sqrt{\pi}} \left( e^{-\left( x_1 - \frac{\Lconti}{2} \right)^2} e^{-\left( x_2 - \frac{\Lconti}{2} \right)^2} \right)$, then
  \begin{equation}
    \tilde f(p,\delta) = \frac{64 \pi  (c-c^*) e^{-\frac{1}{8} (2 (\frac{2 \pi p}{\Lconti})^2 + {\delta}^2 + {\delta^{*}}^2) } {|\delta|^2}  |  {\delta} + c  \erfi ( \frac{{\delta} }{2 \sqrt{2}} ) |^2}{ \Lconti | 4 c + \Lconti \left( c^2 + {\delta}^2 \right)|^2  \left({\delta}^2 - {\delta^{*}}^2 \right)};
\end{equation}
    \item if the initial wave-function is $\Psi_1(x_1,x_2) = \frac{1}{\sqrt{2 \pi}} \left( e^{-(x_1-\frac{\Lconti}{4})^2} e^{-(x_2-3 \frac{\Lconti}{4})^2} + e^{-(x_2-\frac{\Lconti}{4})^2} e^{-(x_1-3 \frac{\Lconti}{4})^2} \right)$  , then
\begin{equation}
    \tilde f(p,\delta) = 
    \frac{32  \pi (c-c^*) e^{-\frac{1}{8} (2 (\frac{2 \pi p }{\Lconti})^2 + \delta^2 + \dc^2)}  |\delta|^2  }{\Lconti |\delta^2 + c^2 |\left(\delta^2 - \dc^2 \right) | 4 c + \Lconti (c^2 + \delta^2) |^2} \left| (-1)^p (c+i\delta) \left( - i \delta+ c \erf{\frac{\Lconti-i \delta}{2 \sqrt{2}}} \right)+ (c-i\delta) \left( i \delta+ c \erf{\frac{\Lconti +i \delta}{2 \sqrt{2}}} \right) \right |^2 ,
 \end{equation} 
\end{enumerate}
where by definition, $\erf(z) := \frac{2}{\sqrt{\pi}} \int_0^z e^{-t^2} dt $ and $\erfi(z) := - i \erf(i z)$.

On the lattice,
\begin{enumerate}[label=(\roman*)]
\item if the initial state is $\ket{\Psi_2} = b_{x_0}^\dag b_{x_0}^\dag  \ket{v}/\sqrt{2}$, then
\begin{equation}
    \tilde f(p,\delta) = \frac{8 |\sin^4(\frac{\delta}{2})| \cos^3(\frac{p \pi}{L}) \left( u - u^*\right)}{L |L u^2 + 2 u \cos(\frac{p \pi}{L}) \cos(\frac{\delta}{2})+ L \cos^2(\frac{p \pi}{L}) \sin^2(\frac{\delta}{2})|^2 \left(\cos(\frac{\delta^*}{2}) -\cos(\frac{\delta}{2})\right)};
\end{equation}
\item if the initial state is $\ket{\Psi_3} = b_{x_0}^\dag b_{x_0+(L-1)/2}^\dag \ket{v}$, then 

  \begin{equation}
    \tilde f(p,\delta) = \frac{8 \cos(\frac{p \pi}{L}) \left( u - u^*\right) |\sin(\frac{\delta}{2})|^2 |u^2 + \cos^2(\frac{p \pi}{L}) \sin^2(\frac{\delta}{2})| \left(  \cos(\frac{\dc - \delta}{4}) + (-1)^p \cos(\frac{\dc + \delta}{4}) \right)}{L |L u^2 + 2 u \cos(\frac{p \pi}{L}) \cos(\frac{\delta}{2})+ L \cos^2(\frac{p \pi}{L}) \sin^2(\frac{\delta}{2})|^2 \left(\cos(\frac{\delta^*}{2}) -\cos(\frac{\delta}{2})\right)},
    \label{Eq:tildef:AppendixG:Psi3}
\end{equation}
\end{enumerate}
where by definition, $\cos(z) := \frac{e^{i z} + e^{- i z}}{2}$ and $\sin(z) := \frac{e^{iz} - e^{- i z}}{2 i }$ for $z \in \mathbb C$.
\end{widetext}

\section{Derivation of Eq.~(\ref{ana1})}
\label{ApSumtoInt}
In this Appendix, we derive Eq.~(\ref{ana1}). Assuming $\Lconti |c| \gg 1$ and $\Lconti \gg 1$, we find  
\begin{equation}
    g_0(\Lconti,c) = \frac{32 (c-c^*)(2 \pi +|c|^2 + \sqrt{2 \pi} ( c+ c^*))}{|c|^4 \Lconti^3 }.
    \label{Eq:g0:Appendix}
\end{equation}

We first rewrite the sum over $p$ in Eq.~(\ref{Nsamepos}) as a sum over even $p$ and a sum over odd $p$ because the expression for the linearized quasimomenta $\delta$ in Eq.~\eqref{dconti} depends only on the parity of $p$ and never on its actual value. This means that instead of defining $\mathcal S_p^{(0)}$, the set of allowed values of $\delta$ for a given $p$, as we did in the main text, we can simply define $\mathcal S_e^{(0)}$ and
$\mathcal S_o^{(0)}$, where $e$ stands for ``even'' and $o$ for ``odd".
The double sum now takes a factorized expression:
\begin{align}
    \mathcal N(t) & \approx  g_0(\Lconti,c) \left(\sum_{p \text{ even}} e^{- \frac 14 \left( \frac{2 \pi p}{\Lconti} \right)^2} \right)
    \left( \sum_{\delta \in \mathcal S^{(0)}_e} 
    \frac{|\delta|^4 e^{- \frac i2 (\delta^2 - \delta^{* 2}) t}}
    {\delta^2- \delta^{*2}}\right) \nonumber \\
    &+ g_0(\Lconti,c) \left(\sum_{p \text{ odd}} e^{- \frac 14 \left( \frac{2 \pi p}{\Lconti} \right)^2} \right)
    \left( \sum_{\delta \in \mathcal S^{(0)}_o} 
    \frac{|\delta|^4 e^{- \frac i2 (\delta^2 - \delta^{* 2}) t}}
    {\delta^2- \delta^{*2}}\right).
\end{align}
 In the large $\ell$ limit, we transform the sums over $p$ into integrals
\begin{equation}
\begin{split}
    \sum_{p \mbox{ \scriptsize{even}}} e^{-\frac{ 1}{4}(\frac{2\pi p }{\Lconti})^2}& \approx \sum_{p \mbox{ \scriptsize{odd}}} e^{-\frac{ 1}{4}(\frac{2\pi p }{\Lconti})^2} \\ & \approx \int_{-\infty}^{\infty}  e^{-\frac{ 1}{4}(\frac{4 \pi  p}{\Lconti})^2 } \,dp = \frac{\Lconti}{2 \sqrt{\pi}}.
\end{split}
\end{equation}
Thus, we have
\begin{equation}
     \mathcal{N}(t) \approx  g_0(\Lconti,c) \frac{\Lconti}{2 \sqrt{\pi} } \sum_{a = e,o}\underbrace{\sum_{\delta \in \mathcal{S}^{(0)}_a}  \frac{|\delta|^4 }{{\delta}^2 - {\delta^{*}}^2} e^{-\frac{i}{2} ({\delta}^2 - {\delta^{*}}^2)t}}_{\mathcal{J}_a}.  \label{15}
 \end{equation}  
We use the definitions of $\mathcal{S}^{(0)}_e$ and $\mathcal{S}^{(0)}_o$ and the fact that $\Lconti|c| \gg 1$ to find the following expression, valid for $a =  e,o$:
 \begin{equation}
\mathcal{J}_a = \frac{i 128 \pi^4}{\tilde \Gamma_c \Lconti} \sum_{q = 0}^{\infty} (q+D_a)^2 e^{ -  (q+D_a)^2 \frac{\tilde{\Gamma}_c}{\Lconti^3} t},
\end{equation}   
with $D_e = 1/2$ and $D_o = 1$; we denote $\tilde{\Gamma}_c = 128 \pi^2 \frac{|\Im[c]|}{|c|^2} = 128 \pi^2 \Gamma_c$. 

 When $| \frac{\tilde{\Gamma}_c}{\Lconti^3} t  |\ll 1$, we can approximate the sum by an integral, because $(q+D_a)^2 e^{ -  (q+D_a)^2 \frac{\tilde{\Gamma}_c}{\Lconti^3} t}$ is slowly varying when $q$ varies by $1$: 
 \begin{equation}
 \mathcal{J}_a = \frac{i 128 \pi^4}{\tilde \Gamma_c \Lconti} \int_{0}^{\infty} dq  (q+D_a)^2  e^{ -  (q+D_a)^2 \frac{\tilde{\Gamma}_c}{\Lconti^3} t}.
\end{equation}   
Changing variable and using 
$y=(q+D_a)^2 \frac{\tilde{\Gamma}_c}{\Lconti^3} t$, we obtain
 \begin{equation}
     \mathcal{J}_a =  \frac{i 64 \pi^4}{  \tilde \Gamma_c \Lconti}  \left(  \frac{\tilde{\Gamma}_c}{\Lconti^3} t \right)^{-3/2} \int_{D_a^2 \frac{\tilde{\Gamma}_c}{\Lconti^3} t}^{\infty} \,dy \sqrt{y} e^{-y} 
\end{equation}   
The lower bound of the integral can be approximated by zero when 
$| \frac{\tilde{\Gamma}_c}{\Lconti^3} t  |\ll 1$; in this limit the integral evaluates to $\sqrt{\pi}/2$. 
This leads us to Eq.~(\ref{ana1}) with
\begin{equation}
    h_0(\Lconti,c) = \frac{\Lconti^{3/2} \left(2 \pi + |c|^2  + 2 \sqrt{2 \pi} \Re[c] \right)  }{  64 \sqrt{2}  \pi |c|^2}.
    \label{eq:h0lc}
\end{equation}
Eq.~\eqref{ana2} can be derived along similar lines;
here,
we just report the explicit expressions of $g_1(\Lconti,c)$ and $h_1(\Lconti)$ that we obtained assuming $\Lconti |c| \gg 1$ and $\Lconti \gg 1$:
\begin{align}\label{eq:g1lc}
    g_1(\Lconti,c) = & \begin{cases} \dfrac{128 \pi (c-c^*)}{ \Lconti^3 |c|^2}, \quad& \mbox{for } p \mbox{ is even}; \qquad \\ \\ 0, \quad& \mbox{for } p \mbox{ is odd};   \end{cases}
\\
h_1(\Lconti) = & \frac{  \sqrt{\Lconti}}{2 \sqrt{2}}.\label{eq:h1l}
\end{align}

\section{Derivation of Eq.~(\ref{ana_far})}
\label{ApDerLatt}
In this Appendix, we derive Eq.~(\ref{ana_far}). 
By using the total momentum and the relative quasimomentum,
Eq.~\eqref{Evol:2} can be rewritten as
\begin{equation}
    \mathcal{N}(t)  = \sum_{p=0}^{L-1}\sum_{\delta \in \mathcal {S}_p'}   \tilde f \left( p, \delta \right)  e^{-\frac{i}{2} ({\delta}^2 - {\delta^{*}}^2) \cos( \frac{ p \pi}{L}) t} ,
\end{equation}
where $\mathcal S'_p$ is the set of solutions of the equations~\eqref{BeLattice2}; in order to avoid double counting of solutions, 
we always take $0<\Re[\delta]<2 \pi $. 
The explicit expression for $\tilde f(p, \delta)$, obtained using the definitions given in Sec.~\ref{SecSpectrum} and an initial state of the form
\begin{equation}
    \ket{\Psi_3}  = b_{x_0}^\dag b_{x_0+(L-1)/2}^\dag \ket{v} ,
\end{equation}
 is reported in Eq.~\eqref{Eq:tildef:AppendixG:Psi3}, see Appendix~\ref{Expr_f}. 
 Similarly to the continuum case, if we consider a sufficiently large time, then the dynamics is well captured by the linearized spectrum, and we have to sum over the $\delta$ in the set $\mathcal S^{\prime (0)}_p$ and $\mathcal S^{\prime (2 \pi)}_p$, corresponding to the values given in Eq.~\eqref{d0} and Eq.~\eqref{d2pi} respectively. 
 We assume the following symmetry between the sums over $\delta$ close to $0$ and over those close to $2 \pi$: 
 \begin{equation}
 \begin{split}
           \sum_{\delta \in \mathcal S_p^{\prime(0)}} &  \tilde f \left( p, \delta \right)  e^{-\frac{i}{2} ({\delta}^2 - {\delta^{*}}^2) \cos( \frac{ p \pi}{L}) t}  \\ & \approx \sum_{\delta \in \mathcal S_p^{\prime (2 \pi)}}   \tilde f \left( p, \delta \right)  e^{-\frac{i}{2} ({\delta}^2 - {\delta^{*}}^2) \cos( \frac{ p \pi}{L}) t} ,
 \end{split}
 \end{equation}
 which we verified numerically. 
 This observation significantly simplifies the calculation, because we can sum only over $\mathcal S_p^{\prime (0)}$ and approximate $\tilde f \left( p, \delta \right)$ by its expansion for $\delta \sim 0$, that we call $\tilde f_3 \left( p, \delta \right)$. 
 Assuming $L|u| \gg 1$, we have
 \begin{equation}
    \tilde f_3 \left( p, \delta \right) = \frac{16 \left( 1 + (-1)^p \right) \left( u - u^* \right) \cos( \frac{ p \pi}{L})}{|u|^2 L^3} \frac{|\delta|^2}{{\delta}^2 - {\delta^*}^2}.
\end{equation}
Note that $\tilde f_3 \left( p, \delta \right) = 0$ if $p$ is odd. This technical simplification does not appear for the initial state consisting of two bosons at the same site and it makes the derivation of Eq.~\eqref{ana_same} more lengthy. 
By using the analytical expression of the linearized quasimomenta Eq.~(\ref{d0}), we find 
\begin{equation}
\label{eq:particle_number_lattice_sum_reduction}
    \mathcal{N}(t) \approx \frac{16}{L^2} S
\end{equation}
with
\begin{equation}
    S = \sum_{p=0}^{(L-1)/2} \sum_{q=0}^{(L-1)/4 -1} e^{- \frac{1}{2} \left( q+ \frac{1}{2}\right)^2 \cos^2(\frac{2\pi p}{L}) \xi},
\end{equation}
and with $\xi = \frac{\Gamma t}{L^3}$. 
Without loss of generality (provided that $L$ is odd) we consider $L \equiv 1\pmod4$.
Since we are interested in the intermediate time regime 
that is located after the initial transient dynamics but before the exponential asymptotic decay, we will consider $1 \ll \xi \ll L^2$ to further simplify the expression for $S$.
Under this assumption, we can perform a series expansion of the cosine at the exponent for $p$ close to $L/4$ and extend the ranges of the two sums towards infinity to arrive at
\begin{equation}
    S \approx \sum_{p=-\infty}^{\infty} \sum_{q=0}^{\infty} e^{- \frac{2\pi^2\xi}{L^2} \left( q+ \frac{1}{2}\right)^2 \left(p-\frac{L}{4} \right)^2}.
\end{equation}
We now change variable $p \rightarrow -p + (L-1)/4$, so that $p-L/4$ becomes $p+1/4$:
\begin{equation}
    S \approx
    \sum_{p=-\infty}^{\infty} \sum_{q=0}^{\infty} e^{- \frac{2\pi^2\xi}{L^2} \left( q+ \frac{1}{2}\right)^2 \left(p+\frac{1}{4}\right)^2}.
\end{equation}
Using the Fourier representation of the Gaussian $e^{-\sigma_q (p+\frac{1}{4})^2}=\sqrt{\frac{\pi}{\sigma_q}} \int_{-\infty}^\infty e^{-\frac{x^2}{4\sigma_q}+i \frac{x}{4}} e^{ipx} \frac{dx}{2\pi}$ with $\sigma_q = \frac{2\pi^2\xi}{L^2} \left( q+ \frac{1}{2}\right)^2$, we introduce a Dirichlet kernel $\sum_{p=-\infty}^{\infty} e^{i p x}=2\pi \sum_{n=-\infty}^{\infty} \delta(x-2\pi n)$: 
\begin{equation}
\begin{split}
    S \approx & \sum_{q=0}^{\infty} \frac{1}{2 \sqrt{\sigma_q \pi}} \int_{-\infty}^{\infty} e^{-\frac{x^2}{4\sigma_q}+i \frac{x}{4}} \sum_{p = - \infty}^{\infty}e^{ipx} dx \\
    = & \sum_{q= 0}^{\infty} \sqrt{\frac{\pi}{\sigma_q}} \sum_{n= -\infty}^{\infty} e^{-\frac{\pi^2 n^2}{\sigma_q}} i^n.
\end{split}
\end{equation}
The terms corresponding to $n$ odd are canceled out by those corresponding to $-n$. 
We obtain
\begin{equation}
    \begin{split}
    S \approx \sum_{q=0}^\infty \underbrace{\sqrt{\frac{1}{2\pi\xi(\frac{q}{L}+\frac{1}{2L})^2}} \theta_4\left(e^{-\frac{1}{\frac{\xi}{2}\left(\frac{q}{L}+\frac{1}{2L}\right)^2}}\right)}_{F(q)}
    \end{split}
\end{equation}
where $\theta_4(q)=\sum_{n=-\infty}^{\infty}(-1)^n q^{n^2}$ is one of the Jacobi theta functions~\cite{Abramowitz_1974}.
In the large $L$ limit, we substitute the summation with an integral and include the first and second Euler-Maclaurin correction~\cite{Abramowitz_1974}:
\begin{equation}
    S \approx \int_{0}^{\infty} F(q) dq + \frac{F(\infty)+ F(0)}{2} + \frac{F'(\infty)- F'(0)}{12}.
\end{equation}
 In the $\xi\ll L^2$ limit, the correction $\frac{F(\infty)+ F(0)}{2} + \frac{F'(\infty)- F'(0)}{12}$ reads $\frac{4}{3}\sqrt{\frac{L^2}{2\pi\xi}}$. Changing the variable and using $x = \frac{L}{\sqrt{\xi} \left( q+1/2\right)}$, we obtain
 \begin{equation}   
    S \approx
    \sqrt{\frac{L^2}{2\pi\xi}}\left(\frac{4}{3}+
    \underbrace{\int_{0}^{2L/\sqrt{\xi}} \frac{dx}{x} \theta_4\left(e^{-2x^2}\right)}_{I}\right).
\end{equation}
Notice that $2L/\sqrt{\xi}\gg1$; at large $x$, $\theta_4\left(e^{-2x^2}\right)\approx1$. The integral is therefore logarithmically divergent.
We can rewrite it by introducing a cutoff scale $1\ll\Lambda\ll 2L/\sqrt{\xi}$ as
\begin{equation}
    \begin{split}
    I    \approx 
    \int_{0}^{\Lambda} \frac{dx}{x} \theta_4\left(e^{-2x^2}\right)
    +    \int_{\Lambda}^{2L/\sqrt{\xi}} \frac{dx}{x}.
    \end{split}
\end{equation}
In the second integral we approximated $\theta_4\left(e^{-2x^2}\right)\approx1$, since $\Lambda\gg1$.
We thus get
\begin{equation}
    \begin{split}
    \int_{0}^{2L/\sqrt{\xi}} \frac{dx}{x} \theta_4\left(e^{-2x^2}\right) \approx
    \zeta + \log\left(\frac{2L}{\sqrt{\xi}}\right)
    \end{split}
\end{equation}
where
\begin{equation}
    \zeta = \lim_{\Lambda\to\infty} \int_{0}^{\Lambda} \frac{dx}{x} \theta_4\left(e^{-2x^2}\right)- \log(\Lambda)
\end{equation}
which is just a real constant. We evaluated it numerically and obtained $\zeta\approx 0.183599$. We finally arrive at 
\begin{equation}
    \begin{split}
    S =
    \sqrt{\frac{L^2}{2\pi\xi}}\left(\underbrace{\frac{4}{3}+\zeta}_{C'}+
    \log\left(\frac{2L}{\sqrt{\xi}}\right)\right)
    \end{split}
    \label{eq:constant_Cprime}
\end{equation}
which, together with Eq.~\eqref{eq:particle_number_lattice_sum_reduction}, gives Eq.~\eqref{ana_far} we reported in the main text.

Equation ~\eqref{ana_same} can be derived along similar lines;
we here just report the constant $C$ we obtained:
\begin{equation}\label{eq:constant_C} 
    \begin{split}
    C = & \lim_{\Lambda \to \infty} \left[\int_0^\Lambda \frac{dx}{x} \theta_4(0,e^{-2 x^2})   -\log \Lambda\right] - \frac{1}{2} \log(2) + \frac{23}{12} \\ \approx  & 1.75.
\end{split}
\end{equation}
%
%
%
\newpage
\bibliography{biblio}

\begin{thebibliography}{79}%
\makeatletter
\providecommand \@ifxundefined [1]{%
 \@ifx{#1\undefined}
}%
\providecommand \@ifnum [1]{%
 \ifnum #1\expandafter \@firstoftwo
 \else \expandafter \@secondoftwo
 \fi
}%
\providecommand \@ifx [1]{%
 \ifx #1\expandafter \@firstoftwo
 \else \expandafter \@secondoftwo
 \fi
}%
\providecommand \natexlab [1]{#1}%
\providecommand \enquote  [1]{``#1''}%
\providecommand \bibnamefont  [1]{#1}%
\providecommand \bibfnamefont [1]{#1}%
\providecommand \citenamefont [1]{#1}%
\providecommand \href@noop [0]{\@secondoftwo}%
\providecommand \href [0]{\begingroup \@sanitize@url \@href}%
\providecommand \@href[1]{\@@startlink{#1}\@@href}%
\providecommand \@@href[1]{\endgroup#1\@@endlink}%
\providecommand \@sanitize@url [0]{\catcode `\\12\catcode `\$12\catcode `\&12\catcode `\#12\catcode `\^12\catcode `\_12\catcode `\%12\relax}%
\providecommand \@@startlink[1]{}%
\providecommand \@@endlink[0]{}%
\providecommand \url  [0]{\begingroup\@sanitize@url \@url }%
\providecommand \@url [1]{\endgroup\@href {#1}{\urlprefix }}%
\providecommand \urlprefix  [0]{URL }%
\providecommand \Eprint [0]{\href }%
\providecommand \doibase [0]{https://doi.org/}%
\providecommand \selectlanguage [0]{\@gobble}%
\providecommand \bibinfo  [0]{\@secondoftwo}%
\providecommand \bibfield  [0]{\@secondoftwo}%
\providecommand \translation [1]{[#1]}%
\providecommand \BibitemOpen [0]{}%
\providecommand \bibitemStop [0]{}%
\providecommand \bibitemNoStop [0]{.\EOS\space}%
\providecommand \EOS [0]{\spacefactor3000\relax}%
\providecommand \BibitemShut  [1]{\csname bibitem#1\endcsname}%
\let\auto@bib@innerbib\@empty
\bibitem [{\citenamefont {Hohenberg}\ and\ \citenamefont {Halperin}(1977)}]{Hohenberg_1977}%
  \BibitemOpen
  \bibfield  {author} {\bibinfo {author} {\bibfnamefont {P.~C.}\ \bibnamefont {Hohenberg}}\ and\ \bibinfo {author} {\bibfnamefont {B.~I.}\ \bibnamefont {Halperin}},\ }\bibfield  {title} {\bibinfo {title} {Theory of dynamic critical phenomena},\ }\href {https://doi.org/10.1103/RevModPhys.49.435} {\bibfield  {journal} {\bibinfo  {journal} {Rev. Mod. Phys.}\ }\textbf {\bibinfo {volume} {49}},\ \bibinfo {pages} {435} (\bibinfo {year} {1977})}\BibitemShut {NoStop}%
\bibitem [{\citenamefont {Goldenfeld}(1992)}]{Goldenfeld_1992}%
  \BibitemOpen
  \bibfield  {author} {\bibinfo {author} {\bibfnamefont {N.}~\bibnamefont {Goldenfeld}},\ }\href {https://doi.org/10.1201/9780429493492} {\emph {\bibinfo {title} {{Lectures On Phase Transitions And The Renormalization Group}}}}\ (\bibinfo  {publisher} {CRC Press},\ \bibinfo {year} {1992})\BibitemShut {NoStop}%
\bibitem [{\citenamefont {Zinn-Justin}(2002)}]{ZinnJustin_2002}%
  \BibitemOpen
  \bibfield  {author} {\bibinfo {author} {\bibfnamefont {J.}~\bibnamefont {Zinn-Justin}},\ }\href {https://doi.org/10.1093/acprof:oso/9780198509233.001.0001} {\emph {\bibinfo {title} {{Quantum Field Theory and Critical Phenomena}}}}\ (\bibinfo  {publisher} {Oxford University Press},\ \bibinfo {year} {2002})\BibitemShut {NoStop}%
\bibitem [{\citenamefont {Sieberer}\ \emph {et~al.}(2016)\citenamefont {Sieberer}, \citenamefont {Buchhold},\ and\ \citenamefont {Diehl}}]{Sieberer_2016}%
  \BibitemOpen
  \bibfield  {author} {\bibinfo {author} {\bibfnamefont {L.~M.}\ \bibnamefont {Sieberer}}, \bibinfo {author} {\bibfnamefont {M.}~\bibnamefont {Buchhold}},\ and\ \bibinfo {author} {\bibfnamefont {S.}~\bibnamefont {Diehl}},\ }\bibfield  {title} {\bibinfo {title} {Keldysh field theory for driven open quantum systems},\ }\href {https://doi.org/10.1088/0034-4885/79/9/096001} {\bibfield  {journal} {\bibinfo  {journal} {Reports on Progress in Physics}\ }\textbf {\bibinfo {volume} {79}},\ \bibinfo {pages} {096001} (\bibinfo {year} {2016})}\BibitemShut {NoStop}%
\bibitem [{\citenamefont {Sieberer}\ \emph {et~al.}(2023)\citenamefont {Sieberer}, \citenamefont {Buchhold}, \citenamefont {Marino},\ and\ \citenamefont {Diehl}}]{sieberer2023universality}%
  \BibitemOpen
  \bibfield  {author} {\bibinfo {author} {\bibfnamefont {L.~M.}\ \bibnamefont {Sieberer}}, \bibinfo {author} {\bibfnamefont {M.}~\bibnamefont {Buchhold}}, \bibinfo {author} {\bibfnamefont {J.}~\bibnamefont {Marino}},\ and\ \bibinfo {author} {\bibfnamefont {S.}~\bibnamefont {Diehl}},\ }\href@noop {} {\bibinfo {title} {Universality in driven open quantum matter}} (\bibinfo {year} {2023}),\ \Eprint {https://arxiv.org/abs/2312.03073} {arXiv:2312.03073 [cond-mat.stat-mech]} \BibitemShut {NoStop}%
\bibitem [{\citenamefont {Altman}\ \emph {et~al.}(2015)\citenamefont {Altman}, \citenamefont {Sieberer}, \citenamefont {Chen}, \citenamefont {Diehl},\ and\ \citenamefont {Toner}}]{altman_2015}%
  \BibitemOpen
  \bibfield  {author} {\bibinfo {author} {\bibfnamefont {E.}~\bibnamefont {Altman}}, \bibinfo {author} {\bibfnamefont {L.~M.}\ \bibnamefont {Sieberer}}, \bibinfo {author} {\bibfnamefont {L.}~\bibnamefont {Chen}}, \bibinfo {author} {\bibfnamefont {S.}~\bibnamefont {Diehl}},\ and\ \bibinfo {author} {\bibfnamefont {J.}~\bibnamefont {Toner}},\ }\bibfield  {title} {\bibinfo {title} {Two-dimensional superfluidity of exciton polaritons requires strong anisotropy},\ }\href {https://doi.org/10.1103/PhysRevX.5.011017} {\bibfield  {journal} {\bibinfo  {journal} {Phys. Rev. X}\ }\textbf {\bibinfo {volume} {5}},\ \bibinfo {pages} {011017} (\bibinfo {year} {2015})}\BibitemShut {NoStop}%
\bibitem [{\citenamefont {Fontaine}\ \emph {et~al.}(2022)\citenamefont {Fontaine}, \citenamefont {Squizzato}, \citenamefont {Baboux}, \citenamefont {Amelio}, \citenamefont {Lemaître}, \citenamefont {Morassi}, \citenamefont {Sagnes}, \citenamefont {Le~Gratiet}, \citenamefont {Harouri}, \citenamefont {Wouters}, \citenamefont {Carusotto}, \citenamefont {Amo}, \citenamefont {Richard}, \citenamefont {Minguzzi}, \citenamefont {Canet}, \citenamefont {Ravets},\ and\ \citenamefont {Bloch}}]{Fontaine_2022}%
  \BibitemOpen
  \bibfield  {author} {\bibinfo {author} {\bibfnamefont {Q.}~\bibnamefont {Fontaine}}, \bibinfo {author} {\bibfnamefont {D.}~\bibnamefont {Squizzato}}, \bibinfo {author} {\bibfnamefont {F.}~\bibnamefont {Baboux}}, \bibinfo {author} {\bibfnamefont {I.}~\bibnamefont {Amelio}}, \bibinfo {author} {\bibfnamefont {A.}~\bibnamefont {Lemaître}}, \bibinfo {author} {\bibfnamefont {M.}~\bibnamefont {Morassi}}, \bibinfo {author} {\bibfnamefont {I.}~\bibnamefont {Sagnes}}, \bibinfo {author} {\bibfnamefont {L.}~\bibnamefont {Le~Gratiet}}, \bibinfo {author} {\bibfnamefont {A.}~\bibnamefont {Harouri}}, \bibinfo {author} {\bibfnamefont {M.}~\bibnamefont {Wouters}}, \bibinfo {author} {\bibfnamefont {I.}~\bibnamefont {Carusotto}}, \bibinfo {author} {\bibfnamefont {A.}~\bibnamefont {Amo}}, \bibinfo {author} {\bibfnamefont {M.}~\bibnamefont {Richard}}, \bibinfo {author} {\bibfnamefont {A.}~\bibnamefont {Minguzzi}}, \bibinfo {author} {\bibfnamefont {L.}~\bibnamefont {Canet}}, \bibinfo {author} {\bibfnamefont {S.}~\bibnamefont
  {Ravets}},\ and\ \bibinfo {author} {\bibfnamefont {J.}~\bibnamefont {Bloch}},\ }\bibfield  {title} {\bibinfo {title} {{Kardar–Parisi–Zhang universality in a one-dimensional polariton condensate}},\ }\href {https://doi.org/10.1038/s41586-022-05001-8} {\bibfield  {journal} {\bibinfo  {journal} {Nature}\ }\textbf {\bibinfo {volume} {608}},\ \bibinfo {pages} {687–691} (\bibinfo {year} {2022})}\BibitemShut {NoStop}%
\bibitem [{\citenamefont {Barontini}\ \emph {et~al.}(2013)\citenamefont {Barontini}, \citenamefont {Labouvie}, \citenamefont {Stubenrauch}, \citenamefont {Vogler}, \citenamefont {Guarrera},\ and\ \citenamefont {Ott}}]{Barontini_2013}%
  \BibitemOpen
  \bibfield  {author} {\bibinfo {author} {\bibfnamefont {G.}~\bibnamefont {Barontini}}, \bibinfo {author} {\bibfnamefont {R.}~\bibnamefont {Labouvie}}, \bibinfo {author} {\bibfnamefont {F.}~\bibnamefont {Stubenrauch}}, \bibinfo {author} {\bibfnamefont {A.}~\bibnamefont {Vogler}}, \bibinfo {author} {\bibfnamefont {V.}~\bibnamefont {Guarrera}},\ and\ \bibinfo {author} {\bibfnamefont {H.}~\bibnamefont {Ott}},\ }\bibfield  {title} {\bibinfo {title} {Controlling the dynamics of an open many-body quantum system with localized dissipation},\ }\href {https://doi.org/10.1103/PhysRevLett.110.035302} {\bibfield  {journal} {\bibinfo  {journal} {Phys. Rev. Lett.}\ }\textbf {\bibinfo {volume} {110}},\ \bibinfo {pages} {035302} (\bibinfo {year} {2013})}\BibitemShut {NoStop}%
\bibitem [{\citenamefont {Perfetto}\ \emph {et~al.}(2023{\natexlab{a}})\citenamefont {Perfetto}, \citenamefont {Carollo}, \citenamefont {Garrahan},\ and\ \citenamefont {Lesanovsky}}]{perfetto_2023}%
  \BibitemOpen
  \bibfield  {author} {\bibinfo {author} {\bibfnamefont {G.}~\bibnamefont {Perfetto}}, \bibinfo {author} {\bibfnamefont {F.}~\bibnamefont {Carollo}}, \bibinfo {author} {\bibfnamefont {J.~P.}\ \bibnamefont {Garrahan}},\ and\ \bibinfo {author} {\bibfnamefont {I.}~\bibnamefont {Lesanovsky}},\ }\bibfield  {title} {\bibinfo {title} {Reaction-limited quantum reaction-diffusion dynamics},\ }\href {https://doi.org/10.1103/PhysRevLett.130.210402} {\bibfield  {journal} {\bibinfo  {journal} {Phys. Rev. Lett.}\ }\textbf {\bibinfo {volume} {130}},\ \bibinfo {pages} {210402} (\bibinfo {year} {2023}{\natexlab{a}})}\BibitemShut {NoStop}%
\bibitem [{\citenamefont {Perfetto}\ \emph {et~al.}(2023{\natexlab{b}})\citenamefont {Perfetto}, \citenamefont {Carollo}, \citenamefont {Garrahan},\ and\ \citenamefont {Lesanovsky}}]{perfetto_2023bis}%
  \BibitemOpen
  \bibfield  {author} {\bibinfo {author} {\bibfnamefont {G.}~\bibnamefont {Perfetto}}, \bibinfo {author} {\bibfnamefont {F.}~\bibnamefont {Carollo}}, \bibinfo {author} {\bibfnamefont {J.~P.}\ \bibnamefont {Garrahan}},\ and\ \bibinfo {author} {\bibfnamefont {I.}~\bibnamefont {Lesanovsky}},\ }\bibfield  {title} {\bibinfo {title} {Quantum reaction-limited reaction-diffusion dynamics of annihilation processes},\ }\href {https://doi.org/10.1103/PhysRevE.108.064104} {\bibfield  {journal} {\bibinfo  {journal} {Phys. Rev. E}\ }\textbf {\bibinfo {volume} {108}},\ \bibinfo {pages} {064104} (\bibinfo {year} {2023}{\natexlab{b}})}\BibitemShut {NoStop}%
\bibitem [{\citenamefont {Rowlands}\ \emph {et~al.}(2024)\citenamefont {Rowlands}, \citenamefont {Lesanovsky},\ and\ \citenamefont {Perfetto}}]{rowlands2023quantum}%
  \BibitemOpen
  \bibfield  {author} {\bibinfo {author} {\bibfnamefont {S.}~\bibnamefont {Rowlands}}, \bibinfo {author} {\bibfnamefont {I.}~\bibnamefont {Lesanovsky}},\ and\ \bibinfo {author} {\bibfnamefont {G.}~\bibnamefont {Perfetto}},\ }\bibfield  {title} {\bibinfo {title} {Quantum reaction-limited reaction–diffusion dynamics of noninteracting bose gases},\ }\href {https://doi.org/10.1088/1367-2630/ad397a} {\bibfield  {journal} {\bibinfo  {journal} {New J. Phys.}\ }\textbf {\bibinfo {volume} {26}},\ \bibinfo {pages} {043010} (\bibinfo {year} {2024})}\BibitemShut {NoStop}%
\bibitem [{\citenamefont {Buchhold}\ \emph {et~al.}(2017)\citenamefont {Buchhold}, \citenamefont {Everest}, \citenamefont {Marcuzzi}, \citenamefont {Lesanovsky},\ and\ \citenamefont {Diehl}}]{buchhold_2017}%
  \BibitemOpen
  \bibfield  {author} {\bibinfo {author} {\bibfnamefont {M.}~\bibnamefont {Buchhold}}, \bibinfo {author} {\bibfnamefont {B.}~\bibnamefont {Everest}}, \bibinfo {author} {\bibfnamefont {M.}~\bibnamefont {Marcuzzi}}, \bibinfo {author} {\bibfnamefont {I.}~\bibnamefont {Lesanovsky}},\ and\ \bibinfo {author} {\bibfnamefont {S.}~\bibnamefont {Diehl}},\ }\bibfield  {title} {\bibinfo {title} {Nonequilibrium effective field theory for absorbing state phase transitions in driven open quantum spin systems},\ }\href {https://doi.org/10.1103/PhysRevB.95.014308} {\bibfield  {journal} {\bibinfo  {journal} {Phys. Rev. B}\ }\textbf {\bibinfo {volume} {95}},\ \bibinfo {pages} {014308} (\bibinfo {year} {2017})}\BibitemShut {NoStop}%
\bibitem [{\citenamefont {Wintermantel}\ \emph {et~al.}(2021)\citenamefont {Wintermantel}, \citenamefont {Buchhold}, \citenamefont {Shevate}, \citenamefont {Morgado}, \citenamefont {Wang}, \citenamefont {Lochead}, \citenamefont {Diehl},\ and\ \citenamefont {Whitlock}}]{Wintermantel_2021}%
  \BibitemOpen
  \bibfield  {author} {\bibinfo {author} {\bibfnamefont {T.~M.}\ \bibnamefont {Wintermantel}}, \bibinfo {author} {\bibfnamefont {M.}~\bibnamefont {Buchhold}}, \bibinfo {author} {\bibfnamefont {S.}~\bibnamefont {Shevate}}, \bibinfo {author} {\bibfnamefont {M.}~\bibnamefont {Morgado}}, \bibinfo {author} {\bibfnamefont {Y.}~\bibnamefont {Wang}}, \bibinfo {author} {\bibfnamefont {G.}~\bibnamefont {Lochead}}, \bibinfo {author} {\bibfnamefont {S.}~\bibnamefont {Diehl}},\ and\ \bibinfo {author} {\bibfnamefont {S.}~\bibnamefont {Whitlock}},\ }\bibfield  {title} {\bibinfo {title} {Epidemic growth and {Griffiths} effects on an emergent network of excited atoms},\ }\href {https://doi.org/10.1038/s41467-020-20333-7} {\bibfield  {journal} {\bibinfo  {journal} {Nature Communications}\ }\textbf {\bibinfo {volume} {12}},\ \bibinfo {pages} {103} (\bibinfo {year} {2021})}\BibitemShut {NoStop}%
\bibitem [{\citenamefont {Cai}\ and\ \citenamefont {Barthel}(2013)}]{cai2013algebraic}%
  \BibitemOpen
  \bibfield  {author} {\bibinfo {author} {\bibfnamefont {Z.}~\bibnamefont {Cai}}\ and\ \bibinfo {author} {\bibfnamefont {T.}~\bibnamefont {Barthel}},\ }\bibfield  {title} {\bibinfo {title} {Algebraic versus exponential decoherence in dissipative many-particle systems},\ }\href {https://doi.org/10.1103/PhysRevLett.111.150403} {\bibfield  {journal} {\bibinfo  {journal} {Phys. Rev. Lett.}\ }\textbf {\bibinfo {volume} {111}},\ \bibinfo {pages} {150403} (\bibinfo {year} {2013})}\BibitemShut {NoStop}%
\bibitem [{\citenamefont {Begg}\ and\ \citenamefont {Hanai}(2024)}]{begg2024quantum}%
  \BibitemOpen
  \bibfield  {author} {\bibinfo {author} {\bibfnamefont {S.~E.}\ \bibnamefont {Begg}}\ and\ \bibinfo {author} {\bibfnamefont {R.}~\bibnamefont {Hanai}},\ }\bibfield  {title} {\bibinfo {title} {Quantum criticality in open quantum spin chains with nonreciprocity},\ }\href {https://doi.org/10.1103/PhysRevLett.132.120401} {\bibfield  {journal} {\bibinfo  {journal} {Phys. Rev. Lett.}\ }\textbf {\bibinfo {volume} {132}},\ \bibinfo {pages} {120401} (\bibinfo {year} {2024})}\BibitemShut {NoStop}%
\bibitem [{\citenamefont {D\"urr}\ \emph {et~al.}(2009)\citenamefont {D\"urr}, \citenamefont {Garc\'{\i}a-Ripoll}, \citenamefont {Syassen}, \citenamefont {Bauer}, \citenamefont {Lettner}, \citenamefont {Cirac},\ and\ \citenamefont {Rempe}}]{Durr_2009}%
  \BibitemOpen
  \bibfield  {author} {\bibinfo {author} {\bibfnamefont {S.}~\bibnamefont {D\"urr}}, \bibinfo {author} {\bibfnamefont {J.~J.}\ \bibnamefont {Garc\'{\i}a-Ripoll}}, \bibinfo {author} {\bibfnamefont {N.}~\bibnamefont {Syassen}}, \bibinfo {author} {\bibfnamefont {D.~M.}\ \bibnamefont {Bauer}}, \bibinfo {author} {\bibfnamefont {M.}~\bibnamefont {Lettner}}, \bibinfo {author} {\bibfnamefont {J.~I.}\ \bibnamefont {Cirac}},\ and\ \bibinfo {author} {\bibfnamefont {G.}~\bibnamefont {Rempe}},\ }\bibfield  {title} {\bibinfo {title} {{Lieb-Liniger model of a dissipation-induced Tonks-Girardeau gas}},\ }\href {https://doi.org/10.1103/PhysRevA.79.023614} {\bibfield  {journal} {\bibinfo  {journal} {Phys. Rev. A}\ }\textbf {\bibinfo {volume} {79}},\ \bibinfo {pages} {023614} (\bibinfo {year} {2009})}\BibitemShut {NoStop}%
\bibitem [{\citenamefont {Garc{\'{\i}}a-Ripoll}\ \emph {et~al.}(2009)\citenamefont {Garc{\'{\i}}a-Ripoll}, \citenamefont {D\"urr}, \citenamefont {Syassen}, \citenamefont {Bauer}, \citenamefont {Lettner}, \citenamefont {Rempe},\ and\ \citenamefont {Cirac}}]{GarciaRipoll_2009}%
  \BibitemOpen
  \bibfield  {author} {\bibinfo {author} {\bibfnamefont {J.~J.}\ \bibnamefont {Garc{\'{\i}}a-Ripoll}}, \bibinfo {author} {\bibfnamefont {S.}~\bibnamefont {D\"urr}}, \bibinfo {author} {\bibfnamefont {N.}~\bibnamefont {Syassen}}, \bibinfo {author} {\bibfnamefont {D.~M.}\ \bibnamefont {Bauer}}, \bibinfo {author} {\bibfnamefont {M.}~\bibnamefont {Lettner}}, \bibinfo {author} {\bibfnamefont {G.}~\bibnamefont {Rempe}},\ and\ \bibinfo {author} {\bibfnamefont {J.~I.}\ \bibnamefont {Cirac}},\ }\bibfield  {title} {\bibinfo {title} {Dissipation-induced hard-core boson gas in an optical lattice},\ }\href {https://doi.org/10.1088/1367-2630/11/1/013053} {\bibfield  {journal} {\bibinfo  {journal} {New J. Phys.}\ }\textbf {\bibinfo {volume} {11}},\ \bibinfo {pages} {013053} (\bibinfo {year} {2009})}\BibitemShut {NoStop}%
\bibitem [{\citenamefont {Baur}\ and\ \citenamefont {Mueller}(2010)}]{Baur_2010}%
  \BibitemOpen
  \bibfield  {author} {\bibinfo {author} {\bibfnamefont {S.~K.}\ \bibnamefont {Baur}}\ and\ \bibinfo {author} {\bibfnamefont {E.~J.}\ \bibnamefont {Mueller}},\ }\bibfield  {title} {\bibinfo {title} {Two-body recombination in a quantum-mechanical lattice gas: Entropy generation and probing of short-range magnetic correlations},\ }\href {https://doi.org/10.1103/physreva.82.023626} {\bibfield  {journal} {\bibinfo  {journal} {Phys. Rev. A}\ }\textbf {\bibinfo {volume} {82}},\ \bibinfo {pages} {023626} (\bibinfo {year} {2010})}\BibitemShut {NoStop}%
\bibitem [{\citenamefont {Foss-Feig}\ \emph {et~al.}(2012)\citenamefont {Foss-Feig}, \citenamefont {Daley}, \citenamefont {Thompson},\ and\ \citenamefont {Rey}}]{FossFeig_2012}%
  \BibitemOpen
  \bibfield  {author} {\bibinfo {author} {\bibfnamefont {M.}~\bibnamefont {Foss-Feig}}, \bibinfo {author} {\bibfnamefont {A.~J.}\ \bibnamefont {Daley}}, \bibinfo {author} {\bibfnamefont {J.~K.}\ \bibnamefont {Thompson}},\ and\ \bibinfo {author} {\bibfnamefont {A.~M.}\ \bibnamefont {Rey}},\ }\bibfield  {title} {\bibinfo {title} {Steady-state many-body entanglement of hot reactive fermions},\ }\href {https://doi.org/10.1103/PhysRevLett.109.230501} {\bibfield  {journal} {\bibinfo  {journal} {Phys. Rev. Lett.}\ }\textbf {\bibinfo {volume} {109}},\ \bibinfo {pages} {230501} (\bibinfo {year} {2012})}\BibitemShut {NoStop}%
\bibitem [{\citenamefont {Gri\ifmmode~\check{s}\else \v{s}\fi{}ins}\ \emph {et~al.}(2016)\citenamefont {Gri\ifmmode~\check{s}\else \v{s}\fi{}ins}, \citenamefont {Rauer}, \citenamefont {Langen}, \citenamefont {Schmiedmayer},\ and\ \citenamefont {Mazets}}]{Grisins_2016}%
  \BibitemOpen
  \bibfield  {author} {\bibinfo {author} {\bibfnamefont {P.}~\bibnamefont {Gri\ifmmode~\check{s}\else \v{s}\fi{}ins}}, \bibinfo {author} {\bibfnamefont {B.}~\bibnamefont {Rauer}}, \bibinfo {author} {\bibfnamefont {T.}~\bibnamefont {Langen}}, \bibinfo {author} {\bibfnamefont {J.}~\bibnamefont {Schmiedmayer}},\ and\ \bibinfo {author} {\bibfnamefont {I.~E.}\ \bibnamefont {Mazets}},\ }\bibfield  {title} {\bibinfo {title} {Degenerate bose gases with uniform loss},\ }\href {https://doi.org/10.1103/PhysRevA.93.033634} {\bibfield  {journal} {\bibinfo  {journal} {Phys. Rev. A}\ }\textbf {\bibinfo {volume} {93}},\ \bibinfo {pages} {033634} (\bibinfo {year} {2016})}\BibitemShut {NoStop}%
\bibitem [{\citenamefont {Johnson}\ \emph {et~al.}(2017)\citenamefont {Johnson}, \citenamefont {Szigeti}, \citenamefont {Schemmer},\ and\ \citenamefont {Bouchoule}}]{Johnson_2017}%
  \BibitemOpen
  \bibfield  {author} {\bibinfo {author} {\bibfnamefont {A.}~\bibnamefont {Johnson}}, \bibinfo {author} {\bibfnamefont {S.~S.}\ \bibnamefont {Szigeti}}, \bibinfo {author} {\bibfnamefont {M.}~\bibnamefont {Schemmer}},\ and\ \bibinfo {author} {\bibfnamefont {I.}~\bibnamefont {Bouchoule}},\ }\bibfield  {title} {\bibinfo {title} {Long-lived nonthermal states realized by atom losses in one-dimensional quasicondensates},\ }\href {https://doi.org/10.1103/PhysRevA.96.013623} {\bibfield  {journal} {\bibinfo  {journal} {Phys. Rev. A}\ }\textbf {\bibinfo {volume} {96}},\ \bibinfo {pages} {013623} (\bibinfo {year} {2017})}\BibitemShut {NoStop}%
\bibitem [{\citenamefont {Yamamoto}\ \emph {et~al.}(2019)\citenamefont {Yamamoto}, \citenamefont {Nakagawa}, \citenamefont {Adachi}, \citenamefont {Takasan}, \citenamefont {Ueda},\ and\ \citenamefont {Kawakami}}]{Yamamoto1_2019}%
  \BibitemOpen
  \bibfield  {author} {\bibinfo {author} {\bibfnamefont {K.}~\bibnamefont {Yamamoto}}, \bibinfo {author} {\bibfnamefont {M.}~\bibnamefont {Nakagawa}}, \bibinfo {author} {\bibfnamefont {K.}~\bibnamefont {Adachi}}, \bibinfo {author} {\bibfnamefont {K.}~\bibnamefont {Takasan}}, \bibinfo {author} {\bibfnamefont {M.}~\bibnamefont {Ueda}},\ and\ \bibinfo {author} {\bibfnamefont {N.}~\bibnamefont {Kawakami}},\ }\bibfield  {title} {\bibinfo {title} {{Theory of Non-Hermitian Fermionic Superfluidity with a Complex-Valued Interaction}},\ }\href {https://doi.org/10.1103/PhysRevLett.123.123601} {\bibfield  {journal} {\bibinfo  {journal} {Phys. Rev. Lett.}\ }\textbf {\bibinfo {volume} {123}},\ \bibinfo {pages} {123601} (\bibinfo {year} {2019})}\BibitemShut {NoStop}%
\bibitem [{\citenamefont {Goto}\ and\ \citenamefont {Danshita}(2020)}]{Goto_2020}%
  \BibitemOpen
  \bibfield  {author} {\bibinfo {author} {\bibfnamefont {S.}~\bibnamefont {Goto}}\ and\ \bibinfo {author} {\bibfnamefont {I.}~\bibnamefont {Danshita}},\ }\bibfield  {title} {\bibinfo {title} {Measurement-induced transitions of the entanglement scaling law in ultracold gases with controllable dissipation},\ }\href {https://doi.org/10.1103/PhysRevA.102.033316} {\bibfield  {journal} {\bibinfo  {journal} {Phys. Rev. A}\ }\textbf {\bibinfo {volume} {102}},\ \bibinfo {pages} {033316} (\bibinfo {year} {2020})}\BibitemShut {NoStop}%
\bibitem [{\citenamefont {Bouchoule}\ \emph {et~al.}(2020)\citenamefont {Bouchoule}, \citenamefont {Doyon},\ and\ \citenamefont {Dubail}}]{Bouchoule_2020b}%
  \BibitemOpen
  \bibfield  {author} {\bibinfo {author} {\bibfnamefont {I.}~\bibnamefont {Bouchoule}}, \bibinfo {author} {\bibfnamefont {B.}~\bibnamefont {Doyon}},\ and\ \bibinfo {author} {\bibfnamefont {J.}~\bibnamefont {Dubail}},\ }\bibfield  {title} {\bibinfo {title} {{The effect of atom losses on the distribution of rapidities in the one-dimensional Bose gas}},\ }\href {https://doi.org/10.21468/SciPostPhys.9.4.044} {\bibfield  {journal} {\bibinfo  {journal} {SciPost Phys.}\ }\textbf {\bibinfo {volume} {9}},\ \bibinfo {pages} {44} (\bibinfo {year} {2020})}\BibitemShut {NoStop}%
\bibitem [{\citenamefont {Rossini}\ \emph {et~al.}(2021)\citenamefont {Rossini}, \citenamefont {Ghermaoui}, \citenamefont {Aguilera}, \citenamefont {Vatr\'e}, \citenamefont {Bouganne}, \citenamefont {Beugnon}, \citenamefont {Gerbier},\ and\ \citenamefont {Mazza}}]{Rossini_2020}%
  \BibitemOpen
  \bibfield  {author} {\bibinfo {author} {\bibfnamefont {D.}~\bibnamefont {Rossini}}, \bibinfo {author} {\bibfnamefont {A.}~\bibnamefont {Ghermaoui}}, \bibinfo {author} {\bibfnamefont {M.~B.}\ \bibnamefont {Aguilera}}, \bibinfo {author} {\bibfnamefont {R.}~\bibnamefont {Vatr\'e}}, \bibinfo {author} {\bibfnamefont {R.}~\bibnamefont {Bouganne}}, \bibinfo {author} {\bibfnamefont {J.}~\bibnamefont {Beugnon}}, \bibinfo {author} {\bibfnamefont {F.}~\bibnamefont {Gerbier}},\ and\ \bibinfo {author} {\bibfnamefont {L.}~\bibnamefont {Mazza}},\ }\bibfield  {title} {\bibinfo {title} {{Strong correlations in lossy one-dimensional quantum gases: From the quantum Zeno effect to the generalized Gibbs ensemble}},\ }\href {https://doi.org/10.1103/PhysRevA.103.L060201} {\bibfield  {journal} {\bibinfo  {journal} {Phys. Rev. A}\ }\textbf {\bibinfo {volume} {103}},\ \bibinfo {pages} {L060201} (\bibinfo {year} {2021})}\BibitemShut {NoStop}%
\bibitem [{\citenamefont {Nakagawa}\ \emph {et~al.}(2020)\citenamefont {Nakagawa}, \citenamefont {Tsuji}, \citenamefont {Kawakami},\ and\ \citenamefont {Ueda}}]{Nakagawa_2020}%
  \BibitemOpen
  \bibfield  {author} {\bibinfo {author} {\bibfnamefont {M.}~\bibnamefont {Nakagawa}}, \bibinfo {author} {\bibfnamefont {N.}~\bibnamefont {Tsuji}}, \bibinfo {author} {\bibfnamefont {N.}~\bibnamefont {Kawakami}},\ and\ \bibinfo {author} {\bibfnamefont {M.}~\bibnamefont {Ueda}},\ }\bibfield  {title} {\bibinfo {title} {{Dynamical Sign Reversal of Magnetic Correlations in Dissipative Hubbard Models}},\ }\href {https://doi.org/10.1103/PhysRevLett.124.147203} {\bibfield  {journal} {\bibinfo  {journal} {Phys. Rev. Lett.}\ }\textbf {\bibinfo {volume} {124}},\ \bibinfo {pages} {147203} (\bibinfo {year} {2020})}\BibitemShut {NoStop}%
\bibitem [{\citenamefont {Booker}\ \emph {et~al.}(2020)\citenamefont {Booker}, \citenamefont {Buča},\ and\ \citenamefont {Jaksch}}]{Booker_2020}%
  \BibitemOpen
  \bibfield  {author} {\bibinfo {author} {\bibfnamefont {C.}~\bibnamefont {Booker}}, \bibinfo {author} {\bibfnamefont {B.}~\bibnamefont {Buča}},\ and\ \bibinfo {author} {\bibfnamefont {D.}~\bibnamefont {Jaksch}},\ }\bibfield  {title} {\bibinfo {title} {Non-stationarity and dissipative time crystals: spectral properties and finite-size effects},\ }\href {https://doi.org/10.1088/1367-2630/ababc4} {\bibfield  {journal} {\bibinfo  {journal} {New J. Phys.}\ }\textbf {\bibinfo {volume} {22}},\ \bibinfo {pages} {085007} (\bibinfo {year} {2020})}\BibitemShut {NoStop}%
\bibitem [{\citenamefont {Bouchoule}\ and\ \citenamefont {Dubail}(2021)}]{Bouchoule_2021}%
  \BibitemOpen
  \bibfield  {author} {\bibinfo {author} {\bibfnamefont {I.}~\bibnamefont {Bouchoule}}\ and\ \bibinfo {author} {\bibfnamefont {J.}~\bibnamefont {Dubail}},\ }\bibfield  {title} {\bibinfo {title} {{Breakdown of Tan's Relation in Lossy One-Dimensional Bose Gases}},\ }\href {https://doi.org/10.1103/PhysRevLett.126.160603} {\bibfield  {journal} {\bibinfo  {journal} {Phys. Rev. Lett.}\ }\textbf {\bibinfo {volume} {126}},\ \bibinfo {pages} {160603} (\bibinfo {year} {2021})}\BibitemShut {NoStop}%
\bibitem [{\citenamefont {Bouchoule}\ \emph {et~al.}(2021)\citenamefont {Bouchoule}, \citenamefont {Dubois},\ and\ \citenamefont {Barbier}}]{Bouchoule_2021PRA}%
  \BibitemOpen
  \bibfield  {author} {\bibinfo {author} {\bibfnamefont {I.}~\bibnamefont {Bouchoule}}, \bibinfo {author} {\bibfnamefont {L.}~\bibnamefont {Dubois}},\ and\ \bibinfo {author} {\bibfnamefont {L.-P.}\ \bibnamefont {Barbier}},\ }\bibfield  {title} {\bibinfo {title} {{Losses in interacting quantum gases: Ultraviolet divergence and its regularization}},\ }\href {https://doi.org/10.1103/PhysRevA.104.L031304} {\bibfield  {journal} {\bibinfo  {journal} {Phys. Rev. A}\ }\textbf {\bibinfo {volume} {104}},\ \bibinfo {pages} {L031304} (\bibinfo {year} {2021})}\BibitemShut {NoStop}%
\bibitem [{\citenamefont {Rosso}\ \emph {et~al.}(2021)\citenamefont {Rosso}, \citenamefont {Rossini}, \citenamefont {Biella},\ and\ \citenamefont {Mazza}}]{Rosso_2021}%
  \BibitemOpen
  \bibfield  {author} {\bibinfo {author} {\bibfnamefont {L.}~\bibnamefont {Rosso}}, \bibinfo {author} {\bibfnamefont {D.}~\bibnamefont {Rossini}}, \bibinfo {author} {\bibfnamefont {A.}~\bibnamefont {Biella}},\ and\ \bibinfo {author} {\bibfnamefont {L.}~\bibnamefont {Mazza}},\ }\bibfield  {title} {\bibinfo {title} {{One-dimensional spin-1/2 fermionic gases with two-body losses: Weak dissipation and spin conservation}},\ }\href {https://doi.org/10.1103/PhysRevA.104.053305} {\bibfield  {journal} {\bibinfo  {journal} {Phys. Rev. A}\ }\textbf {\bibinfo {volume} {104}},\ \bibinfo {pages} {053305} (\bibinfo {year} {2021})}\BibitemShut {NoStop}%
\bibitem [{\citenamefont {Nakagawa}\ \emph {et~al.}(2021)\citenamefont {Nakagawa}, \citenamefont {Kawakami},\ and\ \citenamefont {Ueda}}]{Nakagawa_2021}%
  \BibitemOpen
  \bibfield  {author} {\bibinfo {author} {\bibfnamefont {M.}~\bibnamefont {Nakagawa}}, \bibinfo {author} {\bibfnamefont {N.}~\bibnamefont {Kawakami}},\ and\ \bibinfo {author} {\bibfnamefont {M.}~\bibnamefont {Ueda}},\ }\bibfield  {title} {\bibinfo {title} {{Exact Liouvillian Spectrum of a One-Dimensional Dissipative Hubbard Model}},\ }\href {https://doi.org/10.1103/PhysRevLett.126.110404} {\bibfield  {journal} {\bibinfo  {journal} {Phys. Rev. Lett.}\ }\textbf {\bibinfo {volume} {126}},\ \bibinfo {pages} {110404} (\bibinfo {year} {2021})}\BibitemShut {NoStop}%
\bibitem [{\citenamefont {Yamamoto}\ \emph {et~al.}(2021)\citenamefont {Yamamoto}, \citenamefont {Nakagawa}, \citenamefont {Tsuji}, \citenamefont {Ueda},\ and\ \citenamefont {Kawakami}}]{Yamamoto2_2021}%
  \BibitemOpen
  \bibfield  {author} {\bibinfo {author} {\bibfnamefont {K.}~\bibnamefont {Yamamoto}}, \bibinfo {author} {\bibfnamefont {M.}~\bibnamefont {Nakagawa}}, \bibinfo {author} {\bibfnamefont {N.}~\bibnamefont {Tsuji}}, \bibinfo {author} {\bibfnamefont {M.}~\bibnamefont {Ueda}},\ and\ \bibinfo {author} {\bibfnamefont {N.}~\bibnamefont {Kawakami}},\ }\bibfield  {title} {\bibinfo {title} {Collective excitations and nonequilibrium phase transition in dissipative fermionic superfluids},\ }\href {https://doi.org/10.1103/PhysRevLett.127.055301} {\bibfield  {journal} {\bibinfo  {journal} {Phys. Rev. Lett.}\ }\textbf {\bibinfo {volume} {127}},\ \bibinfo {pages} {055301} (\bibinfo {year} {2021})}\BibitemShut {NoStop}%
\bibitem [{\citenamefont {Iskin}(2021)}]{iskin2021non}%
  \BibitemOpen
  \bibfield  {author} {\bibinfo {author} {\bibfnamefont {M.}~\bibnamefont {Iskin}},\ }\bibfield  {title} {\bibinfo {title} {{Non-Hermitian BCS-BEC evolution with a complex scattering length}},\ }\href {https://doi.org/10.1103/PhysRevA.103.013724} {\bibfield  {journal} {\bibinfo  {journal} {Phys. Rev. A}\ }\textbf {\bibinfo {volume} {103}},\ \bibinfo {pages} {013724} (\bibinfo {year} {2021})}\BibitemShut {NoStop}%
\bibitem [{\citenamefont {Rosso}\ \emph {et~al.}(2022{\natexlab{a}})\citenamefont {Rosso}, \citenamefont {Mazza},\ and\ \citenamefont {Biella}}]{Rosso_2022}%
  \BibitemOpen
  \bibfield  {author} {\bibinfo {author} {\bibfnamefont {L.}~\bibnamefont {Rosso}}, \bibinfo {author} {\bibfnamefont {L.}~\bibnamefont {Mazza}},\ and\ \bibinfo {author} {\bibfnamefont {A.}~\bibnamefont {Biella}},\ }\bibfield  {title} {\bibinfo {title} {{Eightfold way to dark states in SU(3) cold gases with two-body losses}},\ }\href {https://doi.org/10.1103/PhysRevA.105.L051302} {\bibfield  {journal} {\bibinfo  {journal} {Phys. Rev. A}\ }\textbf {\bibinfo {volume} {105}},\ \bibinfo {pages} {L051302} (\bibinfo {year} {2022}{\natexlab{a}})}\BibitemShut {NoStop}%
\bibitem [{\citenamefont {Yamamoto}\ \emph {et~al.}(2022)\citenamefont {Yamamoto}, \citenamefont {Nakagawa}, \citenamefont {Tezuka}, \citenamefont {Ueda},\ and\ \citenamefont {Kawakami}}]{Yamamoto3_2022}%
  \BibitemOpen
  \bibfield  {author} {\bibinfo {author} {\bibfnamefont {K.}~\bibnamefont {Yamamoto}}, \bibinfo {author} {\bibfnamefont {M.}~\bibnamefont {Nakagawa}}, \bibinfo {author} {\bibfnamefont {M.}~\bibnamefont {Tezuka}}, \bibinfo {author} {\bibfnamefont {M.}~\bibnamefont {Ueda}},\ and\ \bibinfo {author} {\bibfnamefont {N.}~\bibnamefont {Kawakami}},\ }\bibfield  {title} {\bibinfo {title} {{Universal properties of dissipative Tomonaga-Luttinger liquids: Case study of a non-Hermitian XXZ spin chain}},\ }\href {https://doi.org/10.1103/PhysRevB.105.205125} {\bibfield  {journal} {\bibinfo  {journal} {Phys. Rev. B}\ }\textbf {\bibinfo {volume} {105}},\ \bibinfo {pages} {205125} (\bibinfo {year} {2022})}\BibitemShut {NoStop}%
\bibitem [{\citenamefont {Wang}\ \emph {et~al.}(2022)\citenamefont {Wang}, \citenamefont {Liu},\ and\ \citenamefont {Shi}}]{wang2022complex}%
  \BibitemOpen
  \bibfield  {author} {\bibinfo {author} {\bibfnamefont {C.}~\bibnamefont {Wang}}, \bibinfo {author} {\bibfnamefont {C.}~\bibnamefont {Liu}},\ and\ \bibinfo {author} {\bibfnamefont {Z.-Y.}\ \bibnamefont {Shi}},\ }\bibfield  {title} {\bibinfo {title} {{Complex contact interaction for systems with short-range two-body losses}},\ }\href {https://doi.org/10.1103/PhysRevLett.129.203401} {\bibfield  {journal} {\bibinfo  {journal} {Phys. Rev. Lett.}\ }\textbf {\bibinfo {volume} {129}},\ \bibinfo {pages} {203401} (\bibinfo {year} {2022})}\BibitemShut {NoStop}%
\bibitem [{\citenamefont {Yoshida}\ and\ \citenamefont {Katsura}(2023)}]{Yoshida_2023}%
  \BibitemOpen
  \bibfield  {author} {\bibinfo {author} {\bibfnamefont {H.}~\bibnamefont {Yoshida}}\ and\ \bibinfo {author} {\bibfnamefont {H.}~\bibnamefont {Katsura}},\ }\bibfield  {title} {\bibinfo {title} {{Liouvillian gap and single spin-flip dynamics in the dissipative Fermi-Hubbard model}},\ }\href {https://doi.org/10.1103/PhysRevA.107.033332} {\bibfield  {journal} {\bibinfo  {journal} {Phys. Rev. A}\ }\textbf {\bibinfo {volume} {107}},\ \bibinfo {pages} {033332} (\bibinfo {year} {2023})}\BibitemShut {NoStop}%
\bibitem [{\citenamefont {Huang}\ \emph {et~al.}(2023)\citenamefont {Huang}, \citenamefont {Giamarchi},\ and\ \citenamefont {Cazalilla}}]{Huang_2023}%
  \BibitemOpen
  \bibfield  {author} {\bibinfo {author} {\bibfnamefont {C.-H.}\ \bibnamefont {Huang}}, \bibinfo {author} {\bibfnamefont {T.}~\bibnamefont {Giamarchi}},\ and\ \bibinfo {author} {\bibfnamefont {M.~A.}\ \bibnamefont {Cazalilla}},\ }\bibfield  {title} {\bibinfo {title} {{Modeling particle loss in open systems using Keldysh path integral and second order cumulant expansion}},\ }\href {https://doi.org/10.1103/PhysRevResearch.5.043192} {\bibfield  {journal} {\bibinfo  {journal} {Phys. Rev. Res.}\ }\textbf {\bibinfo {volume} {5}},\ \bibinfo {pages} {043192} (\bibinfo {year} {2023})}\BibitemShut {NoStop}%
\bibitem [{\citenamefont {Mazza}\ and\ \citenamefont {Schir\`o}(2023)}]{Mazza_2023}%
  \BibitemOpen
  \bibfield  {author} {\bibinfo {author} {\bibfnamefont {G.}~\bibnamefont {Mazza}}\ and\ \bibinfo {author} {\bibfnamefont {M.}~\bibnamefont {Schir\`o}},\ }\bibfield  {title} {\bibinfo {title} {Dissipative dynamics of a fermionic superfluid with two-body losses},\ }\href {https://doi.org/10.1103/PhysRevA.107.L051301} {\bibfield  {journal} {\bibinfo  {journal} {Phys. Rev. A}\ }\textbf {\bibinfo {volume} {107}},\ \bibinfo {pages} {L051301} (\bibinfo {year} {2023})}\BibitemShut {NoStop}%
\bibitem [{\citenamefont {Hamanaka}\ \emph {et~al.}(2023)\citenamefont {Hamanaka}, \citenamefont {Yamamoto},\ and\ \citenamefont {Yoshida}}]{Hamanaka_2023}%
  \BibitemOpen
  \bibfield  {author} {\bibinfo {author} {\bibfnamefont {S.}~\bibnamefont {Hamanaka}}, \bibinfo {author} {\bibfnamefont {K.}~\bibnamefont {Yamamoto}},\ and\ \bibinfo {author} {\bibfnamefont {T.}~\bibnamefont {Yoshida}},\ }\bibfield  {title} {\bibinfo {title} {{Interaction-induced Liouvillian skin effect in a fermionic chain with a two-body loss}},\ }\href {https://doi.org/10.1103/PhysRevB.108.155114} {\bibfield  {journal} {\bibinfo  {journal} {Phys. Rev. B}\ }\textbf {\bibinfo {volume} {108}},\ \bibinfo {pages} {155114} (\bibinfo {year} {2023})}\BibitemShut {NoStop}%
\bibitem [{\citenamefont {Nagao}\ \emph {et~al.}(2023)\citenamefont {Nagao}, \citenamefont {Danshita},\ and\ \citenamefont {Yunoki}}]{nagao2023semiclassical}%
  \BibitemOpen
  \bibfield  {author} {\bibinfo {author} {\bibfnamefont {K.}~\bibnamefont {Nagao}}, \bibinfo {author} {\bibfnamefont {I.}~\bibnamefont {Danshita}},\ and\ \bibinfo {author} {\bibfnamefont {S.}~\bibnamefont {Yunoki}},\ }\href@noop {} {\bibinfo {title} {{Semiclassical descriptions of dissipative dynamics of strongly interacting Bose gases in optical lattices}}} (\bibinfo {year} {2023}),\ \Eprint {https://arxiv.org/abs/2307.16170} {arXiv:2307.16170 [cond-mat.quant-gas]} \BibitemShut {NoStop}%
\bibitem [{\citenamefont {Xiao}\ and\ \citenamefont {Watanabe}(2023)}]{xiao2023local}%
  \BibitemOpen
  \bibfield  {author} {\bibinfo {author} {\bibfnamefont {T.}~\bibnamefont {Xiao}}\ and\ \bibinfo {author} {\bibfnamefont {G.}~\bibnamefont {Watanabe}},\ }\href@noop {} {\bibinfo {title} {{Local Non-Hermitian Hamiltonian Formalism for Dissipative Fermionic Systems and Loss-Induced Population Increase in Fermi Superfluids}}} (\bibinfo {year} {2023}),\ \Eprint {https://arxiv.org/abs/2306.16235} {arXiv:2306.16235 [cond-mat.quant-gas]} \BibitemShut {NoStop}%
\bibitem [{\citenamefont {Tajima}\ \emph {et~al.}(2023)\citenamefont {Tajima}, \citenamefont {Sekino}, \citenamefont {Inotani}, \citenamefont {Dohi}, \citenamefont {Nagataki},\ and\ \citenamefont {Hayata}}]{tajima2023non}%
  \BibitemOpen
  \bibfield  {author} {\bibinfo {author} {\bibfnamefont {H.}~\bibnamefont {Tajima}}, \bibinfo {author} {\bibfnamefont {Y.}~\bibnamefont {Sekino}}, \bibinfo {author} {\bibfnamefont {D.}~\bibnamefont {Inotani}}, \bibinfo {author} {\bibfnamefont {A.}~\bibnamefont {Dohi}}, \bibinfo {author} {\bibfnamefont {S.}~\bibnamefont {Nagataki}},\ and\ \bibinfo {author} {\bibfnamefont {T.}~\bibnamefont {Hayata}},\ }\bibfield  {title} {\bibinfo {title} {{Non-Hermitian topological Fermi superfluid near the $p$-wave unitary limit}},\ }\href {https://doi.org/10.1103/PhysRevA.107.033331} {\bibfield  {journal} {\bibinfo  {journal} {Phys. Rev. A}\ }\textbf {\bibinfo {volume} {107}},\ \bibinfo {pages} {033331} (\bibinfo {year} {2023})}\BibitemShut {NoStop}%
\bibitem [{\citenamefont {Maki}\ \emph {et~al.}(2024)\citenamefont {Maki}, \citenamefont {Rosso}, \citenamefont {Mazza},\ and\ \citenamefont {Biella}}]{maki2024loss}%
  \BibitemOpen
  \bibfield  {author} {\bibinfo {author} {\bibfnamefont {J.}~\bibnamefont {Maki}}, \bibinfo {author} {\bibfnamefont {L.}~\bibnamefont {Rosso}}, \bibinfo {author} {\bibfnamefont {L.}~\bibnamefont {Mazza}},\ and\ \bibinfo {author} {\bibfnamefont {A.}~\bibnamefont {Biella}},\ }\href@noop {} {\bibinfo {title} {{Loss induced collective mode in one-dimensional Bose gases}}} (\bibinfo {year} {2024}),\ \Eprint {https://arxiv.org/abs/2402.05824} {arXiv:2402.05824 [cond-mat.quant-gas]} \BibitemShut {NoStop}%
\bibitem [{\citenamefont {Syassen}\ \emph {et~al.}(2008)\citenamefont {Syassen}, \citenamefont {Bauer}, \citenamefont {Lettner}, \citenamefont {Volz}, \citenamefont {Dietze}, \citenamefont {Garc{\'\i}a-Ripoll}, \citenamefont {Cirac}, \citenamefont {Rempe},\ and\ \citenamefont {D{\"u}rr}}]{Syassen_2008}%
  \BibitemOpen
  \bibfield  {author} {\bibinfo {author} {\bibfnamefont {N.}~\bibnamefont {Syassen}}, \bibinfo {author} {\bibfnamefont {D.~M.}\ \bibnamefont {Bauer}}, \bibinfo {author} {\bibfnamefont {M.}~\bibnamefont {Lettner}}, \bibinfo {author} {\bibfnamefont {T.}~\bibnamefont {Volz}}, \bibinfo {author} {\bibfnamefont {D.}~\bibnamefont {Dietze}}, \bibinfo {author} {\bibfnamefont {J.~J.}\ \bibnamefont {Garc{\'\i}a-Ripoll}}, \bibinfo {author} {\bibfnamefont {J.~I.}\ \bibnamefont {Cirac}}, \bibinfo {author} {\bibfnamefont {G.}~\bibnamefont {Rempe}},\ and\ \bibinfo {author} {\bibfnamefont {S.}~\bibnamefont {D{\"u}rr}},\ }\bibfield  {title} {\bibinfo {title} {Strong dissipation inhibits losses and induces correlations in cold molecular gases},\ }\href {https://doi.org/10.1126/science.1155309} {\bibfield  {journal} {\bibinfo  {journal} {Science}\ }\textbf {\bibinfo {volume} {320}},\ \bibinfo {pages} {1329} (\bibinfo {year} {2008})}\BibitemShut {NoStop}%
\bibitem [{\citenamefont {Yan}\ \emph {et~al.}(2013)\citenamefont {Yan}, \citenamefont {Moses}, \citenamefont {Gadway}, \citenamefont {Covey}, \citenamefont {Hazzard}, \citenamefont {Rey}, \citenamefont {Jin},\ and\ \citenamefont {Ye}}]{Yan_2013}%
  \BibitemOpen
  \bibfield  {author} {\bibinfo {author} {\bibfnamefont {B.}~\bibnamefont {Yan}}, \bibinfo {author} {\bibfnamefont {S.~A.}\ \bibnamefont {Moses}}, \bibinfo {author} {\bibfnamefont {B.}~\bibnamefont {Gadway}}, \bibinfo {author} {\bibfnamefont {J.~P.}\ \bibnamefont {Covey}}, \bibinfo {author} {\bibfnamefont {K.~R.~A.}\ \bibnamefont {Hazzard}}, \bibinfo {author} {\bibfnamefont {A.~M.}\ \bibnamefont {Rey}}, \bibinfo {author} {\bibfnamefont {D.~S.}\ \bibnamefont {Jin}},\ and\ \bibinfo {author} {\bibfnamefont {J.}~\bibnamefont {Ye}},\ }\bibfield  {title} {\bibinfo {title} {Observation of dipolar spin-exchange interactions with lattice-confined polar molecules},\ }\href {https://doi.org/10.1038/nature12483} {\bibfield  {journal} {\bibinfo  {journal} {Nature}\ }\textbf {\bibinfo {volume} {501}},\ \bibinfo {pages} {521–525} (\bibinfo {year} {2013})}\BibitemShut {NoStop}%
\bibitem [{\citenamefont {Zhu}\ \emph {et~al.}(2014)\citenamefont {Zhu}, \citenamefont {Gadway}, \citenamefont {Foss-Feig}, \citenamefont {Schachenmayer}, \citenamefont {Wall}, \citenamefont {Hazzard}, \citenamefont {Yan}, \citenamefont {Moses}, \citenamefont {Covey}, \citenamefont {Jin}, \citenamefont {Ye}, \citenamefont {Holland},\ and\ \citenamefont {Rey}}]{Zhu_2014}%
  \BibitemOpen
  \bibfield  {author} {\bibinfo {author} {\bibfnamefont {B.}~\bibnamefont {Zhu}}, \bibinfo {author} {\bibfnamefont {B.}~\bibnamefont {Gadway}}, \bibinfo {author} {\bibfnamefont {M.}~\bibnamefont {Foss-Feig}}, \bibinfo {author} {\bibfnamefont {J.}~\bibnamefont {Schachenmayer}}, \bibinfo {author} {\bibfnamefont {M.~L.}\ \bibnamefont {Wall}}, \bibinfo {author} {\bibfnamefont {K.~R.~A.}\ \bibnamefont {Hazzard}}, \bibinfo {author} {\bibfnamefont {B.}~\bibnamefont {Yan}}, \bibinfo {author} {\bibfnamefont {S.~A.}\ \bibnamefont {Moses}}, \bibinfo {author} {\bibfnamefont {J.~P.}\ \bibnamefont {Covey}}, \bibinfo {author} {\bibfnamefont {D.~S.}\ \bibnamefont {Jin}}, \bibinfo {author} {\bibfnamefont {J.}~\bibnamefont {Ye}}, \bibinfo {author} {\bibfnamefont {M.}~\bibnamefont {Holland}},\ and\ \bibinfo {author} {\bibfnamefont {A.~M.}\ \bibnamefont {Rey}},\ }\bibfield  {title} {\bibinfo {title} {Suppressing the loss of ultracold molecules via the continuous quantum zeno effect},\ }\href
  {https://doi.org/10.1103/PhysRevLett.112.070404} {\bibfield  {journal} {\bibinfo  {journal} {Phys. Rev. Lett.}\ }\textbf {\bibinfo {volume} {112}},\ \bibinfo {pages} {070404} (\bibinfo {year} {2014})}\BibitemShut {NoStop}%
\bibitem [{\citenamefont {Rauer}\ \emph {et~al.}(2016)\citenamefont {Rauer}, \citenamefont {Gri\ifmmode~\check{s}\else \v{s}\fi{}ins}, \citenamefont {Mazets}, \citenamefont {Schweigler}, \citenamefont {Rohringer}, \citenamefont {Geiger}, \citenamefont {Langen},\ and\ \citenamefont {Schmiedmayer}}]{Rauer_2016}%
  \BibitemOpen
  \bibfield  {author} {\bibinfo {author} {\bibfnamefont {B.}~\bibnamefont {Rauer}}, \bibinfo {author} {\bibfnamefont {P.}~\bibnamefont {Gri\ifmmode~\check{s}\else \v{s}\fi{}ins}}, \bibinfo {author} {\bibfnamefont {I.~E.}\ \bibnamefont {Mazets}}, \bibinfo {author} {\bibfnamefont {T.}~\bibnamefont {Schweigler}}, \bibinfo {author} {\bibfnamefont {W.}~\bibnamefont {Rohringer}}, \bibinfo {author} {\bibfnamefont {R.}~\bibnamefont {Geiger}}, \bibinfo {author} {\bibfnamefont {T.}~\bibnamefont {Langen}},\ and\ \bibinfo {author} {\bibfnamefont {J.}~\bibnamefont {Schmiedmayer}},\ }\bibfield  {title} {\bibinfo {title} {{Cooling of a One-Dimensional Bose Gas}},\ }\href {https://doi.org/10.1103/PhysRevLett.116.030402} {\bibfield  {journal} {\bibinfo  {journal} {Phys. Rev. Lett.}\ }\textbf {\bibinfo {volume} {116}},\ \bibinfo {pages} {030402} (\bibinfo {year} {2016})}\BibitemShut {NoStop}%
\bibitem [{\citenamefont {Tomita}\ \emph {et~al.}(2017)\citenamefont {Tomita}, \citenamefont {Nakajima}, \citenamefont {Danshita}, \citenamefont {Takasu},\ and\ \citenamefont {Takahashi}}]{Tomita_2017}%
  \BibitemOpen
  \bibfield  {author} {\bibinfo {author} {\bibfnamefont {T.}~\bibnamefont {Tomita}}, \bibinfo {author} {\bibfnamefont {S.}~\bibnamefont {Nakajima}}, \bibinfo {author} {\bibfnamefont {I.}~\bibnamefont {Danshita}}, \bibinfo {author} {\bibfnamefont {Y.}~\bibnamefont {Takasu}},\ and\ \bibinfo {author} {\bibfnamefont {Y.}~\bibnamefont {Takahashi}},\ }\bibfield  {title} {\bibinfo {title} {{Observation of the Mott insulator to superfluid crossover of a driven-dissipative Bose-Hubbard system}},\ }\href {https://doi.org/10.1126/sciadv.1701513} {\bibfield  {journal} {\bibinfo  {journal} {Sci. Adv.}\ }\textbf {\bibinfo {volume} {3}},\ \bibinfo {pages} {e1701513} (\bibinfo {year} {2017})}\BibitemShut {NoStop}%
\bibitem [{\citenamefont {Schemmer}\ and\ \citenamefont {Bouchoule}(2018)}]{Schemmer_2018}%
  \BibitemOpen
  \bibfield  {author} {\bibinfo {author} {\bibfnamefont {M.}~\bibnamefont {Schemmer}}\ and\ \bibinfo {author} {\bibfnamefont {I.}~\bibnamefont {Bouchoule}},\ }\bibfield  {title} {\bibinfo {title} {Cooling a bose gas by three-body losses},\ }\href {https://doi.org/10.1103/PhysRevLett.121.200401} {\bibfield  {journal} {\bibinfo  {journal} {Phys. Rev. Lett.}\ }\textbf {\bibinfo {volume} {121}},\ \bibinfo {pages} {200401} (\bibinfo {year} {2018})}\BibitemShut {NoStop}%
\bibitem [{\citenamefont {Sponselee}\ \emph {et~al.}(2019)\citenamefont {Sponselee}, \citenamefont {Freystatzky}, \citenamefont {Abeln}, \citenamefont {Diem}, \citenamefont {Hundt}, \citenamefont {Kochanke}, \citenamefont {Ponath}, \citenamefont {Santra}, \citenamefont {Mathey}, \citenamefont {Sengstock},\ and\ \citenamefont {Becker}}]{Sponselee_2018}%
  \BibitemOpen
  \bibfield  {author} {\bibinfo {author} {\bibfnamefont {K.}~\bibnamefont {Sponselee}}, \bibinfo {author} {\bibfnamefont {L.}~\bibnamefont {Freystatzky}}, \bibinfo {author} {\bibfnamefont {B.}~\bibnamefont {Abeln}}, \bibinfo {author} {\bibfnamefont {M.}~\bibnamefont {Diem}}, \bibinfo {author} {\bibfnamefont {B.}~\bibnamefont {Hundt}}, \bibinfo {author} {\bibfnamefont {A.}~\bibnamefont {Kochanke}}, \bibinfo {author} {\bibfnamefont {T.}~\bibnamefont {Ponath}}, \bibinfo {author} {\bibfnamefont {B.}~\bibnamefont {Santra}}, \bibinfo {author} {\bibfnamefont {L.}~\bibnamefont {Mathey}}, \bibinfo {author} {\bibfnamefont {K.}~\bibnamefont {Sengstock}},\ and\ \bibinfo {author} {\bibfnamefont {C.}~\bibnamefont {Becker}},\ }\bibfield  {title} {\bibinfo {title} {{Dynamics of ultracold quantum gases in the dissipative Fermi-Hubbard model}},\ }\href {https://doi.org/10.1088/2058-9565/aadccd} {\bibfield  {journal} {\bibinfo  {journal} {Quantum Sci. Technol.}\ }\textbf {\bibinfo {volume} {4}},\ \bibinfo {pages} {014002}
  (\bibinfo {year} {2019})}\BibitemShut {NoStop}%
\bibitem [{\citenamefont {Tomita}\ \emph {et~al.}(2019)\citenamefont {Tomita}, \citenamefont {Nakajima}, \citenamefont {Takasu},\ and\ \citenamefont {Takahashi}}]{Tomita_2019}%
  \BibitemOpen
  \bibfield  {author} {\bibinfo {author} {\bibfnamefont {T.}~\bibnamefont {Tomita}}, \bibinfo {author} {\bibfnamefont {S.}~\bibnamefont {Nakajima}}, \bibinfo {author} {\bibfnamefont {Y.}~\bibnamefont {Takasu}},\ and\ \bibinfo {author} {\bibfnamefont {Y.}~\bibnamefont {Takahashi}},\ }\bibfield  {title} {\bibinfo {title} {{Dissipative Bose-Hubbard system with intrinsic two-body loss}},\ }\href {https://doi.org/10.1103/PhysRevA.99.031601} {\bibfield  {journal} {\bibinfo  {journal} {Phys. Rev. A}\ }\textbf {\bibinfo {volume} {99}},\ \bibinfo {pages} {031601(R)} (\bibinfo {year} {2019})}\BibitemShut {NoStop}%
\bibitem [{\citenamefont {Bouchoule}\ and\ \citenamefont {Schemmer}(2020)}]{Bouchoule_Schemmer_2020}%
  \BibitemOpen
  \bibfield  {author} {\bibinfo {author} {\bibfnamefont {I.}~\bibnamefont {Bouchoule}}\ and\ \bibinfo {author} {\bibfnamefont {M.}~\bibnamefont {Schemmer}},\ }\bibfield  {title} {\bibinfo {title} {{Asymptotic temperature of a lossy condensate}},\ }\href {https://doi.org/10.21468/SciPostPhys.8.4.060} {\bibfield  {journal} {\bibinfo  {journal} {SciPost Phys.}\ }\textbf {\bibinfo {volume} {8}},\ \bibinfo {pages} {060} (\bibinfo {year} {2020})}\BibitemShut {NoStop}%
\bibitem [{\citenamefont {Honda}\ \emph {et~al.}(2023)\citenamefont {Honda}, \citenamefont {Taie}, \citenamefont {Takasu}, \citenamefont {Nishizawa}, \citenamefont {Nakagawa},\ and\ \citenamefont {Takahashi}}]{Honda_2023}%
  \BibitemOpen
  \bibfield  {author} {\bibinfo {author} {\bibfnamefont {K.}~\bibnamefont {Honda}}, \bibinfo {author} {\bibfnamefont {S.}~\bibnamefont {Taie}}, \bibinfo {author} {\bibfnamefont {Y.}~\bibnamefont {Takasu}}, \bibinfo {author} {\bibfnamefont {N.}~\bibnamefont {Nishizawa}}, \bibinfo {author} {\bibfnamefont {M.}~\bibnamefont {Nakagawa}},\ and\ \bibinfo {author} {\bibfnamefont {Y.}~\bibnamefont {Takahashi}},\ }\bibfield  {title} {\bibinfo {title} {{Observation of the Sign Reversal of the Magnetic Correlation in a Driven-Dissipative Fermi Gas in Double Wells}},\ }\href {https://doi.org/10.1103/PhysRevLett.130.063001} {\bibfield  {journal} {\bibinfo  {journal} {Phys. Rev. Lett.}\ }\textbf {\bibinfo {volume} {130}},\ \bibinfo {pages} {063001} (\bibinfo {year} {2023})}\BibitemShut {NoStop}%
\bibitem [{\citenamefont {Rosso}\ \emph {et~al.}(2023)\citenamefont {Rosso}, \citenamefont {Biella}, \citenamefont {De~Nardis},\ and\ \citenamefont {Mazza}}]{Rosso_2023}%
  \BibitemOpen
  \bibfield  {author} {\bibinfo {author} {\bibfnamefont {L.}~\bibnamefont {Rosso}}, \bibinfo {author} {\bibfnamefont {A.}~\bibnamefont {Biella}}, \bibinfo {author} {\bibfnamefont {J.}~\bibnamefont {De~Nardis}},\ and\ \bibinfo {author} {\bibfnamefont {L.}~\bibnamefont {Mazza}},\ }\bibfield  {title} {\bibinfo {title} {{Dynamical theory for one-dimensional fermions with strong two-body losses: Universal non-Hermitian Zeno physics and spin-charge separation}},\ }\href {https://doi.org/10.1103/PhysRevA.107.013303} {\bibfield  {journal} {\bibinfo  {journal} {Phys. Rev. A}\ }\textbf {\bibinfo {volume} {107}},\ \bibinfo {pages} {013303} (\bibinfo {year} {2023})}\BibitemShut {NoStop}%
\bibitem [{\citenamefont {Gerbino}\ \emph {et~al.}(2023)\citenamefont {Gerbino}, \citenamefont {Lesanovsky},\ and\ \citenamefont {Perfetto}}]{gerbino2023largescale}%
  \BibitemOpen
  \bibfield  {author} {\bibinfo {author} {\bibfnamefont {F.}~\bibnamefont {Gerbino}}, \bibinfo {author} {\bibfnamefont {I.}~\bibnamefont {Lesanovsky}},\ and\ \bibinfo {author} {\bibfnamefont {G.}~\bibnamefont {Perfetto}},\ }\href@noop {} {\bibinfo {title} {{Large-scale universality in Quantum Reaction-Diffusion from Keldysh field theory}}} (\bibinfo {year} {2023}),\ \Eprint {https://arxiv.org/abs/2307.14945} {arXiv:2307.14945 [cond-mat.stat-mech]} \BibitemShut {NoStop}%
\bibitem [{\citenamefont {Rosso}\ \emph {et~al.}(2022{\natexlab{b}})\citenamefont {Rosso}, \citenamefont {Biella},\ and\ \citenamefont {Mazza}}]{Rosso_2021bis}%
  \BibitemOpen
  \bibfield  {author} {\bibinfo {author} {\bibfnamefont {L.}~\bibnamefont {Rosso}}, \bibinfo {author} {\bibfnamefont {A.}~\bibnamefont {Biella}},\ and\ \bibinfo {author} {\bibfnamefont {L.}~\bibnamefont {Mazza}},\ }\bibfield  {title} {\bibinfo {title} {{The one-dimensional Bose gas with strong two-body losses: the effect of the harmonic confinement}},\ }\href {https://doi.org/10.21468/SciPostPhys.12.1.044} {\bibfield  {journal} {\bibinfo  {journal} {SciPost Phys.}\ }\textbf {\bibinfo {volume} {12}},\ \bibinfo {pages} {44} (\bibinfo {year} {2022}{\natexlab{b}})}\BibitemShut {NoStop}%
\bibitem [{\citenamefont {Riggio}\ \emph {et~al.}(2024)\citenamefont {Riggio}, \citenamefont {Rosso}, \citenamefont {Karevski},\ and\ \citenamefont {Dubail}}]{Riggio_2024}%
  \BibitemOpen
  \bibfield  {author} {\bibinfo {author} {\bibfnamefont {F.}~\bibnamefont {Riggio}}, \bibinfo {author} {\bibfnamefont {L.}~\bibnamefont {Rosso}}, \bibinfo {author} {\bibfnamefont {D.}~\bibnamefont {Karevski}},\ and\ \bibinfo {author} {\bibfnamefont {J.}~\bibnamefont {Dubail}},\ }\bibfield  {title} {\bibinfo {title} {Effects of atom losses on a one-dimensional lattice gas of hard-core bosons},\ }\href {https://doi.org/10.1103/PhysRevA.109.023311} {\bibfield  {journal} {\bibinfo  {journal} {Phys. Rev. A}\ }\textbf {\bibinfo {volume} {109}},\ \bibinfo {pages} {023311} (\bibinfo {year} {2024})}\BibitemShut {NoStop}%
\bibitem [{\citenamefont {Ashida}\ \emph {et~al.}(2020)\citenamefont {Ashida}, \citenamefont {Gong},\ and\ \citenamefont {Ueda}}]{Ashida_2020}%
  \BibitemOpen
  \bibfield  {author} {\bibinfo {author} {\bibfnamefont {Y.}~\bibnamefont {Ashida}}, \bibinfo {author} {\bibfnamefont {Z.}~\bibnamefont {Gong}},\ and\ \bibinfo {author} {\bibfnamefont {M.}~\bibnamefont {Ueda}},\ }\bibfield  {title} {\bibinfo {title} {Non-hermitian physics},\ }\href {https://doi.org/10.1080/00018732.2021.1876991} {\bibfield  {journal} {\bibinfo  {journal} {Advances in Physics}\ }\textbf {\bibinfo {volume} {69}},\ \bibinfo {pages} {249–435} (\bibinfo {year} {2020})}\BibitemShut {NoStop}%
\bibitem [{\citenamefont {Breuer}\ and\ \citenamefont {Petruccione}(2007)}]{Breuer_2007}%
  \BibitemOpen
  \bibfield  {author} {\bibinfo {author} {\bibfnamefont {H.-P.}\ \bibnamefont {Breuer}}\ and\ \bibinfo {author} {\bibfnamefont {F.}~\bibnamefont {Petruccione}},\ }\href {https://doi.org/10.1093/acprof:oso/9780199213900.001.0001} {\emph {\bibinfo {title} {{The Theory of Open Quantum Systems}}}}\ (\bibinfo  {publisher} {Oxford University Press},\ \bibinfo {year} {2007})\BibitemShut {NoStop}%
\bibitem [{\citenamefont {Lieb}\ and\ \citenamefont {Liniger}(1963)}]{LiebLiniger1963}%
  \BibitemOpen
  \bibfield  {author} {\bibinfo {author} {\bibfnamefont {E.~H.}\ \bibnamefont {Lieb}}\ and\ \bibinfo {author} {\bibfnamefont {W.}~\bibnamefont {Liniger}},\ }\bibfield  {title} {\bibinfo {title} {{Exact Analysis of an Interacting Bose Gas. I. The General Solution and the Ground State}},\ }\href {https://doi.org/10.1103/PhysRev.130.1605} {\bibfield  {journal} {\bibinfo  {journal} {Phys. Rev.}\ }\textbf {\bibinfo {volume} {130}},\ \bibinfo {pages} {1605} (\bibinfo {year} {1963})}\BibitemShut {NoStop}%
\bibitem [{\citenamefont {Franchini}(2017)}]{Franchini_2017}%
  \BibitemOpen
  \bibfield  {author} {\bibinfo {author} {\bibfnamefont {F.}~\bibnamefont {Franchini}},\ }\href {https://doi.org/10.1007/978-3-319-48487-7} {\emph {\bibinfo {title} {An Introduction to Integrable Techniques for One-Dimensional Quantum Systems}}}\ (\bibinfo  {publisher} {Springer International Publishing},\ \bibinfo {year} {2017})\BibitemShut {NoStop}%
\bibitem [{\citenamefont {Gaudin}(1983)}]{Gaudin}%
  \BibitemOpen
  \bibfield  {author} {\bibinfo {author} {\bibfnamefont {M.}~\bibnamefont {Gaudin}},\ }\href@noop {} {\emph {\bibinfo {title} {La fonction d'onde de {B}ethe}}}\ (\bibinfo  {publisher} {Masson, Paris},\ \bibinfo {year} {1983})\ \bibinfo {note} {{\it The Bethe Wavefunction} (translation by J.-S. Caux), Cambridge University Press, 2014.}\BibitemShut {Stop}%
\bibitem [{\citenamefont {Korepin}\ \emph {et~al.}(1997)\citenamefont {Korepin}, \citenamefont {Bogoliubov},\ and\ \citenamefont {Izergin}}]{korepin1997quantum}%
  \BibitemOpen
  \bibfield  {author} {\bibinfo {author} {\bibfnamefont {V.~E.}\ \bibnamefont {Korepin}}, \bibinfo {author} {\bibfnamefont {N.~M.}\ \bibnamefont {Bogoliubov}},\ and\ \bibinfo {author} {\bibfnamefont {A.~G.}\ \bibnamefont {Izergin}},\ }\href@noop {} {\emph {\bibinfo {title} {Quantum inverse scattering method and correlation functions}}},\ Vol.~\bibinfo {volume} {3}\ (\bibinfo  {publisher} {Cambridge University Press},\ \bibinfo {year} {1997})\BibitemShut {NoStop}%
\bibitem [{\citenamefont {Zvonarev}(2010)}]{Zvonarev_2010}%
  \BibitemOpen
  \bibfield  {author} {\bibinfo {author} {\bibfnamefont {M.}~\bibnamefont {Zvonarev}},\ }\href {http://cmt.harvard.edu/demler/TEACHING/Physics284/LectureZvonarev.pdf} {\emph {\bibinfo {title} {{Notes on Bethe Ansatz}}}}\ (\bibinfo {year} {2010})\BibitemShut {NoStop}%
\bibitem [{\citenamefont {{\v{S}}amaj}\ and\ \citenamefont {Bajnok}(2013)}]{vsamaj2013introduction}%
  \BibitemOpen
  \bibfield  {author} {\bibinfo {author} {\bibfnamefont {L.}~\bibnamefont {{\v{S}}amaj}}\ and\ \bibinfo {author} {\bibfnamefont {Z.}~\bibnamefont {Bajnok}},\ }\href@noop {} {\emph {\bibinfo {title} {Introduction to the statistical physics of integrable many-body systems}}}\ (\bibinfo  {publisher} {Cambridge University Press},\ \bibinfo {year} {2013})\BibitemShut {NoStop}%
\bibitem [{\citenamefont {Choy}\ and\ \citenamefont {Haldane}(1982)}]{choy1982failure}%
  \BibitemOpen
  \bibfield  {author} {\bibinfo {author} {\bibfnamefont {T.}~\bibnamefont {Choy}}\ and\ \bibinfo {author} {\bibfnamefont {F.}~\bibnamefont {Haldane}},\ }\bibfield  {title} {\bibinfo {title} {{Failure of Bethe-Ansatz solutions of generalisations of the Hubbard chain to arbitrary permutation symmetry}},\ }\href {https://doi.org/10.1016/0375-9601(82)90057-3} {\bibfield  {journal} {\bibinfo  {journal} {Phys. Lett. A}\ }\textbf {\bibinfo {volume} {90}},\ \bibinfo {pages} {83} (\bibinfo {year} {1982})}\BibitemShut {NoStop}%
\bibitem [{\citenamefont {Medvedyeva}\ \emph {et~al.}(2016)\citenamefont {Medvedyeva}, \citenamefont {Essler},\ and\ \citenamefont {Prosen}}]{medvedyeva2016exact}%
  \BibitemOpen
  \bibfield  {author} {\bibinfo {author} {\bibfnamefont {M.~V.}\ \bibnamefont {Medvedyeva}}, \bibinfo {author} {\bibfnamefont {F.~H.~L.}\ \bibnamefont {Essler}},\ and\ \bibinfo {author} {\bibfnamefont {T.}~\bibnamefont {Prosen}},\ }\bibfield  {title} {\bibinfo {title} {{Exact Bethe ansatz spectrum of a tight-binding chain with dephasing noise}},\ }\href {https://doi.org/10.1103/PhysRevLett.117.137202} {\bibfield  {journal} {\bibinfo  {journal} {Phys. Rev. Lett.}\ }\textbf {\bibinfo {volume} {117}},\ \bibinfo {pages} {137202} (\bibinfo {year} {2016})}\BibitemShut {NoStop}%
\bibitem [{\citenamefont {Alba}(2023)}]{alba2023free}%
  \BibitemOpen
  \bibfield  {author} {\bibinfo {author} {\bibfnamefont {V.}~\bibnamefont {Alba}},\ }\href@noop {} {\bibinfo {title} {Free fermions with dephasing and boundary driving: Bethe ansatz results}} (\bibinfo {year} {2023}),\ \Eprint {https://arxiv.org/abs/2309.12978} {arXiv:2309.12978 [cond-mat.stat-mech]} \BibitemShut {NoStop}%
\bibitem [{\citenamefont {Essler}\ \emph {et~al.}(2005)\citenamefont {Essler}, \citenamefont {Frahm}, \citenamefont {G{\"o}hmann}, \citenamefont {Kl{\"u}mper},\ and\ \citenamefont {Korepin}}]{essler_2005}%
  \BibitemOpen
  \bibfield  {author} {\bibinfo {author} {\bibfnamefont {F.~H.~L.}\ \bibnamefont {Essler}}, \bibinfo {author} {\bibfnamefont {H.}~\bibnamefont {Frahm}}, \bibinfo {author} {\bibfnamefont {F.}~\bibnamefont {G{\"o}hmann}}, \bibinfo {author} {\bibfnamefont {A.}~\bibnamefont {Kl{\"u}mper}},\ and\ \bibinfo {author} {\bibfnamefont {V.~E.}\ \bibnamefont {Korepin}},\ }\href {https://doi.org/10.1017/CBO9780511534843} {\emph {\bibinfo {title} {The one-dimensional Hubbard model}}}\ (\bibinfo  {publisher} {Cambridge University Press},\ \bibinfo {year} {2005})\BibitemShut {NoStop}%
\bibitem [{\citenamefont {Zhi}\ and\ \citenamefont {Hong}(1991)}]{zhi1991pseudospin}%
  \BibitemOpen
  \bibfield  {author} {\bibinfo {author} {\bibfnamefont {S.}~\bibnamefont {Zhi}}\ and\ \bibinfo {author} {\bibfnamefont {S.}~\bibnamefont {Hong}},\ }\bibfield  {title} {\bibinfo {title} {{Pseudospin Wave State in One-Dimensional Half-Filled Hubbard Model}},\ }\href {https://doi.org/10.1088/0253-6102/16/4/485} {\bibfield  {journal} {\bibinfo  {journal} {Commun. Theor. Phys.}\ }\textbf {\bibinfo {volume} {16}},\ \bibinfo {pages} {485} (\bibinfo {year} {1991})}\BibitemShut {NoStop}%
\bibitem [{\citenamefont {Zhang}\ and\ \citenamefont {Song}(2021)}]{zhang2021eta}%
  \BibitemOpen
  \bibfield  {author} {\bibinfo {author} {\bibfnamefont {X.~Z.}\ \bibnamefont {Zhang}}\ and\ \bibinfo {author} {\bibfnamefont {Z.}~\bibnamefont {Song}},\ }\bibfield  {title} {\bibinfo {title} {{$\eta$-pairing ground states in the non-Hermitian Hubbard model}},\ }\href {https://doi.org/10.1103/PhysRevB.103.235153} {\bibfield  {journal} {\bibinfo  {journal} {Phys. Rev. B}\ }\textbf {\bibinfo {volume} {103}},\ \bibinfo {pages} {235153} (\bibinfo {year} {2021})}\BibitemShut {NoStop}%
\bibitem [{\citenamefont {Oelkers}\ and\ \citenamefont {Links}(2007)}]{Oelkers_2007}%
  \BibitemOpen
  \bibfield  {author} {\bibinfo {author} {\bibfnamefont {N.}~\bibnamefont {Oelkers}}\ and\ \bibinfo {author} {\bibfnamefont {J.}~\bibnamefont {Links}},\ }\bibfield  {title} {\bibinfo {title} {{Ground-state properties of the attractive one-dimensional Bose-Hubbard model}},\ }\href {https://doi.org/10.1103/PhysRevB.75.115119} {\bibfield  {journal} {\bibinfo  {journal} {Phys. Rev. B}\ }\textbf {\bibinfo {volume} {75}},\ \bibinfo {pages} {115119} (\bibinfo {year} {2007})}\BibitemShut {NoStop}%
\bibitem [{\citenamefont {Li}\ \emph {et~al.}(2022)\citenamefont {Li}, \citenamefont {Schneble},\ and\ \citenamefont {Wei}}]{Li_2022}%
  \BibitemOpen
  \bibfield  {author} {\bibinfo {author} {\bibfnamefont {Y.}~\bibnamefont {Li}}, \bibinfo {author} {\bibfnamefont {D.}~\bibnamefont {Schneble}},\ and\ \bibinfo {author} {\bibfnamefont {T.-C.}\ \bibnamefont {Wei}},\ }\bibfield  {title} {\bibinfo {title} {{Two-particle states in one-dimensional coupled Bose-Hubbard models}},\ }\href {https://doi.org/10.1103/PhysRevA.105.053310} {\bibfield  {journal} {\bibinfo  {journal} {Phys. Rev. A}\ }\textbf {\bibinfo {volume} {105}},\ \bibinfo {pages} {053310} (\bibinfo {year} {2022})}\BibitemShut {NoStop}%
\bibitem [{\citenamefont {Tanaka}\ \emph {et~al.}(2013)\citenamefont {Tanaka}, \citenamefont {Yonezawa},\ and\ \citenamefont {Cheon}}]{Tanaka_2013}%
  \BibitemOpen
  \bibfield  {author} {\bibinfo {author} {\bibfnamefont {A.}~\bibnamefont {Tanaka}}, \bibinfo {author} {\bibfnamefont {N.}~\bibnamefont {Yonezawa}},\ and\ \bibinfo {author} {\bibfnamefont {T.}~\bibnamefont {Cheon}},\ }\bibfield  {title} {\bibinfo {title} {{Exotic quantum holonomy and non-Hermitian degeneracies in the two-body Lieb–Liniger model}},\ }\href {https://doi.org/10.1088/1751-8113/46/31/315302} {\bibfield  {journal} {\bibinfo  {journal} {J. Phys. A: Math. Theor.}\ }\textbf {\bibinfo {volume} {46}},\ \bibinfo {pages} {315302} (\bibinfo {year} {2013})}\BibitemShut {NoStop}%
\bibitem [{\citenamefont {Yang}(1989)}]{Yang_1989}%
  \BibitemOpen
  \bibfield  {author} {\bibinfo {author} {\bibfnamefont {C.~N.}\ \bibnamefont {Yang}},\ }\bibfield  {title} {\bibinfo {title} {{\ensuremath{\eta} pairing and off-diagonal long-range order in a Hubbard model}},\ }\href {https://doi.org/10.1103/PhysRevLett.63.2144} {\bibfield  {journal} {\bibinfo  {journal} {Phys. Rev. Lett.}\ }\textbf {\bibinfo {volume} {63}},\ \bibinfo {pages} {2144} (\bibinfo {year} {1989})}\BibitemShut {NoStop}%
\bibitem [{\citenamefont {Giamarchi}(2003)}]{giam}%
  \BibitemOpen
  \bibfield  {author} {\bibinfo {author} {\bibfnamefont {T.}~\bibnamefont {Giamarchi}},\ }\href {https://doi.org/10.1093/acprof:oso/9780198525004.001.0001} {\emph {\bibinfo {title} {{{Quantum Physics in One Dimension}}}}}\ (\bibinfo  {publisher} {Oxford University Press},\ \bibinfo {year} {2003})\BibitemShut {NoStop}%
\bibitem [{\citenamefont {Ogata}\ and\ \citenamefont {Shiba}(1990)}]{ogata_1990}%
  \BibitemOpen
  \bibfield  {author} {\bibinfo {author} {\bibfnamefont {M.}~\bibnamefont {Ogata}}\ and\ \bibinfo {author} {\bibfnamefont {H.}~\bibnamefont {Shiba}},\ }\bibfield  {title} {\bibinfo {title} {{Bethe-ansatz wave function, momentum distribution, and spin correlation in the one-dimensional strongly correlated Hubbard model}},\ }\href {https://doi.org/10.1103/PhysRevB.41.2326} {\bibfield  {journal} {\bibinfo  {journal} {Phys. Rev. B}\ }\textbf {\bibinfo {volume} {41}},\ \bibinfo {pages} {2326} (\bibinfo {year} {1990})}\BibitemShut {NoStop}%
\bibitem [{\citenamefont {Abramowitz}(1974)}]{Abramowitz_1974}%
  \BibitemOpen
  \bibfield  {author} {\bibinfo {author} {\bibfnamefont {M.}~\bibnamefont {Abramowitz}},\ }\href@noop {} {\emph {\bibinfo {title} {Handbook of Mathematical Functions, With Formulas, Graphs, and Mathematical Tables,}}}\ (\bibinfo  {publisher} {Dover Publications, Inc.},\ \bibinfo {address} {USA},\ \bibinfo {year} {1974})\BibitemShut {NoStop}%
\end{thebibliography}%
\end{document}